\documentclass[showpacs,preprintnumbers,amsmath,amssymb]{revtex4}

\usepackage{graphicx}
\usepackage{dcolumn}
\usepackage{bm}
\usepackage{xcolor}

\newcommand{\bq}{\begin{equation}}
\newcommand{\eq}{\end{equation}}
\newcommand{\bqa}{\begin{eqnarray}}
\newcommand{\eqa}{\end{eqnarray}}
\newcommand{\nn}{\nonumber \\}

\def\be     {\begin{equation}}
\def\ee     {\end{equation}}
\def\bea        {\begin{eqnarray}}
\def\eea        {\end{eqnarray}}
\def\bnn    {\begin{eqnarray*}}
\def\enn    {\end{eqnarray*}}

\begin{document}

\title{\boldmath Monotonicity of RG flow in emergent dual holography of worldsheet nonlinear $\sigma$ model}

\author{Ki-Seok Kim$^{a,b}$, Arpita Mitra$^{a}$, Debangshu Mukherjee$^{b}$, and Shinsei Ryu$^{c}$}
	
	\affiliation{$^{a}$Department of Physics, POSTECH, Pohang, Gyeongbuk 37673, Korea \\ $^{b}$Asia Pacific Center for Theoretical Physics (APCTP), Pohang, Gyeongbuk 37673, Korea \\ $^{c}$Department of Physics, Princeton University, Princeton, New Jersey 08540, USA}
	
	\email[Ki-Seok Kim: ]{tkfkd@postech.ac.kr}
	\email[Arpita Mitra: ]{arpitamitra89@gmail.com}
	\email[Debangshu Mukherjee:]{debangshu0@gmail.com}	
	\email[Shinsei Ryu: ]{shinseir@princeton.edu}

\date{\today}

\begin{abstract}
Based on the renormalization group (RG) flow of worldsheet bosonic string theory, we construct an effective holographic dual description of the target space theory identifying the RG scale with the emergent extra dimension. This results in an effective dilaton-gravity-gauge theory, analogous to the low-energy description of bosonic M-theory. We argue that this holographic dual effective field theory is non-perturbative in the $\alpha'$ expansion, where a class of string quantum fluctuations are resummed to all orders. To investigate the monotonicity of
the RG flow of the target space metric in the emergent spacetime, we consider entropy production along the RG flow. We construct a microscopic entropy functional based on the probability distribution function of the holographic dual effective field theory, regarded as Gibbs- or Shannon-type entropy. Given that the Ricci flow represents the 1-loop RG flow equation of the target space metric for the 2D non-linear sigma model, and motivated by Perelman's proof of the monotonicity of Ricci flow, we propose a Perelman's entropy functional for the holographic dual effective field theory. This entropy functional is also non-perturbative in the $\alpha'$ expansion, and thus, generalizes the 1-loop result to the all-loop order. Furthermore, utilizing the equivalence between the Hamilton-Jacobi equation and the local RG equation, we suggest that the RG flow of holographic Perelman's entropy functional is the Weyl anomaly. This eventually reaffirms the monotonicity of RG flow for the emergent target spacetime but in a non-perturbative way. Interestingly, we find that the microscopic entropy production rate can be determined by integrating the rate of change of the holographic Perelman's entropy functional over all possible metric configurations along the flow.
\end{abstract}


\maketitle

	\section{Introduction}
	
%
%
One of the fundamental challenges in quantum field theories (QFTs) is finding an effective low-energy description for strongly coupled theories.
Strong-weak duality, which transforms `electric' degrees of freedom into `magnetic' ones essentially, serves as a promising way for addressing this problem \cite{SW_Theory,SW_Theory_Review}. 
Another type of strong-weak duality is the holographic dual description \cite{Holographic_Duality_I,Holographic_Duality_II,Holographic_Duality_III,Holographic_Duality_IV}, where strong interactions are introduced by the renormalization group (RG) flows of collective dual fields \cite{Holographic_Duality_V,Holographic_Duality_VI,Holographic_Duality_VII}. 
These RG flow equations can be made manifest at the level of an effective action, where an emergent extra dimension is identified as an RG scale parameter \cite{SungSik_Holography_I,SungSik_Holography_II,SungSik_Holography_III, RG_Flow_Holography_Monotonicity, Emergent_AdS2_BH_RG,Nonperturbative_Wilson_RG_Disorder,Nonperturbative_Wilson_RG,Einstein_Klein_Gordon_RG_Kim,Einstein_Dirac_RG_Kim,RG_GR_Geometry_I_Kim,RG_GR_Geometry_II_Kim,Kondo_Holography_Kim,Kitaev_Entanglement_Entropy_Kim,RG_Holography_First_Kim}. 
As a result, such collective dual fields serve as semiclassical backgrounds for original degrees of freedom, i.e., UV quantum fields. 
Diagrammatically, not only self-energy corrections given by the RG flows of either scalar or electromagnetic fields but also vertex corrections described by the RG flows of gravitational fields are self-consistently introduced to form coupled differential equations in the semiclassical (strong coupling) regime of dual holography (QFTs).
	
In general, to verify this fascinating strong-weak duality conjecture, entropy has been investigated to count the number of all accessible quantum microstates and to show the correspondence between the microstates of QFTs and those of 
their holographic duals.
In this regard, the black hole entropy problem has been studied extensively \cite{BH_entropy_0,BH_entropy_I,BH_entropy_II,BH_entropy_III,BH_entropy_IV,BH_entropy_V,BH_entropy_VI}. 
The Bekenstein-Hawking entropy formula, i.e., the area law of the black hole entropy \cite{BH_entropy_0,BH_entropy_I,BH_entropy_II,BH_entropy_III,BH_entropy_IV} is modified by quantum gravity corrections \cite{BH_entropy_string_theory_Review_I,BH_entropy_string_theory_Review_II}. 
Not only bulk gravitational path integrals in anti-de Sitter space (AdS) but also conformal field theory (CFT) calculations are shown to give the same entropy formula, that is, the leading area law with the universal subleading logarithmic correction, higher-order perturbative corrections in the $1/N$ expansion, and even non-perturbative quantum corrections \cite{BH_entropy_string_theory_I,BH_entropy_string_theory_II,BH_entropy_string_theory_III,BH_entropy_string_theory_IV,BH_entropy_string_theory_V,BH_entropy_string_theory_VI,BH_entropy_string_theory_VIII}.
	%
	%

Entanglement entropy has also been playing a central role in understanding the holographic duality conjecture \cite{Entanglement_Entropy_Calabrese_Cardy_I,Entanglement_Entropy_Calabrese_Cardy_II,Entanglement_Entropy_Ryu_Takayanagi_I,Entanglement_Entropy_Ryu_Takayanagi_II,Entanglement_Entropy_Review_III,Entanglement_Entropy_Review_IV}. 
The so-called Ryu-Takayanagi formula was shown to match the CFT entanglement entropy \cite{Entanglement_Entropy_Ryu_Takayanagi_I,Entanglement_Entropy_Ryu_Takayanagi_II,Entanglement_Entropy_Review_III,Entanglement_Entropy_Review_IV}, while the Ryu-Takayanagi formula itself has been proved rigorously within the holographic duality conjecture \cite{Entanglement_Entropy_Proof_I,Entanglement_Entropy_Proof_II}.

%
%
The RG flow is a crucial tool that helps in the understanding of the effective low-energy description of generic QFTs. 
The most fundamental property of the RG flow is monotonicity from UV to IR under relevant perturbations at the UV fixed point. The monotonicity of the RG flow was proven in the context of two-dimensional conformal field theories by A. B. Zamolodchikov, referred to as 
{\it $c$-theorem} \cite{c_theorem}.
This $c$-theorem has been generalized 
to {\it $a$-theorem} in four dimensions, conjectured by J. Cardy  \cite{a_theorem}. 
Subsequently, the $a$-theorem was proven in ref.\ \cite{a_f_theorem_i} from the physics perspectives. 
These RG monotonicity theorems have also been extended into three dimensions, referred to as {\it $F$-theorem} \cite{f_theorem_SUSY,f_theorem_noSUSY,a_f_theorem_ii}. Rather interestingly, the RG monotonicity property in general dimensions has been understood in the perspectives of entanglement entropy \cite{a_f_theorem_EE_i,a_f_theorem_EE_ii,a_f_theorem_EE_iii,a_f_theorem_EE_iv,a_f_theorem_EE_v,c_theorem_holography_i,c_theorem_holography_ii}. 
More recently, the RG flow has been investigated from the perspectives of relative entropy \cite{RG_Flow_Relative_Entropy_I,RG_Flow_Relative_Entropy_II,RG_Flow_Relative_Entropy_Gradient_Flow}, although the connection between the relative entropy and the RG flow had been discussed sometime before \cite{RG_Flow_Relative_Entropy_0}. 
In particular, ref.\ \cite{RG_Flow_Relative_Entropy_Gradient_Flow} proposed that Polchinski's equation for exact RG flow is equivalent to the optimal transport gradient flow of a field-theoretic relative entropy.

In this paper, we revisit the monotonicity of the RG flow based on non-linear sigma models (NLSMs). In fact, NLSMs are a rich playground where the coupling space is parametrized by a Riemannian manifold. 
The 1-loop RG flow equations 
for NLSMs \cite{Ricci_Flow_I} have been shown to be equivalent to the Ricci flow equations introduced by Hamilton \cite{Ricci_Flow_0}. Subsequently, with the advent of the AdS/CFT correspondence, the RG flow of QFTs with gravity duals and their beta functions were studied extensively. It was demonstrated that the holographic RG flow is equivalent to the Ricci flow where the extradimensional coordinate of the bulk theory is dual to the energy scale of the boundary QFT and plays the role of time in the evolution of the geometry from UV to IR \cite{Holographic_Duality_V, Holographic_Duality_VI, Holographic_Duality_VII,Holographic_RG_Flow_Ricci_Flow_I, Holographic_RG_Flow_Ricci_Flow_II}.
	%
	%
Moreover, the Ricci flow 
\cite{Ricci_Flow_0, Ricci_Flow_I, Ricci_Flow_II, Yu_Nakamura_Ricci_Flow} 
has been shown to be a  gradient flow \cite{Ricci_Flow_Monotonicity}, where the evolution of a Riemannian metric is given by a gradient of a functional. 
Indeed, Perelman constructed the so-called `entropy' functional and demonstrated that the Ricci flow is a gradient flow with positive definite metric, extremizing his entropy functional \cite{Ricci_Flow_Monotonicity, Ricci_NLsM_Gradient_i, Ricci_NLsM_Gradient_ii}. 
Here, Perelman's entropy functional is quite analogous to Zamolodchikov's $c$-functional (function), where the RG flow of a coupling constant is given by a gradient flow of this $c$-functional. Perelman was also able to show the monotonicity of the Ricci flow based on the above-mentioned entropy functional.
			
In this study, we propose an effective description of a holographic dual for worldsheet bosonic string theory, 
where the RG flows of the $D$-dimensional target spacetime metric, the two-form Kalb-Ramond gauge field, and the dilaton field are manifested at the level of an effective action \cite{SungSik_Holography_I,SungSik_Holography_II,SungSik_Holography_III, RG_Flow_Holography_Monotonicity, Emergent_AdS2_BH_RG,Nonperturbative_Wilson_RG_Disorder,Nonperturbative_Wilson_RG,Einstein_Klein_Gordon_RG_Kim,Einstein_Dirac_RG_Kim,RG_GR_Geometry_I_Kim,RG_GR_Geometry_II_Kim,Kondo_Holography_Kim,Kitaev_Entanglement_Entropy_Kim,RG_Holography_First_Kim}. 
In essence, we suggest a $(D+1)$-dimensional dilaton-gravity-gauge effective theory 
for the dynamics of an emergent target spacetime, 
regarded as a low-energy effective field theory at strong string couplings, where the extra dimension is identified with an RG scale. 
This might 
potentially 
be related to the so-called bosonic $M$ theory in 27 dimensions, where a compactification to 26 dimensions reproduces the low-energy effective action of bosonic string theory \cite{Bosonic_M_Theory}. 
We argue that this holographic dual effective field theory is non-perturbative in nature in the $\alpha'$ expansion. More precisely, a class of string quantum fluctuations is organized and resummed to the all-loop order of the $\alpha'$ expansion based on the 1-loop RG $\beta$-functions and the 1-loop low-energy effective action. It turns out that our holographic dual effective field theory is nothing but the Wilsonian effective action of the target spacetime. 
			
We would like to point out that although the interpretation of the bulk radial coordinate as the RG scale of the boundary theory has been made robust over the years through numerous investigations, it has been generalized further. There has been further investigations into generalized holographic duals where the boundary theory is a more general QFT (i.e. \textit{not} a CFT) \cite{Holographic_RG_Flow_Ricci_Flow_I, Ghosh:2017big, Ghosh:2018qtg} and the interpretation of radial bulk coordinate as the dual energy scale continues to hold. The formalism described in this manuscript will be applicable in such contexts.
			
To investigate the monotonicity of RG flows within the holographic dual effective field theory, we introduce two types of entropy functionals and discuss the monotonicity of RG flows, inspired by the monotonicity of the Ricci flow given by Perelman \cite{Ricci_Flow_Monotonicity}. Based on the $(D+1)$-dimensional dilaton-gravity-gauge effective theory, 
we first obtain the probability distribution functional for macroscopic events and construct a Shannon-type microscopic entropy functional in the grand-canonical ensemble. Benchmarking the entropy production description of Seifert in the Langevin system \cite{Entropy_Production}, we show the monotonicity of this first type of entropy functional. 
%
			%
In addition to this microscopic entropy functional, we construct a macroscopic entropy functional from our holographic dual effective field theory as the second type. 
This construction proceeds through several steps.
First, we derive an on-shell IR effective action from the holographic dual effective field theory. Second, we point out that this renormalized effective action functional has to satisfy the Hamilton-Jacobi equation of the holographic dual effective field theory. See ref. \cite{HJ_Review} for the Hamilton-Jacobi equation in the context of holographic renormalization. Third, we assume that the effective IR boundary action derived from the holographic dual effective field theory satisfies a local RG equation. See refs. \cite{Local_RG_I,Local_RG_II,Local_RG_III} for introduction to local RG equations. Based on the equivalence between the Hamilton-Jacobi equation and the local RG equation \cite{Vasudev_Shyam_III,Vasudev_Shyam_IV}, fourth, we extract out the Weyl anomaly \cite{Local_RG_I,Local_RG_II,Local_RG_III} from the Hamilton-Jacobi equation, where the IR boundary condition is applied consistently. Finally, we obtain the holographic Perelman's entropy functional from the IR on-shell effective action. This macroscopic entropy functional gives rise to the gradient flow equation for the RG flow as the Perelman's entropy functional does. We find that the RG flow of the holographic Perelman's entropy functional coincides with the Weyl anomaly as the Perelman's entropy functional does \cite{Ricci_NLsM_Gradient_i,Ricci_NLsM_Gradient_ii}. This leads us to the monotonicity of the RG flow of the emergent target spacetime. Furthermore, we discover that these two types of entropy production rates are deeply related. We show that the microscopic entropy production rate can be determined by integrating the rate of change of the holographic Perelman's entropy functional over all possible metric configurations along the flow. Our logical flow serves as a way of the proof for the monotonicity of RG flow, distinct from 
previous studies  \cite{c_theorem,a_theorem,a_f_theorem_i,f_theorem_SUSY,f_theorem_noSUSY,a_f_theorem_ii,a_f_theorem_EE_i,a_f_theorem_EE_ii,a_f_theorem_EE_iii,a_f_theorem_EE_iv,a_f_theorem_EE_v,
			c_theorem_holography_i,c_theorem_holography_ii}.

			%
			%
			
						
The main summary of the results and the organization of the paper are as follows:
\begin{itemize}
\item In section \ref{Cohomological_Construction_HDEFT} we first construct a holographic dual effective field theory (EFT) of target spacetime of 2D NLSM by utilizing its RG flow equations within the Hamiltonian formulation. More precisely, based on the 1-loop RG $\beta-$functions and the 1-loop effective action, we construct the holographic dual EFT as a Wilsonian effective action, where a class of string quantum fluctuations are resummed to 
all orders in the $\alpha'$ expansion. 
By introducing a Gaussian noise we `modify' the RG flow equation which is equivalent to the Langevin equation. 
Here, the introduction of noise fluctuations corresponds to incorporating irrelevant/marginal deformations, such as the $T\bar{T}$ deformation for the metric or the $JJ$ deformation for the two-form gauge field on the worldsheet. For further details, see Appendix A.
In this framework, the 1-loop RG flow equation is introduced as a `gauge' constraint within the Fadeev-Poppov procedure and the RG energy scale is introduced as a holographic bulk dimension. Due to the involvement of recursive RG transformations to reach the IR from UV in the presence of the 1-loop low-energy effective action functional, we claim that the resulting holographic dual EFT is non-perturbative in nature. For convenience, we have chosen Gaussian normal coordinates in the path integral formulation to construct the Wilsonian effective action and as a result, Lorentz covariance is manifestly broken. Thus our result may not be the most general one.

\item Our main motivation is to understand the monotonicity of  RG flow based on the holographic dual EFT. In this study, we consider both the microscopic and macroscopic entropy functionals. In section \ref{Entropy_Production_Gibbs_Type}, we first construct a probability distribution function that satisfies the Fokker-Planck equation in the field space. Thereafter we have constructed a path-dependent microscopic entropy functional from the probability distribution function. We establish that this entropy satisfies a monotonicity condition through the involvement of a conserved rank 2 symmetric current associated with the conservation of the probability distribution function. 
				
\item In section \ref{HDEFT_Entropy_Gradient_Flow}, we consider that all the RG $\beta$-functions of the holographic dual EFT can be expressed as the gradient of the effective potential resulting from integrating out a class of string quantum fluctuations in the all-loop order of the $\alpha'$ expansion. We construct the holographic Perelman's entropy functional to understand the monotonicity from the perspective of gradient flow. We emphasize that this holographic Perelman's entropy functional generalizes the original one based on the 1-loop level `Wilsonian' effective action. We discover a remarkable relation between the rate of change for both the microscopic entropy functional discussed in section \ref{Entropy_Production_Gibbs_Type} and the holographic Perelman's entropy functional.
				
\item We conclude in section \ref{Summary_Discussion}. In addition to a rather long summary, we discuss potential deeper relations among the Perelman's $F$-functional or the Zamolodchikov's $c$-functional, the relative entropy, and `exact' functional RG equations. Moreover, we point out that the tachyon condensation problem of bosonic string theory may be resolved in our holographic dual EFT as a future direction. Following that in appendix \ref{NP}, we discussed the non-perturbative nature of the holographic EFT 
in detail. Here, we argue that the construction of section \ref{Cohomological_Construction_HDEFT} is consistent with the Wilsonian RG scheme in a `brute force' way. 
In appendix \ref{Review:Langevin}, we review stochastic thermodynamics, focusing on entropy production during Brownian motion \cite{Entropy_Production}. This section would be helpful in understanding the holographic construction of section \ref{Cohomological_Construction_HDEFT} and section \ref{Entropy_Production_Gibbs_Type}. In appendix \ref{mono}, we reviewed the monotonicity of RG low in NLSM considered in \cite{Ricci_NLsM_Gradient_i, Ricci_NLsM_Gradient_ii}. This review clarifies the role of the Perelman's entropy functional or the Zamolodchikov's $c$-functional in the monotonicity of the RG flow.
\end{itemize}

			
			%
			%
			
\section{Emergent dual holography from the worldsheet nonlinear $\sigma$-model for general curved spacetimes} \label{Cohomological_Construction_HDEFT}
			
Recently, we proposed a general prescription for the holographic dual EFT \cite{RG_Flow_Holography_Monotonicity,Emergent_AdS2_BH_RG,Nonperturbative_Wilson_RG_Disorder,Nonperturbative_Wilson_RG} beyond the explicit microscopic implementation of Wilsonian RG transformations \cite{Einstein_Klein_Gordon_RG_Kim,Einstein_Dirac_RG_Kim,RG_GR_Geometry_I_Kim,RG_GR_Geometry_II_Kim,Kondo_Holography_Kim,Kitaev_Entanglement_Entropy_Kim,RG_Holography_First_Kim}. 
The general prescription has many similarities with a cohomological-type topological field theory formulation a la Witten \cite{TQFT_Cohomology,TQFT_Witten_Type}.  
Given 1-loop RG flow equations, one may reformulate the resulting effective field theory for all the coupling fields by imposing the RG flow equations as `gauge' constraints based on the Faddeev-Popov procedure \cite{QFT_textbook}. Here, Faddeev–Popov ghosts are introduced to constrain the paths of all the coupling fields into the RG flow equations. 
As a result, Becchi-Rouet-Stora-Tyutin (BRST) symmetries play a central role in the construction of an effective field theory, referred to as the BRST cohomology \cite{TQFT_Cohomology,TQFT_Witten_Type}. 
This cohomological-type topological field theory formulation has many applications in the context of non-equilibrium thermodynamics \cite{MSR_Formulation_SUSY_i,MSR_Formulation_SUSY_ii,MSR_Formulation_SUSY_iii,MSR_Formulation_SUSY_iv,MSR_Formulation_SUSY_v,MSR_Formulation_SUSY_vi} and effective hydrodynamics with fluctuations \cite{Schwinger_Keldysh_Symmetries_i,Schwinger_Keldysh_Symmetries_ii,Schwinger_Keldysh_Symmetries_iii,Schwinger_Keldysh_Symmetries_iv,Schwinger_Keldysh_Symmetries_v,Schwinger_Keldysh_Symmetries_vi}. 
Recently, refs.\ \cite{Horava_I, Horava_II} introduced 
a non-relativistic topological quantum field theory, specifically of the Lifshitz type, associated with a generalized family of Ricci flow equations.

Interestingly, our construction of the holographic dual EFT extends beyond the cohomological-type topological field theory formulation by introducing a Wilsonian effective potential into the holographic dual EFT 
\cite{RG_Flow_Holography_Monotonicity,Emergent_AdS2_BH_RG,Nonperturbative_Wilson_RG_Disorder,Nonperturbative_Wilson_RG}. 
This Wilsonian effective potential results from quantum fluctuations of high energy or heavy or fast degrees of freedom in the Wilsonian RG transformation. It turns out that the 1-loop Wilsonian effective potential gives rise to the 1-loop RG flow equation, where the RG $\beta$-function is given by the gradient of the effective potential. For example, the RG flow of a coupling constant is given by the gradient of the Zamolodchikov's $c$-function in conformal perturbation theory \cite{c_theorem}. Based on these two building blocks near the UV fixed point, we construct the holographic dual EFT, where the emergent extradimension is identified with an RG scale parameter. If we do not take into account the Wilsonian effective potential in our holographic dual EFT, the resulting EFT becomes purely a cohomological-type topological field theory. Here, all the information of the path integral expression are `localized' into the 1-loop RG $\beta$-function, and there exist no fluctuation effects in any observable.
			
It turns out that the introduction of the 1-loop effective potential into the holographic dual EFT takes into account a class of quantum fluctuations in the all-loop order \cite{RG_Flow_Holography_Monotonicity,Emergent_AdS2_BH_RG,Nonperturbative_Wilson_RG_Disorder,Nonperturbative_Wilson_RG}. 
Suppose the first step of the Wilsonian RG transformation to integrate out high energy quantum fluctuations. Then, we obtain an effective potential for the coupling constant in the 1-loop level. This Wilsonian effective potential generates an RG flow for the coupling constant as its gradient flow. As a result, the original coupling constant is updated into a renormalized one. Now, we perform the second step of the Wilsonian RG transformation, where the cut-off scale is lowered from that of the first RG transformation. Then, we obtain an effective potential in terms of the renormalized coupling constant. Again, this second effective potential gives rise to an RG flow as its gradient one. As a result, the first renormalized coupling constant is further updated. The complete RG flow is determined by the variational principle of the resulting EFT, where the recursive Wilsonian RG transformations result in the emergent extra dimension identified with the RG scale. Finally, the evolution of the renormalized coupling constant is described by the second-order differential equation with the accumulated effective potential along the extra-dimensional space, where both UV and IR boundary conditions are self-consistently determined. This recursive Wilsonian RG transformation procedure allows us to take quantum corrections in a non-perturbative way. It is remarkable to show that the cohomological-type field theory construction with the Wilsonian effective potential is completely consistent with the recursive RG transformation as mentioned above \cite{RG_Flow_Holography_Monotonicity,Emergent_AdS2_BH_RG,Nonperturbative_Wilson_RG_Disorder,Nonperturbative_Wilson_RG}.
			
In this section we show only the cohomological-type field theory construction for the holographic dual EFT. A deeper relation between this construction and the recursive Wilsonian RG transformations will be clarified in appendix A. The non-perturbative nature of the holographic dual EFT will be also clarified in appendix A.
			
We start by writing the path integral representation of the worldsheet nonlinear $\sigma$-model in a generic curved background $G_{\mu\nu}(x)$ and a Kalb-Ramond two-form gauge field $B_{\mu\nu}(x)$ for the target spacetime, 
\begin{align} 
& Z = \int D x^{\mu}(\sigma) D b_{ab}(\sigma) D c^{a}(\sigma)
				\nn &\qquad 
				\exp\Big[ - \frac{1}{4 \pi \alpha'} \int_{M} d^{2} \sigma \sqrt{g(\sigma)} \Big\{ \Big( g^{ab}(\sigma) G_{\mu\nu}(x) + i \epsilon^{ab} B_{\mu\nu}(x) \Big) \partial_{a} x^{\mu}(\sigma) \partial_{b} x^{\nu}(\sigma) + \alpha' R^{(2)}(\sigma) \Phi(x) \Big\} 
				\nn & \hspace*{6.5cm}\quad \quad \quad \quad \quad \quad \quad \quad \quad - \frac{1}{2 \pi} \int_{M} d^{2} \sigma \sqrt{g(\sigma)} b_{ab}(\sigma) \nabla^{a} c^{b}(\sigma) \Big]\ . 
\end{align}
Here, $\frac{1}{2 \pi} \int_{M} d^{2} \sigma \sqrt{g(\sigma)} b_{ab}(\sigma) \nabla^{a} c^{b}(\sigma)$ is the ghost action to implement the Jacobian factor involved with the gauge fixing of the worldsheet metric $g_{ab}(\sigma)$ in the Faddeev-Popov procedure \cite{Polchinski}. The worldsheet covariant derivative for the ghost field is
\begin{align} & \nabla^{a} c^{b}(\sigma) = g^{ac}(\sigma) \nabla_{c} c^{b}(\sigma) = g^{ac}(\sigma) \Big( \partial_{c} c^{b}(\sigma) + {\Gamma}^{b}_{cd}(\sigma) c^{d}(\sigma) \Big) , \end{align} 
where ${\Gamma}^{b}_{cd}(\sigma)$ is the worldsheet connection. $\epsilon^{ab}$ is the antisymmetric tensor in the worldsheet. $\Phi(x)$ is the dilaton field to manage the conformal anomaly, coupled to the worldsheet scalar curvature $R^{(2)}(\sigma)$.
			
Specifically, we consider the conformal gauge
where 
\cite{Polchinski}
\begin{align} & \hat{g}^{ab}(\sigma) = e^{2 \omega(\sigma)} \delta^{ab} . \end{align}
Accordingly, the worldsheet connection is given by
\begin{align} & \hat{\Gamma}_{cd}^{b} = \frac{1}{2} \hat{g}^{be} \Big(\partial_{c} \hat{g}_{e d} + \partial_{d} \hat{g}_{ce} - \partial_{e} \hat{g}_{cd}\Big) = - \Big(\partial_{c} \omega(\sigma) \delta^{b}_{d} + \partial_{d} \omega(\sigma) \delta^{b}_{c} - \delta^{be} \partial_{e} \omega(\sigma) \delta_{cd}\Big) . \end{align}
The worldsheet Riemann tensor is given by
\begin{align} & \hat{R}_{abcd}^{(2)}(\sigma) = \frac{1}{2} \Big(\hat{g}_{ac}(\sigma) \hat{g}_{bd}(\sigma) - \hat{g}_{ad}(\sigma) \hat{g}_{bc}(\sigma)\Big) \hat{R}^{(2)}(\sigma) , \end{align}
where the Ricci scalar is 
\begin{align} & \hat{R}^{(2)}(\sigma) = - 2 e^{- 2 \omega(\sigma)} \delta^{ab} \partial_{a} \partial_{b} \omega(\sigma) \end{align}
in this conformal gauge.
			
			%
			%
			%
			%
			%
			
It is straightforward to perform the RG transformation at the one-loop level in $\alpha'$ \cite{RG_Flow_NLsM_I,RG_Flow_NLsM_II}. 
One separates $x^{\mu}(\sigma)$ into its slow (classical or background) and fast (quantum) degrees of freedom, and expands the resulting worldsheet string action in terms of the fast degrees of freedom up to the second-order in the $\alpha'$ expansion. 
One can do this task in the Riemann normal coordinate, 
where the connection coefficient of the target spacetime vanishes locally to allow for the expansion to be written conveniently. Carrying out the Gaussian integral for the fast degrees of freedom in the conformal gauge, one finds the RG flow equations for the target spacetime metric, the Kalb-Ramond gauge field, and the dilaton field, respectively, as follows \cite{Polchinski}
\begin{align}\label{polRG} \partial_{z} G_{\mu\nu}(x) &= - \beta_{\mu\nu}^{G} = - \alpha' R_{\mu\nu}(x) - 2 \alpha' \nabla_{\mu} \nabla_{\nu} \Phi(x) + \frac{\alpha'}{4} H_{\mu\lambda\omega}(x) H_{\nu}^{\lambda\omega}(x) , 
			\\ \label{KR_RG} \partial_{z} B_{\mu\nu}(x) &= - \beta_{\mu\nu}^{B} = \frac{\alpha'}{2} \nabla^{\omega} H_{\omega\mu\nu}(x) - \alpha' \partial^{\omega} \Phi(x) H_{\omega\mu\nu}(x) ,
				\\ \label{Dilaton_RG} \partial_{z} \Phi(x) &= - \beta^{\Phi} = - \frac{D - 26}{6} + \frac{\alpha'}{2} \nabla^{2} \Phi(x) - \alpha' \partial_{\mu} \Phi(x) \partial^{\mu} \Phi(x) + \frac{\alpha'}{24} H_{\mu\nu\lambda}(x) H^{\mu\nu\lambda}(x) . \end{align}
Here, $z$ is an RG scale for the RG transformation, where $z = 0$ is identified with the UV fixed point. $H_{\omega\mu\nu}(x) = \partial_{\omega} B_{\mu\nu}(x) + \partial_{\mu} B_{\nu\omega}(x) + \partial_{\nu} B_{\omega\mu}(x)$ is the field strength tensor for the two-form gauge field $B_{\mu\nu}(x)$.

			%
			%
			
Our task is to 
make these RG flow equations explicit
at the level of an effective action \cite{Emergent_AdS2_BH_RG,Nonperturbative_Wilson_RG_Disorder,Nonperturbative_Wilson_RG, Einstein_Klein_Gordon_RG_Kim,Einstein_Dirac_RG_Kim,RG_GR_Geometry_I_Kim,RG_GR_Geometry_II_Kim,Kondo_Holography_Kim,Kitaev_Entanglement_Entropy_Kim,RG_Holography_First_Kim}. Based on the Faddeev-Popov procedure, we establish that the holographic dual effective field theory has essentially the same structure as a generating functional of nonequilibrium statistical systems. Here, the RG flow equation corresponds to the Langevin equation and the RG scale parameter gives rise to the emergent extra dimension \cite{RG_Flow_Holography_Monotonicity}. 
			
Considering the target spacetime metric, we modify the RG flow equation \eqref{polRG} as
\begin{align} & \partial_{z} G_{\mu\nu}(x,z) = - \alpha' R_{\mu\nu}(x,z) - 2 \alpha' \nabla_{\mu} \nabla_{\nu} \Phi(x,z) + \frac{\alpha'}{4} H_{\mu\lambda\omega}(x,z) H_{\nu}^{\lambda\omega}(x,z) + \xi_{\mu\nu}(x,z) . \label{RG_Flow_Metric_Only} \end{align}  
Here, $\xi_{\mu\nu}(x,z)$ plays the role of the Gaussian random noise, given by
\begin{align}\label{noisecorrelation}
				& \langle \xi_{\mu\nu}(x,z) \mathcal{G}^{\mu\nu\rho\lambda}(x,z) \xi_{\rho\lambda}(x',z') \rangle = \lambda \delta(x-x') \delta(z-z') , \end{align}
where its average value vanishes, i.e., $\langle \xi_{\mu\nu}(x,z) \rangle = 0$. $\mathcal{G}_{\mu\nu\rho\lambda}(x,z)$ 
			is the Wheeler-DeWitt metric \cite{DeWitt_Metric}, given by
			\begin{align} & \mathcal{G}_{\mu\nu\rho\gamma}(x,z) \equiv \frac{1}{2} G_{\mu\rho}(x,z) G_{\nu\gamma}(x,z) + \frac{1}{2} G_{\nu\rho}(x,z) G_{\mu\gamma}(x,z) - \frac{1}{D-1} G_{\mu\nu}(x,z) G_{\rho\gamma}(x,z) , \end{align}
			and $\mathcal{G}^{\mu\nu\rho\gamma}(x,z)$ is the inverse Wheeler-DeWitt metric, satisfying $\mathcal{G}_{\alpha\beta\gamma\delta} \mathcal{G}^{\gamma\delta\mu\nu} = \delta_{(\alpha}{}^{\mu}\delta_{\beta)}{}^{\nu}$ and given by
			\begin{align} \mathcal{G}^{\mu\nu\rho\gamma}(x,z) = \frac{1}{2} \Big(G^{\mu\rho}(x,z) G^{\nu\gamma}(x,z) +  G^{\nu\rho}(x,z) G^{\mu\gamma}(x,z)\Big) - G^{\mu\nu}(x,z) G^{\rho\gamma}(x,z) . \end{align}
			In Eq. \eqref{noisecorrelation} $\lambda$ is the strength of noise, analogous to the diffusion constant. The RG flow equation given in Eq. (\ref{RG_Flow_Metric_Only}) now can be regarded as the Langevin equation. 
%
%
We point out that the introduction of random noise fluctuations in the RG flow corresponds to a $T \bar{T}$ deformation \cite{TTbar_Deformation} in the worldsheet nonlinear $\sigma$ model, which we clarify in Eq. (\ref{TT_bar_Deform_f1}) of appendix \ref{NP}. We further note that the noise field is simply the auxiliary field in the superspace formulation \cite{Horava_I,Horava_II}, also explicitly demonstrated in our previous work \cite{RG_Flow_Holography_Monotonicity}, but not performed in the current study. 

In a similar fashion, we also introduce random noise fluctuations $\zeta_{\mu\nu}(x,z)$ into the RG flow of the two-form gauge field, and modify Eq. (\ref{KR_RG}) as
\begin{align} & \partial_{z} B_{\mu\nu}(x,z) = \frac{\alpha'}{2} \nabla^{\omega} H_{\omega\mu\nu}(x,z) - \alpha' \partial^{\omega} \Phi(x,z) H_{\omega\mu\nu}(x,z) + \zeta_{\mu\nu}(x,z) . \label{RG_Flow_KR_Gauge_Field_Only} \end{align}  
Here, the random field $\zeta_{\mu\nu}(x,z)$ with $\langle \zeta_{\mu\nu}(x,z) \rangle = 0$ satisfies
\begin{align}\label{noisecorrelation_KR_Gauge_Field}
				& \langle \zeta_{\mu\nu}(x,z) \zeta_{\rho\lambda}(x',z') \rangle = q \delta_{\mu\rho} \delta_{\nu\lambda} \delta(x-x') \delta(z-z') , \end{align}
where $q$ is the variance of their fluctuations. As $\xi_{\mu\nu}(x,z)$ introduces the $T \bar{T}$ deformation into the worldsheet nonlinear $\sigma$ model, $\zeta_{\mu\nu}(x,z)$ gives rise to the $J_{\mu\nu} J^{\mu\nu}$ deformation into the bosonic string theory, where $J_{\mu\nu} = i \epsilon^{ab} \partial_{a} x^{\mu}(\sigma) \partial_{b} x^{\nu}(\sigma)$ is the conserved current minimally coupled to the two-form gauge field $B_{\mu\nu}(x)$. This will be clarified in Eq. (\ref{TT_bar_Deform_f1}) of appendix \ref{NP}.
One may also introduce random noise fluctuations into the RG flow of the dilaton field. Here, we do not consider it.
			
Next, we can construct a generating functional for this Langevin-type equation, following the strategy discussed in appendix \ref{Review:Langevin}. Recalling the Faddeev-Popov procedure \cite{QFT_textbook}, we introduce the following identity 
			%
				%
\begin{align} 1 = & \int_{G_{\mu\nu}(x,0)}^{G_{\mu\nu}(x,z_{f})} D G_{\mu\nu}(x,z) \int_{B_{\mu\nu}(x,0)}^{B_{\mu\nu}(x,z_{f})} D B_{\mu\nu}(x,z) \int_{\Phi(x,0)}^{\Phi(x,z_{f})} D \Phi(x,z) 
\nn & \times \delta\Big(\partial_{z} G_{\mu\nu}(x,z) 
+ \beta_{\mu\nu}^G 
- \xi_{\mu\nu}(x,z)\Big) 
 \delta\Big(\partial_{z} B_{\mu\nu}(x,z) 
+\beta_{\mu\nu}^B - \zeta_{\mu\nu}(x,z)\Big) 
 \delta\Big(\partial_{z} \Phi(x,z) 
+\beta^{\Phi} \Big) 
\nn & \times \mathcal{J}\Big(\frac{\partial}{\partial G_{\mu\nu}(x,z)},\frac{\partial}{\partial B_{\mu\nu}(x,z)},\frac{\partial}{\partial \Phi(x,z)}\Big) . \end{align}
%

Here, the Jacobian factor is given by
\begin{align} \mathcal{J}\Big(\frac{\partial}{\partial G_{\mu\nu}(x,z)},\frac{\partial}{\partial B_{\mu\nu}(x,z)},\frac{\partial}{\partial \Phi(x,z)}\Big) & \equiv \frac{\partial \Big(\partial_{z} G_{\omega\lambda} + \beta_{\omega\lambda}^{G}, \partial_{z} B_{\omega\lambda} + \beta_{\omega\lambda}^{B}, \partial_{z} \Phi + \beta^{\Phi}\Big)}{\partial \Big(G_{\mu\nu}, B_{\mu\nu}, \Phi\Big)} \nn
& = \mbox{Det} \begin{pmatrix} \frac{\partial [\partial_{z} G_{\omega\lambda} + \beta_{\omega\lambda}^{G}]}{\partial G_{\mu\nu}} & \frac{\partial [\partial_{z} G_{\omega\lambda} + \beta_{\omega\lambda}^{G}]}{\partial B_{\mu\nu}} & \frac{\partial [\partial_{z} G_{\omega\lambda} + \beta_{\omega\lambda}^{G}]}{\partial \Phi} \\ \frac{\partial [\partial_{z} B_{\omega\lambda} + \beta_{\omega\lambda}^{B}]}{\partial G_{\mu\nu}} & \frac{\partial [\partial_{z} B_{\omega\lambda} + \beta_{\omega\lambda}^{B}]}{\partial B_{\mu\nu}} & \frac{\partial [\partial_{z} B_{\omega\lambda} + \beta_{\omega\lambda}^{B}]}{\partial \Phi} \\ \frac{\partial [\partial_{z} \Phi + \beta^{\Phi}]}{\partial G_{\mu\nu}} & \frac{\partial [\partial_{z} \Phi + \beta^{\Phi}]}{\partial B_{\mu\nu}} & \frac{\partial [\partial_{z} \Phi + \beta^{\Phi}]}{\partial \Phi}\end{pmatrix} , \label{Jacobian} \end{align}
where all the one-loop $\beta$-functions were introduced before.  
			
Resorting to this identity, it is straightforward to construct the generating functional for such modified RG flows as follows
\begin{align} 
Z = & \int D \xi_{\mu\nu}(x,z) D \zeta_{\mu\nu}(x,z) D G_{\mu\nu}(x,z) D B_{\mu\nu}(x,z) D \Phi(x,z) ~ \mathcal{J}\Big(\frac{\partial}{\partial G_{\mu\nu}(x,z)},\frac{\partial}{\partial B_{\mu\nu}(x,z)},\frac{\partial}{\partial \Phi(x,z)}\Big) \nn &
\times
\delta\Big( \partial_{z} G_{\mu\nu}(x,z) + \beta^G_{\mu\nu} - \xi_{\mu\nu}(x,z) \Big) 
\delta\Big(\partial_{z} B_{\mu\nu}(x,z) +\beta^B_{\mu\nu}
- \zeta_{\mu\nu}(x,z)\Big) 
\delta\Big(\partial_{z} \Phi(x,z) + \beta^{\Phi}
\Big) \nn & 
\times \exp\Big[ - \frac{1}{\alpha'} \int_{0}^{z_{f}} d z \int d^{D} x \sqrt{G(x,z)} e^{- 2 \Phi(x,z)} \Big\{ - \frac{1}{2 \lambda} \xi_{\mu\nu}(x,z) \mathcal{G}^{\mu\nu\rho\lambda}(x,z) \xi_{\rho\lambda}(x,z) - \frac{1}{2 q} \zeta_{\mu\nu}(x,z) \zeta^{\mu\nu}(x,z) \nn & ~~~~~~~~~~~~~~~~~~~~~~~ ~~~~~~~~~~~~~~ + \frac{\alpha'}{4} R(x,z) - \frac{D - 26}{6} - \frac{\alpha'}{48} H_{\mu\nu\lambda}(x,z) H^{\mu\nu\lambda}(x,z) + \alpha' \partial_{\mu} \Phi(x,z) \partial^{\mu} \Phi(x,z) \Big\} \Big] 
			. \end{align}
Here, the ensemble average with respect to Gaussian noise fluctuations is taken into account by the introduction of $\int D \xi_{\mu\nu}(x,z) D \zeta_{\mu\nu}(x,z) \exp\Big[ - \frac{1}{\alpha'} \int_{0}^{z_{f}} d z \int d^{D} x \sqrt{G(x,z)} e^{- 2 \Phi(x,z)} \Big\{ - \frac{1}{2 \lambda} \xi_{\mu\nu}(x,z) \mathcal{G}^{\mu\nu\rho\lambda}(x,z) \xi_{\rho\lambda}(x,z) - \frac{1}{2 q} \zeta_{\mu\nu}(x,z) \zeta^{\mu\nu}(x,z) \Big\} \Big]$. $\alpha'$ has been introduced to consider the string coupling constant. Furthermore, the Einstein-Hilbert action with the Kalb-Ramond gauge field and the dilaton field, $\frac{\alpha'}{4} R(x,z) - \frac{D - 26}{6} - \frac{\alpha'}{48} H_{\mu\nu\lambda}(x,z) H^{\mu\nu\lambda}(x,z) + \alpha' \partial_{\mu} \Phi(x,z) \partial^{\mu} \Phi(x,z)$ was taken into account, which results from quantum fluctuations of a bosonic string \cite{Polchinski}, regarded to be the `Wilsonian' effective potential. 
The introduction of this Wilsonian effective potential plays a central role in a non-perturbative description of the dynamics of the $D$-dimensional target spacetime. This aspect will be clarified in Eq.\ (\ref{Nonperturbative_RG_f2}) of appendix \ref{NP}. 
All the RG $\beta$-functions of the target spacetime metric $\beta_{\mu\nu}^{G} = \alpha' R_{\mu\nu}(x,z) + 2 \alpha' \nabla_{\mu} \nabla_{\nu} \Phi(x,z) - \frac{\alpha'}{4} H_{\mu\lambda\omega}(x,z) H_{\nu}^{\lambda\omega}(x,z)$, the Kalb-Ramond gauge field $\beta_{\mu\nu}^{B} = - \frac{\alpha'}{2} \nabla^{\omega} H_{\omega\mu\nu}(x,z) + \alpha' \nabla^{\omega} \Phi(x,z) H_{\omega\mu\nu}(x,z)$, and the dilaton field $\beta^{\Phi} = \frac{D - 26}{6} - \frac{\alpha'}{2} \nabla^{2} \Phi(x,z) + \alpha' \partial_{\mu} \Phi(x,z) \partial^{\mu} \Phi(x,z) - \frac{\alpha'}{24} H_{\mu\nu\lambda}(x,z) H^{\mu\nu\lambda}(x,z)$ are given by the `gradients' of the effective action $\int_{0}^{z_{f}} d z \int d^{D} x \sqrt{G(x,z)} e^{- 2 \Phi(x,z)} \Big\{ \frac{\alpha'}{4} R(x,z) - \frac{D - 26}{6} - \frac{\alpha'}{48} H_{\mu\nu\lambda}(x,z) H^{\mu\nu\lambda}(x,z) + \alpha' \partial_{\mu} \Phi(x,z) \partial^{\mu} \Phi(x,z) \Big\}$ with respect to $G^{\mu\nu}(x,z)$, $B^{\mu\nu}(x,z)$, and $\Phi(x,z)$, respectively \cite{Polchinski}. This will be further discussed in section \ref{HDEFT_Entropy_Gradient_Flow}.
			
We emphasize that this is not an ad hoc construction for the holographic dual EFT. Actually, this construction is fully consistent with the Wilsonian RG transformation scheme. See appendix A for our concrete demonstration.
			
Resorting to the Lagrange multiplier field $(\Pi_{G}^{\mu\nu}(x,z),\Pi_{B}^{\mu\nu}(x,z),\Pi_{\Phi}(x,z))$ dual to the target spacetime metric, the Kalb-Ramond gauge field, and the dilaton field, respectively, we can exponentiate all the $\delta$-function constraints to obtain an effective action as follows
\begin{align} 
Z = & \int D G_{\mu\nu}(x,z) D \Pi^{\mu\nu}_{G}(x,z) D B_{\mu\nu}(x,z) D \Pi^{\mu\nu}_{B}(x,z) D \Phi(x,z) D \Pi_{\Phi}(x,z) ~ \mathcal{J}\Big(\frac{\partial}{\partial G_{\mu\nu}(x,z)},\frac{\partial}{\partial B_{\mu\nu}(x,z)},\frac{\partial}{\partial \Phi(x,z)}\Big) \nn & \int D \xi_{\mu\nu}(x,z) D \zeta_{\mu\nu}(x,z) ~ \exp\Big[ - \frac{1}{\alpha'} \int_{0}^{z_{f}} d z \int d^{D} x \sqrt{G(x,z)} e^{- 2 \Phi(x,z)} \Big\{ - \frac{1}{2 \lambda} \xi_{\mu\nu}(x,z) \mathcal{G}^{\mu\nu\rho\lambda}(x,z) \xi_{\rho\lambda}(x,z) \nn 
& - \frac{1}{2 q} \zeta_{\mu\nu}(x,z) \zeta^{\mu\nu}(x,z) \Big\} \Big] \exp\Big[ - \frac{1}{\alpha'} \int_{0}^{z_{f}} d z \int d^{D} x \sqrt{G(x,z)} e^{- 2 \Phi(x,z)} \Big\{ 
\nn & ~~~~ 
\Pi_{G}^{\mu\nu}(x,z) \Big(\partial_{z} G_{\mu\nu}(x,z) + \beta_{\mu\nu}^G
- \xi_{\mu\nu}(x,z)\Big) 
 + \Pi_{B}^{\mu\nu}(x,z) \Big(\partial_{z} B_{\mu\nu}(x,z) + \beta^B_{\mu\nu}
 - \zeta_{\mu\nu}(x,z)\Big) 
 + \Pi_{\Phi}(x,z) \Big(\partial_{z} \Phi(x,z) 
+\beta^{\Phi}
\Big) 
\nn
& + \frac{\alpha'}{4} R(x,z) - \frac{D - 26}{6} - \frac{\alpha'}{48} H_{\mu\nu\lambda}(x,z) H^{\mu\nu\lambda}(x,z) + \alpha' \partial_{\mu} \Phi(x,z) \partial^{\mu} \Phi(x,z) \Big\} \Big] . \end{align}
Here, $(\Pi_{G}^{\mu\nu}(x,z),\Pi_{B}^{\mu\nu}(x,z),\Pi_{\Phi}(x,z))$ is the canonical momentum of $(G_{\mu\nu}(x,z),B_{\mu\nu}(x,z),\Phi(x,z))$, respectively.
			
Performing the ensemble average for random noise fluctuations, we obtain
\begin{align} 
Z = & \int D G_{\mu\nu}(x,z) D \Pi_{G}^{\mu\nu}(x,z) D B_{\mu\nu}(x,z) D \Pi^{\mu\nu}_{B}(x,z) D \Phi(x,z) D \Pi_{\Phi}(x,z) ~ \mathcal{J}\Big(\frac{\partial}{\partial G_{\mu\nu}(x,z)},\frac{\partial}{\partial B_{\mu\nu}(x,z)},\frac{\partial}{\partial \Phi(x,z)}\Big) \nn & \exp\Big[ - \frac{1}{\alpha'} \int_{0}^{z_{f}} d z \int d^{D} x \sqrt{G(x,z)} e^{- 2 \Phi(x,z)}
\Big\{ \Pi_{G}^{\mu\nu}(x,z) 
 \Big(\partial_{z} G_{\mu\nu}(x,z) +
 \beta^G_{\mu\nu}
 \Big) 
 + \frac{\lambda}{2} \Pi_{G}^{\mu\nu}(x,z) \mathcal{G}_{\mu\nu\rho\lambda}(x,z) \Pi_{G}^{\rho\lambda}(x,z) \nn & + \Pi_{B}^{\mu\nu}(x,z) \Big(\partial_{z} B_{\mu\nu}(x,z)+
 \beta^B_{\mu\nu}
 \Big) + \frac{q}{2} \Pi_{B,\mu\nu}(x,z) \Pi_{B}^{\mu\nu}(x,z)
+ \Pi_{\Phi}(x,z) \Big(\partial_{z} \Phi(x,z) 
+\beta^{\Phi}
\Big) 
\nn & + \frac{\alpha'}{4} R(x,z) - \frac{D - 26}{6} - \frac{\alpha'}{48} H_{\mu\nu\lambda}(x,z) H^{\mu\nu\lambda}(x,z) + \alpha' \partial_{\mu} \Phi(x,z) \partial^{\mu} \Phi(x,z) \Big\} \Big] . \label{HDEFT} \end{align}
Here, both $\frac{\lambda}{2} \Pi_{G}^{\mu\nu}(x,z) \mathcal{G}_{\mu\nu\rho\lambda}(x,z) \Pi_{G}^{\rho\lambda}(x,z)$ and $\frac{q}{2} \Pi_{B,\mu\nu}(x,z) \Pi_{B}^{\mu\nu}(x,z)$ result from the ensemble average of randomly distributed noise configurations. The RG scale parameter $z$ appears as an extra dimension in the similar spirit of AdS/CFT holography. As a result, we obtain an effective dilaton-gravity-gauge theory in $(D+1)$ spacetime dimensions.

A key feature of this holographic dual effective field theory is that the RG flow of the first order differential equation is now promoted to be the second order differential equation, being interpreted as the `bulk' evolution. To see this point more explicitly, we perform the path integral for all the canonical momentum fields and obtain an effective Lagrangian formulation as follows
\begin{align} 
Z = & \int D G_{\mu\nu}(x,z) D B_{\mu\nu}(x,z) D \Phi(x,z) ~ \mathcal{J}\Big(\frac{\partial}{\partial G_{\mu\nu}(x,z)},\frac{\partial}{\partial B_{\mu\nu}(x,z)},\frac{\partial}{\partial \Phi(x,z)}\Big)  \delta\Big(\partial_{z} \Phi(x,z) +\beta^{\Phi}
\Big) 
\nn & \exp\Big[ - \frac{1}{\alpha'} \int_{0}^{z_{f}} d z \int d^{D} x \sqrt{G(x,z)} e^{- 2 \Phi(x,z)} \Big\{ - \frac{1}{2 \lambda} 
\Big(\partial_{z} G_{\mu\nu}(x,z) + \beta^G_{\mu\nu} 
\Big)
\mathcal{G}^{\mu\nu\rho\lambda}(x,z) \Big(\partial_{z} G_{\rho\lambda}(x,z) +
\beta^G_{\rho\lambda}
\Big) 
\nn & - \frac{1}{2 q} \Big(\partial_{z} B_{\mu\nu}(x,z) +\beta^B_{\mu\nu}
\Big) \Big(\partial_{z} B^{\mu\nu}(x,z) +
\beta^{B\mu\nu}
\Big)
\nn
&
+ \frac{\alpha'}{4} R(x,z) - \frac{D - 26}{6} - \frac{\alpha'}{48} H_{\mu\nu\lambda}(x,z) H^{\mu\nu\lambda}(x,z) + \alpha' \partial_{\mu} \Phi(x,z) \partial^{\mu} \Phi(x,z) \Big\} \Big] . \label{HDEFT_Lagrangian} \end{align}
%
Applying the variational principle to this effective action functional with respect to $G_{\mu\nu}(x,z)$ and $B_{\mu\nu}(x,z)$, we obtain the coupled second-order differential equations in the Gaussian normal coordinate system $d s^{2}(x,z) = d z^{2} + G_{\mu\nu}(x,z) d x^{\mu} d x^{\nu}$. Here, the lapse function $\mathcal{N}$ and the shift vector $\mathcal{N}^{\mu}$ are set to be $\mathcal{N}=1$ and $\mathcal{N}^{\mu}=0$, respectively, in the ADM decomposition \cite{ADM_Hamiltonian_Formulation} of the bulk metric, $d s^{2} = \Big( \mathcal{N}^{2}(x,z) + \mathcal{N}_{\mu}(x,z) \mathcal{N}^{\mu}(x,z) \Big) d z^{2} + 2 \mathcal{N}_{\mu}(x,z) d x^{\mu} d z + G_{\mu\nu}(x,z) d x^{\mu} d x^{\nu}$. On the other hand, the RG flow of the dilaton field remains to be the first-order differential equation in our construction. In appendix \ref{NP}, we discuss how the non-perturbative dynamics can be introduced into our second-order differential equations.
			
			%
			%
			
To solve the second-order differential equation, of course, we need two boundary conditions, here UV ($z = 0$) and IR ($z = z_{f}$) ones. As will be verified below and section \ref{HDEFT_Entropy_Gradient_Flow}, it turns out that both the UV and IR boundary conditions are essentially given by the RG flow equations (\ref{polRG}), (\ref{KR_RG}), and (\ref{Dilaton_RG}). As a result, the conformal anomaly cancellation gives rise to the Neumann boundary condition at both UV and IR. This is rather interesting because our holographic dual effective field theory can describe an RG flow from one conformally invariant bosonic string theory in the $D$-spacetime dimension to the other consistent theory in the same dimension. If we consider asymptotically flat spacetimes only, it has been well known that the vanishing $\beta$-function conditions cause two types of solutions: 
(1) $G_{\mu\nu}(x,0) = \delta_{\mu\nu}$, $B_{\mu\nu}(x,0) = 0$, and $\Phi(x,0) = \Phi_{UV}$ 
($G_{\mu\nu}(x,z_{f}) = \delta_{\mu\nu}$, $B_{\mu\nu}(x,z_{f}) = 0$, and $\Phi(x,z_{f}) = \Phi_{IR}$) 
and (2) $G_{\mu\nu}(x,0) = \delta_{\mu\nu}$, $B_{\mu\nu}(x,0) = 0$, and $\Phi(x,0) = v_{\mu}^{UV} x^{\mu}$ ($G_{\mu\nu}(x,z_{f}) = \delta_{\mu\nu}$, $B_{\mu\nu}(x,z_{f}) = 0$, and $\Phi(x,z_{f}) = v_{\mu}^{IR} x^{\mu}$). Here, our target spacetime is Euclidean. The first type solution is realized only at $D = 26$, i.e., the critical dimension of the bosonic string theory. On the other hand, the second solution can be realized in any dimension as long as 
\begin{align}
v_{\mu}^{UV (IR)} v^{UV (IR), \mu} = \frac{26 - D}{6 \alpha'} 
\end{align}
is satisfied \cite{Polchinski,Bosonic_String_Theory_Any_D_I,Bosonic_String_Theory_Any_D_II}. 
We emphasize that this condition is responsible for the conformal anomaly cancellation in any spacetime dimensions, thus which holds in the all loop order of $\alpha'$ \cite{Bosonic_String_Theory_Any_D_I,Bosonic_String_Theory_Any_D_II}. In this respect the present framework may shed light on the issue of how a noncritical string theory at UV evolves into that at IR, resolving the tachyon condensation problem. We comment on this point in conclusion.
			
Recently, refs.\ \cite{Horava_I,Horava_II} reformulated the Ricci-flow equation in the path integral representation, given by a cohomological topological field theory. Here, the supersymmetry breaking term given by the Wilsonian effective potential was not taken into account, and only the Ricci-flow equation was considered. In other words, the effective field theory is purely topological of the cohomological type. 
On the other hand, the introduction of the Wilsonian effective potential gives rise to dynamics in target spacetime fluctuations beyond the Ricci flow equation. It would be interesting to reformulate our holographic dual effective field theory in the superspace. To obtain the superspace construction for the holographic dual effective field theory, firstly we need to exponentiate the Jacobian factor in the path integral representation as follows
 \begin{align} & \mathcal{J}\Big(\frac{\partial}{\partial G_{\mu\nu}(x,z)},\frac{\partial}{\partial B_{\mu\nu}(x,z)},\frac{\partial}{\partial \Phi(x,z)}\Big) 
				\nn = & \int D \psi_{\mu\nu}^{G}(x,z) D \bar{\psi}^{\mu\nu}_{G}(x,z) D \psi_{\mu\nu}^{B}(x,z) D \bar{\psi}^{\mu\nu}_{B}(x,z) D \psi^{\Phi}(x,z) D \bar{\psi}_{\Phi}(x,z) \nn & \exp\Bigg\{ - \frac{1}{\alpha'} \int_{0}^{z_{f}} d z \int d^{D} x \sqrt{G(x,z)} e^{- 2 \Phi(x,z)}
				\nn & \begin{pmatrix} \bar{\psi}^{\omega\lambda}_{G}(x,z) & \bar{\psi}^{\omega\lambda}_{B}(x,z) & \bar{\psi}_{\Phi}(x,z) \end{pmatrix} \begin{pmatrix} \frac{\partial [\partial_{z} G_{\omega\lambda} + \beta_{\omega\lambda}^{G}]}{\partial G_{\mu\nu}} & \frac{\partial [\partial_{z} G_{\omega\lambda} + \beta_{\omega\lambda}^{G}]}{\partial B_{\mu\nu}} & \frac{\partial [\partial_{z} G_{\omega\lambda} + \beta_{\omega\lambda}^{G}]}{\partial \Phi} \\ \frac{\partial [\partial_{z} B_{\omega\lambda} + \beta_{\omega\lambda}^{B}]}{\partial G_{\mu\nu}} & \frac{\partial [\partial_{z} B_{\omega\lambda} + \beta_{\omega\lambda}^{B}]}{\partial B_{\mu\nu}} & \frac{\partial [\partial_{z} B_{\omega\lambda} + \beta_{\omega\lambda}^{B}]}{\partial \Phi} \\ \frac{\partial [\partial_{z} \Phi + \beta^{\Phi}]}{\partial G_{\mu\nu}} & \frac{\partial [\partial_{z} \Phi + \beta^{\Phi}]}{\partial B_{\mu\nu}} & \frac{\partial [\partial_{z} \Phi + \beta^{\Phi}]}{\partial \Phi} \end{pmatrix} \begin{pmatrix} \psi_{\mu\nu}^{G}(x,z) \\ \psi_{\mu\nu}^{B}(x,z) \\ \psi^{\Phi}(x,z) \end{pmatrix} \Bigg\} . \label{Jacobian_Path_Integral} \end{align}
			Here, $\psi_{\mu\nu}^{G}(x,z)$, $\psi_{\mu\nu}^{B}(x,z)$, and $\psi^{\Phi}(x,z)$ ($\bar{\psi}^{\mu\nu}_{G}(x,z)$, $\bar{\psi}^{\mu\nu}_{B}(x,z)$, and $\bar{\psi}_{\Phi}(x,z)$) are fermionic ghost fields. Previously, we performed a superspace formulation for generic RG flows in emergent dual holography, where two types of BRST symmetries manifest in the holographic dual EFT \cite{RG_Flow_Holography_Monotonicity}. As a result of the BRST symmetry breaking given by an effective potential in the RG flow, we found generalized fluctuation-dissipation theorems for the generic RG flows, to be discussed below in more detail.
			
We would like to emphasize that this cohomological-type field theory construction with the effective potential for the RG flow seems to be consistent with a brute force derivation of the Wilsonian RG transformation for generic QFTs. Our previous work \cite{Einstein_Klein_Gordon_RG_Kim} constructed a holographic dual effective field theory by applying Wilsonian RG transformations recursively, which turns out to be consistent with a cohomological-type field theory construction with an effective potential \cite{Nonperturbative_Wilson_RG,RG_GR_Geometry_I_Kim,RG_GR_Geometry_II_Kim}. We explicitly demonstrated that this holographic dual effective field theory gives rise to non-perturbative physics, being not accessible by perturbative QFT calculations \cite{Emergent_AdS2_BH_RG,Nonperturbative_Wilson_RG_Disorder,Kondo_Holography_Kim}. One important subtle point is that RG flows in NLSMs are realized for the target spacetime metric instead of the worldsheet one, which should be distinguished by those of QFTs. In this respect, we should be careful in figuring out the procedure of a brute force derivation of the holographic dual EFT based on the Wilsonian RG transformation.

We emphasized that the saddle-point approximation for the bulk effective action Eq. (\ref{HDEFT}) or Eq. (\ref{HDEFT_Lagrangian}) with respect to $G_{\mu\nu}(x,z)$ and $B_{\mu\nu}(x,z)$ results in the second order differential equation in the extra-dimensional coordinate $z$ instead of the first order one near the UV fixed point while the dilaton field $\Phi(x,z)$ is still governed by the first order differential equation in $z$. The solution of these coupled equations has to match with the IR boundary conditions. These IR boundary conditions are determined by the IR boundary effective action of the worldsheet theory in a self-consistent way,
\begin{align} 
				Z_{IR} = & \int D x^{\mu}(\sigma) D b_{ab}(\sigma) D c^{a}(\sigma) \nn & \exp\Big[ - \frac{1}{4 \pi \alpha'} \int_{M} d^{2} \sigma \sqrt{g(\sigma)} \Big\{ \Big( g^{ab}(\sigma) G_{\mu\nu}(x,z_{f}) + i \epsilon^{ab}B_{\mu\nu}(x,z_{f}) \Big) \partial_{a} x^{\mu}(\sigma) \partial_{b} x^{\nu}(\sigma) + \alpha' R^{(2)}(\sigma) \Phi(x,z_{f}) \Big\} 
				\nn & \hspace*{6.5cm}\quad \quad \quad \quad \quad \quad \quad \quad \quad \quad \quad ~ - \frac{1}{2 \pi} \int_{M} d^{2} \sigma \sqrt{g(\sigma)} b_{ab}(\sigma) \nabla^{a} c^{b}(\sigma) \Big] 
				\nn \approx & \exp\Big[ - \frac{1}{\alpha'} \int d^{D} x \sqrt{G(x,z_{f})} e^{- 2 \Phi(x,z_{f})} \Big\{ \frac{\alpha'}{4} R(x,z_{f}) - \frac{D - 26}{6} - \frac{\alpha'}{48} H_{\mu\nu\lambda}(x,z_{f}) H^{\mu\nu\lambda}(x,z_{f}) \nn & \quad \quad \quad \quad \quad \quad \quad \quad \quad \quad \quad \quad \quad \quad \quad \quad \quad \quad \quad \quad \quad \quad \quad \quad \quad \quad \quad \quad + \alpha' \partial_{\mu} \Phi(x,z_{f}) \partial^{\mu} \Phi(x,z_{f}) \Big\} \Big] , \label{String_IRBC}
			\end{align}
where the path integral $\int D x^{\mu}(\sigma) D b_{ab}(\sigma) D c^{a}(\sigma)$ has been taken to give the low-energy effective action at the IR boundary. Considering the Wilsonian RG transformations, this residual string action exists at the IR boundary as long as quantum fluctuations of a string are not taken over completely. In this respect this string action or the resulting low-energy effective action after the integration of residual string fluctuations may be regarded as the IR boundary action of the bulk effective action Eq.\ (\ref{HDEFT}). Note that $z_f$ is introduced as a cutoff near the IR thus providing a nonzero $R(x,z_{f})$. Here, $\approx$ implies that the equivalence has not been demonstrated explicitly as far as we know \cite{RG_Flow_NLsM_I}. We recall that the low energy effective action for the target spacetime has been constructed or proposed to reproduce the vanishing RG $\beta$-function conditions \cite{RG_Flow_NLsM_II}.
			
To determine the IR boundary condition for ($G_{\mu\nu}(x,z_{f})$,$B_{\mu\nu}(x,z_{f})$,$\Phi(x,z_{f})$) and the UV one for ($G_{\mu\nu}(x,0)$,$B_{\mu\nu}(x,0)$,$\Phi(x,0)$), we have to consider an effective on-shell action functional,
			\begin{align} 
				\mathcal{S}_{\text{eff}} & = \frac{1}{\alpha'} \int d^{D} x \sqrt{G(x,z_{f})} e^{- 2 \Phi(x,z_{f})} \Big\{ \Pi_{G}^{\mu\nu}(x,z_{f}) G_{\mu\nu}(x,z_{f}) + \Pi_{B}^{\mu\nu}(x,z_{f}) B_{\mu\nu}(x,z_{f}) + \Pi_{\Phi}(x,z_{f}) \Phi(x,z_{f}) 
				\nn & + \frac{\alpha'}{4} R(x,z_{f}) - \frac{D - 26}{6} - \frac{\alpha'}{48} H_{\mu\nu\lambda}(x,z_{f}) H^{\mu\nu\lambda}(x,z_{f}) + \alpha' \partial_{\mu} \Phi(x,z_{f}) \partial^{\mu} \Phi(x,z_{f}) \Big\} \nn & - \frac{1}{\alpha'} \int d^{D} x \sqrt{G(x,0)} e^{- 2 \Phi(x,0)} \Big\{ \Pi_{G}^{\mu\nu}(x,0) G_{\mu\nu}(x,0) + \Pi_{B}^{\mu\nu}(x,0) B_{\mu\nu}(x,0) + \Pi_{\Phi}(x,0) \Phi(x,0) \Big\} .
				\label{OS}
			\end{align}
Here, $\Pi_{G}^{\mu\nu}(x,z_{f}) G_{\mu\nu}(x,z_{f}) + \Pi_{B}^{\mu\nu}(x,z_{f}) B_{\mu\nu}(x,z_{f}) + \Pi_{\Phi}(x,z_{f}) \Phi(x,z_{f})$ and $\Pi_{G}^{\mu\nu}(x,0) G_{\mu\nu}(x,0) + \Pi_{B}^{\mu\nu}(x,0) B_{\mu\nu}(x,0) + \Pi_{\Phi}(x,0) \Phi(x,0)$ result from the bulk action functional in Eq.\ (\ref{HDEFT}), given by the integration by parts, while the Wilsonian effective potential comes from the path integral in Eq.\ (\ref{String_IRBC}). Applying the variational principle for ($G_{\mu\nu}(x,z_{f})$,$B_{\mu\nu}(x,z_{f})$,$\Phi(x,z_{f})$) and for ($G_{\mu\nu}(x,0)$,$B_{\mu\nu}(x,0)$,$\Phi(x,0)$), we will obtain the IR boundary condition for ($\Pi_{G}^{\mu\nu}(x,z_{f})$,$\Pi_{B}^{\mu\nu}(x,z_{f})$,$\Pi_{\Phi}(x,z_{f})$) and the UV one for ($\Pi_{G}^{\mu\nu}(x,0)$,$\Pi_{B}^{\mu\nu}(x,0)$,$\Pi_{\Phi}(x,0)$), respectively. Resorting to the Hamilton's equations of motion for these canonical momenta, we derive both the boundary conditions for ($G_{\mu\nu}(x,z_{f})$,$B_{\mu\nu}(x,z_{f})$,$\Phi(x,z_{f})$) and for ($G_{\mu\nu}(x,0)$,$B_{\mu\nu}(x,0)$,$\Phi(x,0)$), respectively, which will be discussed further in section \ref{HDEFT_Entropy_Gradient_Flow}.
			
The above on-shell effective action is expected to satisfy the Hamilton-Jacobi equation \cite{Holographic_Duality_V,Holographic_Duality_VI,Holographic_Duality_VII}, which results from the RG invariance of the partition function,
\begin{align} 
			& \frac{d }{d z_{f}} \ln \mathcal{Z} = 0 ,
				\label{HJ}
			\end{align}
where we recall that the full partition function is 
			\begin{align} 
				\mathcal{Z} & = \int D G_{\mu\nu}(x,z) D \Pi_{G}^{\mu\nu}(x,z) D B_{\mu\nu}(x,z) D \Pi^{\mu\nu}_{B}(x,z) D \Phi(x,z) D \Pi_{\Phi}(x,z) ~ \mathcal{J}\Big(\frac{\partial}{\partial G_{\mu\nu}(x,z)},\frac{\partial}{\partial B_{\mu\nu}(x,z)},\frac{\partial}{\partial \Phi(x,z)}\Big)
				\nn & \exp\Big[ - \frac{1}{\alpha'} \int d^{D} x \sqrt{G(x,z_{f})} e^{- 2 \Phi(x,z_{f})} \Big\{ \frac{\alpha'}{4} R(x,z_{f}) - \frac{D - 26}{6} - \frac{\alpha'}{48} H_{\mu\nu\lambda}(x,z_{f}) H^{\mu\nu\lambda}(x,z_{f}) \nn & \quad \quad \quad \quad \quad \quad \quad \quad \quad \quad \quad \quad \quad \quad \quad \quad \quad \quad \quad \quad \quad \quad \quad \quad \quad \quad \quad \quad + \alpha' \partial_{\mu} \Phi(x,z_{f}) \partial^{\mu} \Phi(x,z_{f}) \Big\} \Big]
				\nn 
				& \exp\Big[ - \frac{1}{\alpha'} \int_{0}^{z_{f}} d z \int d^{D} x \sqrt{G(x,z)} e^{- 2 \Phi(x,z)}
				\Big\{ \Pi_{G}^{\mu\nu}(x,z) \Big(\partial_{z} G_{\mu\nu}(x,z) + \alpha' R_{\mu\nu}(x,z) + 2 \alpha' \nabla_{\mu} \nabla_{\nu} \Phi(x,z) \nn & - \frac{\alpha'}{4} H_{\mu\lambda\omega}(x,z) H_{\nu}^{\lambda\omega}(x,z) \Big) + \frac{\lambda}{2} \Pi_{G}^{\mu\nu}(x,z) \mathcal{G}_{\mu\nu\rho\lambda}(x,z) \Pi_{G}^{\rho\lambda}(x,z) \nn & + \Pi_{B}^{\mu\nu}(x,z) \Big(\partial_{z} B_{\mu\nu}(x,z) - \frac{\alpha'}{2} \nabla^{\omega} H_{\omega\mu\nu}(x,z) + \alpha' \nabla^{\omega} \Phi(x,z) H_{\omega\mu\nu}(x,z) \Big) + \frac{q}{2} \Pi_{B,\mu\nu}(x,z) \Pi_{B}^{\mu\nu}(x,z) \nn &
				+ \Pi_{\Phi}(x,z) \Big(\partial_{z} \Phi(x,z) + \frac{D - 26}{6} - \frac{\alpha'}{2} \nabla^{2} \Phi(x,z) + \alpha' \partial_{\mu} \Phi(x,z) \partial^{\mu} \Phi(x,z) - \frac{\alpha'}{24} H_{\mu\nu\lambda}(x,z) H^{\mu\nu\lambda}(x,z)\Big) 
				\nn & + \frac{\alpha'}{4} R(x,z) - \frac{D - 26}{6} - \frac{\alpha'}{48} H_{\mu\nu\lambda}(x,z) H^{\mu\nu\lambda}(x,z) + \alpha' \partial_{\mu} \Phi(x,z) \partial^{\mu} \Phi(x,z) \Big\} \Big] . \label{Full_Partition_Function} \end{align}
			%
			%
We point out that the IR boundary effective action has been introduced explicitly. As a result, we obtain from \eqref{HJ}
			\begin{align} 0 & = 
				\frac{\lambda}{2} \Pi_{G}^{\mu\nu}(x,z_{f}) \mathcal{G}_{\mu\nu\rho\lambda}(x,z_{f}) \Pi_{G}^{\rho\lambda}(x,z_{f}) + \Pi_{G}^{\mu\nu}(x,z_{f}) \Big(\alpha' R_{\mu\nu}(x,z) + 2 \alpha' \nabla_{\mu} \nabla_{\nu} \Phi(x,z_{f}) - \frac{\alpha'}{4} H_{\mu\lambda\omega}(x,z_{f}) H_{\nu}^{\lambda\omega}(x,z_{f}) \Big) \nn & + \frac{q}{2} \Pi_{B}^{\mu\nu}(x,z_{f}) \Pi_{B,\mu\nu}(x,z_{f}) + \Pi_{B}^{\mu\nu}(x,z_{f}) \Big( - \frac{\alpha'}{2} \nabla^{\omega} H_{\omega\mu\nu}(x,z_{f}) + \alpha' \nabla^{\omega} \Phi(x,z_{f}) H_{\omega\mu\nu}(x,z_{f}) \Big)
				\nn & + \Pi_{\Phi}(x,z_{f}) \Big( \frac{D - 26}{6} - \frac{\alpha'}{2} \nabla^{2} \Phi(x,z_{f}) + \alpha' \partial_{\mu} \Phi(x,z_{f}) \partial^{\mu} \Phi(x,z_{f}) - \frac{\alpha'}{24} H_{\mu\nu\lambda}(x,z_{f}) H^{\mu\nu\lambda}(x,z_{f})\Big) 
				\nn & + \frac{\alpha'}{4} R(x,z_{f}) - \frac{D - 26}{6} - \frac{\alpha'}{48} H_{\mu\nu\lambda}(x,z_{f}) H^{\mu\nu\lambda}(x,z_{f}) + \alpha' \partial_{\mu} \Phi(x,z_{f}) \partial^{\mu} \Phi(x,z_{f}) . \end{align}
			We recall that both of $\Pi_{G}^{\mu\nu} (x,z_{f})\mathcal{G}_{\mu\nu\rho\lambda}(x,z_{f})\Pi_{G}^{\rho\lambda} (x,z_{f})$ and $\frac{q}{2} \Pi_{B}^{\mu\nu}(x,z_{f}) \Pi_{B,\mu\nu}(x,z_{f})$ originate from random noise fluctuations or equivalently, the corresponding $T \bar{T}$ and $J J$ deformations, respectively \cite{TTbar_Deformation}.
			
In the above expression Eq.\ (\ref{Full_Partition_Function}), we did not introduce the contribution from the ghost action explicitly into the holographic dual EFT, which takes into account the Jacobian factor in the Faddeev-Popov procedure. As expected, it turns out that the introduction of the ghost action gives rise to BRST symmetries for the gauge-fixed action \cite{MSR_Formulation_SUSY_i,MSR_Formulation_SUSY_ii,MSR_Formulation_SUSY_iii,MSR_Formulation_SUSY_iv,MSR_Formulation_SUSY_v,MSR_Formulation_SUSY_vi, Schwinger_Keldysh_Symmetries_i,Schwinger_Keldysh_Symmetries_ii,Schwinger_Keldysh_Symmetries_iii,Schwinger_Keldysh_Symmetries_iv,Schwinger_Keldysh_Symmetries_v,Schwinger_Keldysh_Symmetries_vi}. Recently, we have shown that such BRST symmetries are responsible for novel Ward identities in some correlation functions of the target spacetime metric and the canonical momentum \cite{RG_Flow_Holography_Monotonicity}. One of the Ward identities corresponds to a generalized fluctuation-dissipation theorem (FDT) \cite{Jarzynski_i,Jarzynski_ii,Crooks_i,Crooks_ii,Crooks_iii}, where the `equilibrium' FDT is modified by the RG flow equations, regarded to be a generalization into `non-equilibrium' given by the existence of the RG flow \cite{RG_Flow_Holography_Monotonicity}. These Ward identities give some constraints for the dynamics of dual collective fields (composite particles). 

It is important to figure out how renormalization effects are introduced into the holographic dual effective field theory. In particular, it is necessary to understand the role of random noise fluctuations in the RG flow. In appendix \ref{NP}, we have discussed in detail the non-perturbative nature of the constructed holographic dual effective field theory. This discussion serves as our fundamental motivation for the present EFT formulation.

\section{Entropy production rate in the emergent holographic dual effective field theory} \label{Entropy_Production_Gibbs_Type}
			
%
%
To better understand the properties of the holographic dual effective field theory Eq.\ (\ref{HDEFT}) we calculate 
path-dependent entropy associated with the RG flows of the metric and gauge field. Here, we follow the procedure pioneered by Seifert \cite{Entropy_Production}, reviewed in appendix \ref{Review:Langevin}. We will first construct a probability distribution function associated with the Brownian motion in the target spacetime and express the Fokker-Planck equation in the phase space constructed from all the fields and their conjugate momenta.
			
We recall the partition function from \eqref{HDEFT},
\begin{align} 
				Z = & \int D G_{\mu\nu}(x,z) D \Pi_{G}^{\mu\nu}(x,z) D B_{\mu\nu}(x,z) D \Pi^{\mu\nu}_{B}(x,z) D \Phi(x,z) ~ \mathcal{J}\Big(\frac{\partial}{\partial G_{\mu\nu}(x,z)},\frac{\partial}{\partial B_{\mu\nu}(x,z)},\frac{\partial}{\partial \Phi(x,z)}\Big) \nn & \delta\Big(\partial_{z} \Phi(x,z) + \frac{D - 26}{6} - \frac{\alpha'}{2} \nabla^{2} \Phi(x,z) + \alpha' \partial_{\mu} \Phi(x,z) \partial^{\mu} \Phi(x,z) - \frac{\alpha'}{24} H_{\mu\nu\lambda}(x,z) H^{\mu\nu\lambda}(x,z)\Big) \nn & \exp\Big[ - \frac{1}{\alpha'} \int_{0}^{z_{f}} d z \int d^{D} x \sqrt{G(x,z)} e^{- 2 \Phi(x,z)}
				\Big\{ \Pi_{G}^{\mu\nu}(x,z) \Big(\partial_{z} G_{\mu\nu}(x,z) + \alpha' R_{\mu\nu}(x,z) + 2 \alpha' \nabla_{\mu} \nabla_{\nu} \Phi(x,z) \nn & - \frac{\alpha'}{4} H_{\mu\lambda\omega}(x,z) H_{\nu}^{\lambda\omega}(x,z) \Big) + \frac{\lambda}{2} \Pi_{G}^{\mu\nu}(x,z) \mathcal{G}_{\mu\nu\rho\lambda}(x,z) \Pi_{G}^{\rho\lambda}(x,z) \nn & + \Pi_{B}^{\mu\nu}(x,z) \Big(\partial_{z} B_{\mu\nu}(x,z) - \frac{\alpha'}{2} \nabla^{\omega} H_{\omega\mu\nu}(x,z) + \alpha' \nabla^{\omega} \Phi(x,z) H_{\omega\mu\nu}(x,z) \Big) + \frac{q}{2} \Pi_{B,\mu\nu}(x,z) \Pi_{B}^{\mu\nu}(x,z)
				\nn & + \frac{\alpha'}{4} R(x,z) - \frac{D - 26}{6} - \frac{\alpha'}{48} H_{\mu\nu\lambda}(x,z) H^{\mu\nu\lambda}(x,z) + \alpha' \partial_{\mu} \Phi(x,z) \partial^{\mu} \Phi(x,z) \Big\} \Big] . \label{HDEFT_Bulk_Constraint} \end{align}
where the canonical momentum of the dilaton field has been integrated out. 
			
We now focus on the bulk part to derive the Fokker-Planck equation. The bulk effective Lagrangian is given by
\begin{align}
\mathcal{L}_{\text{eff}} &= \Pi_{G}^{\mu\nu}(x,z) \Big(\partial_{z} G_{\mu\nu}(x,z) + \alpha' R_{\mu\nu}(x,z) + 2 \alpha' \nabla_{\mu} \nabla_{\nu} \Phi(x,z) - \frac{\alpha'}{4} H_{\mu\lambda\omega}(x,z) H_{\nu}^{\lambda\omega}(x,z) \Big) \nn & + \frac{\lambda}{2} \Pi_{G}^{\mu\nu}(x,z) \mathcal{G}_{\mu\nu\rho\lambda}(x,z) \Pi_{G}^{\rho\lambda}(x,z) \nn & + \Pi_{B}^{\mu\nu}(x,z) \Big(\partial_{z} B_{\mu\nu}(x,z) - \frac{\alpha'}{2} \nabla^{\omega} H_{\omega\mu\nu}(x,z) + \alpha' \nabla^{\omega} \Phi(x,z) H_{\omega\mu\nu}(x,z) \Big) + \frac{q}{2} \Pi_{B,\mu\nu}(x,z) \Pi_{B}^{\mu\nu}(x,z)
				\nn & + \frac{\alpha'}{4} R(x,z) - \frac{D - 26}{6} - \frac{\alpha'}{48} H_{\mu\nu\lambda}(x,z) H^{\mu\nu\lambda}(x,z) + \alpha' \partial_{\mu} \Phi(x,z) \partial^{\mu} \Phi(x,z) \ .   
\end{align}
Considering the Legendre transformation,
			\begin{align} & \mathcal{L}_{\text{eff}} = \Pi_{G}^{\mu\nu}(x,z) \partial_{z} G_{\mu\nu}(x,z) + \Pi_{B}^{\mu\nu}(x,z) \partial_{z} B_{\mu\nu}(x,z) + \mathcal{H}_{\text{eff}} \end{align}
in Euclidean spacetime, we obtain the effective bulk Hamiltonian as
			\begin{align} 
				\mathcal{H}_{\text{eff}} &= \Pi_{G}^{\mu\nu}(x,z) \Big( \alpha' R_{\mu\nu}(x,z) + 2 \alpha' \nabla_{\mu} \nabla_{\nu} \Phi(x,z) - \frac{\alpha'}{4} H_{\mu\lambda\omega}(x,z) H_{\nu}^{\lambda\omega}(x,z) \Big) + \frac{\lambda}{2} \Pi_{G}^{\mu\nu}(x,z) \mathcal{G}_{\mu\nu\rho\lambda}(x,z) \Pi_{G}^{\rho\lambda}(x,z) \nn & + \Pi_{B}^{\mu\nu}(x,z) \Big( - \frac{\alpha'}{2} \nabla^{\omega} H_{\omega\mu\nu}(x,z) + \alpha' \nabla^{\omega} \Phi(x,z) H_{\omega\mu\nu}(x,z) \Big) + \frac{q}{2} \Pi_{B,\mu\nu}(x,z) \Pi_{B}^{\mu\nu}(x,z)
				\nn & + \frac{\alpha'}{4} R(x,z) - \frac{D - 26}{6} - \frac{\alpha'}{48} H_{\mu\nu\lambda}(x,z) H^{\mu\nu\lambda}(x,z) + \alpha' \partial_{\mu} \Phi(x,z) \partial^{\mu} \Phi(x,z) . \label{Effective_Bulk_Hamiltonian} \end{align}
			
To obtain an effective Fokker-Planck equation, we introduce the identification
\begin{align} & \Pi_{G}^{\mu\nu}(x,z) \equiv - \frac{\partial}{\partial G_{\mu\nu}(x,z)} , ~~~~~ \Pi_{B}^{\mu\nu}(x,z) \equiv - \frac{\partial}{\partial B_{\mu\nu}(x,z)} \label{Canonical_Momentum} \end{align}
into the effective Hamiltonian, in a similar manner discussed in the review discussed in appendix \ref{Review:Langevin}. Then, we construct the corresponding Fokker-Planck equation for the Langevin-type RG flow equation (modified by noise fluctuations) as follows 
\begin{align}
				\label{eq:FokkerPlanck}
				&\Big( \partial_{z} - \frac{\alpha'}{4} R(x,z) + \frac{D - 26}{6} + \frac{\alpha'}{48} H_{\mu\nu\lambda}(x,z) H^{\mu\nu\lambda}(x,z) - \alpha' \partial_{\mu} \Phi(x,z) \partial^{\mu} \Phi(x,z) \Big) \mathcal{P}(G_{\mu\nu},B_{\mu\nu};z) \notag\\ = &- \frac{\partial}{\partial G_{\mu\nu}(x,z)} \Big\{ \Big( \alpha' R_{\mu\nu}(x,z) + 2 \alpha' \nabla_{\mu} \nabla_{\nu} \Phi(x,z) - \frac{\alpha'}{4} H_{\mu\lambda\omega}(x,z) H_{\nu}^{\lambda\omega}(x,z) - \frac{\lambda}{2} \mathcal{G}_{\mu\nu\rho\gamma}(x,z) \frac{\partial}{\partial G_{\rho\gamma}(x,z)} \Big) \mathcal{P}(G_{\mu\nu},B_{\mu\nu};z) \Big\} \nn &- \frac{\partial}{\partial B_{\mu\nu}(x,z)} \Big\{ \Big( - \frac{\alpha'}{2} \nabla^{\omega} H_{\omega\mu\nu}(x,z) + \alpha' \nabla^{\omega} \Phi(x,z) H_{\omega\mu\nu}(x,z) - \frac{q}{2} \frac{\partial}{\partial B_{\mu\nu}(x,z)} \Big) \mathcal{P}(G_{\mu\nu},B_{\mu\nu};z) \Big\} , 
\end{align}
where $\mathcal{P}(G_{\mu\nu},B_{\mu\nu};z)$ is the probability distribution function associated with both the RG flows of the metric and two-form gauge field along $z$ direction. 
			
One can also derive this Fokker-Planck equation from 
\begin{align}
						\mathcal{P}(G_{\mu\nu},B_{\mu\nu};z) & = {\cal{N}}\int D \xi_{\mu\nu}(x,l) D \zeta_{\mu\nu}(x,l) ~ \exp\Big[ - \frac{1}{\alpha'} \int_{0}^{z} d l \int d^{D} x \sqrt{G(x,l)} e^{- 2 \Phi(x,l)} \Big\{ \nn & - \frac{1}{2 \lambda} \xi_{\mu\nu}(x,l) \mathcal{G}^{\mu\nu\rho\lambda}(x,l) \xi_{\rho\lambda}(x,l) - \frac{1}{2 q} \zeta_{\mu\nu}(x,l) \zeta^{\mu\nu}(x,l) \nn & + \frac{\alpha'}{4} R(x,l) - \frac{D - 26}{6} - \frac{\alpha'}{48} H_{\mu\nu\lambda}(x,l) H^{\mu\nu\lambda}(x,l) + \alpha' \partial_{\mu} \Phi(x,l) \partial^{\mu} \Phi(x,l) \Big\} \Big] \nn & ~ \delta\Big(G_{\mu\nu}-G_{\mu\nu}(x,z)\Big) \delta\Big(B_{\mu\nu}-B_{\mu\nu}(x,z)\Big) \ , \label{eq:FokkerPlanck_Probability_Def}
			\end{align}
where
the RG flow equations for both 
the metric and the gauge field
have to be used. 
$\cal{N}$ is a normalization constant given by $Z^{-1}$ of Eq.\ (\ref{HDEFT_Bulk_Constraint}). 
Applying the Faddeev-Popov procedure 
to Eq.\ (\ref{eq:FokkerPlanck_Probability_Def}) and performing the ensemble averaging with respect to noise fluctuations, we obtain the path integral expression for the probability distribution function as follows 
\begin{align} 
				\mathcal{P}(G_{\mu\nu},B_{\mu\nu};z) = & \frac{1}{Z} \int^{[G_{\mu\nu}(x,z),B_{\mu\nu}(x,z),\Phi(x,z)]} D G_{\mu\nu}(x,l) D \Pi_{G}^{\mu\nu}(x,l) D B_{\mu\nu}(x,l) D \Pi^{\mu\nu}_{B}(x,l) D \Phi(x,l) \nn & \mathcal{J}\Big(\frac{\partial}{\partial G_{\mu\nu}(x,l)},\frac{\partial}{\partial B_{\mu\nu}(x,l)},\frac{\partial}{\partial \Phi(x,l)}\Big) \nn & \delta\Big(\partial_{l} \Phi(x,l) + \frac{D - 26}{6} - \frac{\alpha'}{2} \nabla^{2} \Phi(x,l) + \alpha' \partial_{\mu} \Phi(x,l) \partial^{\mu} \Phi(x,l) - \frac{\alpha'}{24} H_{\mu\nu\lambda}(x,l) H^{\mu\nu\lambda}(x,l)\Big) \nn & \exp\Big[ - \frac{1}{\alpha'} \int_{0}^{z} d l \int d^{D} x \sqrt{G(x,l)} e^{- 2 \Phi(x,l)}
				\Big\{ \Pi_{G}^{\mu\nu}(x,l) \Big(\partial_{l} G_{\mu\nu}(x,l) + \alpha' R_{\mu\nu}(x,l) + 2 \alpha' \nabla_{\mu} \nabla_{\nu} \Phi(x,l) \nn & - \frac{\alpha'}{4} H_{\mu\lambda\omega}(x,l) H_{\nu}^{\lambda\omega}(x,l) \Big) + \frac{\lambda}{2} \Pi_{G}^{\mu\nu}(x,l) \mathcal{G}_{\mu\nu\rho\lambda}(x,l) \Pi_{G}^{\rho\lambda}(x,l) \nn & + \Pi_{B}^{\mu\nu}(x,l) \Big(\partial_{l} B_{\mu\nu}(x,l) - \frac{\alpha'}{2} \nabla^{\omega} H_{\omega\mu\nu}(x,l) + \alpha' \nabla^{\omega} \Phi(x,l) H_{\omega\mu\nu}(x,l) \Big) + \frac{q}{2} \Pi_{B,\mu\nu}(x,l) \Pi_{B}^{\mu\nu}(x,l)
				\nn & + \frac{\alpha'}{4} R(x,l) - \frac{D - 26}{6} - \frac{\alpha'}{48} H_{\mu\nu\lambda}(x,l) H^{\mu\nu\lambda}(x,l) + \alpha' \partial_{\mu} \Phi(x,l) \partial^{\mu} \Phi(x,l) \Big\} \Big] , \label{eq:FokkerPlanck_Path_Integral} \end{align}
where the RG flow of the dilaton field has been introduced. 
In this respect Eq.\ (\ref{eq:FokkerPlanck_Path_Integral}) is a formal path-integral solution of the Fokker-Planck equation (\ref{eq:FokkerPlanck}). See appendix \ref{Review:Langevin} for the Langevin system. 

In the effective Fokker-Planck equation (\ref{eq:FokkerPlanck}), both the RG $\beta$-functions of $\beta_{\mu\nu}^{G}$ and $\beta_{\mu\nu}^{B}$ play the role of an external force while the Wilsonian effective potential $\frac{\alpha'}{4} R(x,z) - \frac{D - 26}{6} - \frac{\alpha'}{48} H_{\mu\nu\lambda}(x,z) H^{\mu\nu\lambda}(x,z) + \alpha' \partial_{\mu} \Phi(x,z) \partial^{\mu} \Phi(x,z)$ acts as an external potential. The conserved (generalized) symmetric current 
associated with the conservation of the probability in the metric and Kalb-Ramond
field is given by 
\begin{align}
\label{eq:conservedJ}
				J_{\mu\nu}^{G}(x,z) & = \Big( \alpha' R_{\mu\nu}(x,z) + 2 \alpha' \nabla_{\mu} \nabla_{\nu} \Phi(x,z) - \frac{\alpha'}{4} H_{\mu\lambda\omega}(x,z) H_{\nu}^{\lambda\omega}(x,z) - \frac{\lambda}{2} \mathcal{G}_{\mu\nu\rho\gamma}(x,z) \frac{\partial}{\partial G_{\rho\gamma}(x,z)} \Big) \mathcal{P}(G_{\mu\nu},B_{\mu\nu};z) , \\ \label{eq:conservedJB} J_{\mu\nu}^{B}(x,z) & = \Big( - \frac{\alpha'}{2} \nabla^{\omega} H_{\omega\mu\nu}(x,z) + \alpha' \nabla^{\omega} \Phi(x,z) H_{\omega\mu\nu}(x,z) - \frac{q}{2} \frac{\partial}{\partial B^{\mu\nu}(x,z)} \Big) \mathcal{P}(G_{\mu\nu},B_{\mu\nu};z) .    
			\end{align}
			%
			%
One can find a similar rank 2 current in gravity theories with ADM foliation \cite{Kim:2024rhw}. 
Computing the Hamiltonian equation of motion for the canonical momentum $\frac{\partial\mathcal{H}_{\text{eff}}}{\partial\Pi_{G}^{\mu\nu}(x,z)}=-\partial_z G_{\mu\nu}(x,z)$ and $\frac{\partial\mathcal{H}_{\text{eff}}}{\partial\Pi_{B}^{\mu\nu}(x,z)}=-\partial_z B_{\mu\nu}(x,z)$ from Eq.\ (\ref{Effective_Bulk_Hamiltonian}), given by
			\begin{align} \Pi_{G}^{\mu\nu}(x,z)  & = - \frac{1}{\lambda} \mathcal{G}^{\mu\nu\rho\lambda}(x,z) \Big(\partial_{z} G_{\rho\lambda}(x,z) + \alpha' R_{\rho\lambda}(x,z) + 2 \alpha' \nabla_{\rho} \nabla_{\lambda} \Phi(x,z) - \frac{\alpha'}{4} H_{\rho\delta\omega}(x,z) H_{\lambda}^{\delta\omega}(x,z) \Big) , \\ \Pi_{B}^{\mu\nu}(x,z)  & = - \frac{1}{q} \Big(\partial_{z} B^{\mu\nu}(x,z) - \frac{\alpha'}{2} \nabla^{\omega} H_{\omega}^{\mu\nu}(x,z) + \alpha' \nabla^{\omega} \Phi(x,z) H_{\omega}^{\mu\nu}(x,z) \Big) , \end{align}
we can express the current as $J_{\mu\nu}^{G}(x,z) \sim \mathcal{P}(G_{\mu\nu},B_{\mu\nu};z) \partial_{z} G_{\mu\nu}(x,z)$ and $J_{\mu\nu}^{B}(x,z) \sim \mathcal{P}(G_{\mu\nu},B_{\mu\nu};z) \partial_{z} B_{\mu\nu}(x,z)$, 
which is conceptually the same as the conserved current of the Brownian motion discussed in appendix \ref{Review:Langevin}.
			
Based on these currents, the Fokker-Planck equation (\ref{eq:FokkerPlanck}) is formally expressed as
\begin{align} 
				& \partial_{z} \mathcal{P}(G_{\mu\nu},B_{\mu\nu};z) + \partial_{G_{\mu\nu}} J_{\mu\nu}^{G}(x,z) + \partial_{B_{\mu\nu}} J_{\mu\nu}^{B}(x,z) = \mathcal{V}_{eff}(x,z) \mathcal{P}(G_{\mu\nu},B_{\mu\nu};z) , \label{FP_Eq_Formal}
\end{align}
where $\mathcal{V}_{eff}(x,z)$ is the Wilsonian effective potential,
\begin{align} 
				& \mathcal{V}_{eff}(x,z) = \frac{\alpha'}{4} R(x,z) - \frac{D - 26}{6} - \frac{\alpha'}{48} H_{\mu\nu\lambda}(x,z) H^{\mu\nu\lambda}(x,z) + \alpha' \partial_{\mu} \Phi(x,z) \partial^{\mu} \Phi(x,z) .  
\end{align}
Since there is an external `potential' term in the RG flow, the conservation law is modified.

At this point, a few remarks are in order.
\begin{itemize}
\item We emphasize that this intuitive construction of the Fokker-Planck equation for the metric RG flow is completely parallel to that for the Langevin equation, where we have used the replacement given in Eq.\ \eqref{Canonical_Momentum}. In essence, this is the traditional method of \emph{quantization} where the canonically conjugate momenta is identified with the derivative operator with respect to the dynamical variable (which in this case corresponds to the metric $G_{\mu \nu}(x,z)$ and the Kalb-Ramond gauge field $B_{\mu\nu}(x,z)$). 

\item The formal solution to the Fokker-Planck equation is given by the above path integral expression of the holographic dual EFT. In this respect, the Fokker-Planck equation is analogous to the Wheeler-DeWitt equation for quantum gravity \cite{DeWitt_Metric}. 

\item It would be interesting to figure out the role of this conserved current in correlation functions of dual collective fields, i.e., symmetries and Ward identities, not addressed in this study.
In particular, 
the right-hand side of Eq. (\ref{FP_Eq_Formal}) may lead us to interpret it as a quantum anomaly in the RG flow beyond the cohomological-type topological field theory construction.
\end{itemize}

			%
			%
			
			%
			%
			
Following Seifert \cite{Entropy_Production}, we introduce a path-dependent microscopic entropy in the target space as
\begin{align}\label{eq:microalaSeifert} & s_{\text{sys}}(G_{\mu\nu},B_{\mu\nu}) = - \ln \mathcal{P}(G_{\mu\nu},B_{\mu\nu};z) . \end{align}
Then, the observable `system' entropy is given by the average of the path-dependent entropy with respect to the probability distribution in the following way
\begin{align} & S_{\text{sys}}(z) = \langle s_{\text{sys}}(G_{\mu\nu},B_{\mu\nu};z) \rangle = - \int_{[G_{\mu\nu}(x,0),B_{\mu\nu}(x,0)]}^{[G_{\mu\nu}(x,z_{f}),B_{\mu\nu}(x,z_{f})]} d [G_{\mu\nu},B_{\mu\nu}] \mathcal{P}(G_{\mu\nu},B_{\mu\nu};z) \ln \mathcal{P}(G_{\mu\nu},B_{\mu\nu};z) . \end{align}
Here, $\int_{[G_{\mu\nu}(x,0),B_{\mu\nu}(x,0)]}^{[G_{\mu\nu}(x,z_{f}),B_{\mu\nu}(x,z_{f})]} d [G_{\mu\nu},B_{\mu\nu}]$ means the line integral, 
given the dilaton field as a functional of $G_{\mu\nu}(x,z)$ and $B_{\mu\nu}(x,z)$ in the $\delta-$function constraint of Eq. (\ref{HDEFT_Bulk_Constraint}).
We suspect that this Gibbs entropy would reproduce the black hole entropy at high temperatures. 
			
From Eq.\ \eqref{eq:microalaSeifert}, it follows that
\begin{align} & \partial_{z} s_{\text{sys}}(G_{\mu\nu},B_{\mu\nu};z) = - \frac{\partial_{z} \mathcal{P}(G_{\mu\nu},B_{\mu\nu};z)}{\mathcal{P}(G_{\mu\nu},B_{\mu\nu};z)} - \frac{\partial_{G_{\mu\nu}} \mathcal{P}(G_{\mu\nu},B_{\mu\nu};z)}{\mathcal{P}(G_{\mu\nu},B_{\mu\nu};z)} [\partial_{z} G_{\mu\nu}(x,z)] - \frac{\partial_{B_{\mu\nu}} \mathcal{P}(G_{\mu\nu},B_{\mu\nu};z)}{\mathcal{P}(G_{\mu\nu},B_{\mu\nu};z)} [\partial_{z} B_{\mu\nu}(x,z)]\ , \end{align}
where the chain rule has been used in the second and third terms. Using the Fokker-Planck equation \eqref{eq:FokkerPlanck} along with the conserved current defined in \eqref{eq:conservedJ} and \eqref{eq:conservedJB}, we obtain
\begin{align} \partial_{z} s_{\text{sys}}(G_{\mu\nu},B_{\mu\nu};z) & = \frac{\partial_{G_{\mu\nu}} J_{\mu\nu}^{G}(x,z)}{\mathcal{P}(G_{\mu\nu},B_{\mu\nu};z)} + \frac{\partial_{B_{\mu\nu}} J_{\mu\nu}^{B}(x,z)}{\mathcal{P}(G_{\mu\nu},B_{\mu\nu};z)} \nn & - \frac{\alpha'}{4} R(x,z) + \frac{D - 26}{6} + \frac{\alpha'}{48} H_{\mu\nu\lambda}(x,z) H^{\mu\nu\lambda}(x,z) - \alpha' \partial_{\mu} \Phi(x,z) \partial^{\mu} \Phi(x,z)
				\nn & - \frac{2}{\lambda} \Big(\alpha' R^{\mu\nu}(x,z) + 2 \alpha' \nabla^{\mu} \nabla^{\nu} \Phi(x,z) - \frac{\alpha'}{4} H^{\mu\lambda\omega}(x,z) H^{\nu}_{\lambda\omega}(x,z) - \frac{J^{G,\mu\nu}(x,z)}{\mathcal{P}(G_{\mu\nu},B_{\mu\nu};z)} \Big) [\partial_{z} G_{\mu\nu}(x,z)] \nn & - \frac{2}{q} \Big(- \frac{\alpha'}{2} \nabla_{\omega} H^{\omega\mu\nu}(x,z) + \alpha' \nabla_{\omega} \Phi(x,z) H^{\omega\mu\nu}(x,z)  - \frac{J^{B,\mu\nu}(x,z)}{\mathcal{P}(G_{\mu\nu},B_{\mu\nu};z)} \Big) [\partial_{z} B_{\mu\nu}(x,z)] . \end{align} 
			
Benchmarking ref.\ \cite{Entropy_Production}, we define the \emph{environmental entropy} production rate as follows
\begin{align} \partial_{z} s_{\text{env}}(G_{\mu\nu},B_{\mu\nu};z) & = \frac{\alpha'}{4} R(x,z) - \frac{D - 26}{6} - \frac{\alpha'}{48} H_{\mu\nu\lambda}(x,z) H^{\mu\nu\lambda}(x,z) + \alpha' \partial_{\mu} \Phi(x,z) \partial^{\mu} \Phi(x,z)
				\nn & + \frac{2}{\lambda} \Big(\alpha' R^{\mu\nu}(x,z) + 2 \alpha' \nabla^{\mu} \nabla^{\nu} \Phi(x,z) - \frac{\alpha'}{4} H^{\mu\lambda\omega}(x,z) H^{\nu}_{\lambda\omega}(x,z) \Big) [\partial_{z} G_{\mu\nu}(x,z)] \nn & + \frac{2}{q} \Big(- \frac{\alpha'}{2} \nabla_{\omega} H^{\omega\mu\nu}(x,z) + \alpha' \nabla_{\omega} \Phi(x,z) H^{\omega\mu\nu}(x,z) \Big) [\partial_{z} B_{\mu\nu}(x,z)]  . \end{align}
In fact, this identification is quite natural since the RG $\beta$-function and the Wilsonian effective action play the roles of external force and potential, respectively. Thus, the production rate of the path-dependent \emph{microscopic total entropy} ($s_{\text{tot}}\equiv s_{\text{sys}}+s_{\text{env}}$) is given by
\begin{align} 
				\partial_{z} s_{\text{tot}}(G_{\mu\nu},B_{\mu\nu};z) &= \partial_{z} s_{\text{env}}(G_{\mu\nu},B_{\mu\nu};z) + \partial_{z} s_{\text{sys}}(G_{\mu\nu},B_{\mu\nu};z)
				\nn 
				& = \frac{\partial_{G_{\mu\nu}} J_{\mu\nu}^{G}(x,z)}{\mathcal{P}(G_{\mu\nu},B_{\mu\nu};z)} + \frac{2}{\lambda} \frac{J^{G,\mu\nu}(x,z)}{\mathcal{P}(G_{\mu\nu},B_{\mu\nu};z)} [\partial_{z} G_{\mu\nu}(x,z)] \nn & + \frac{\partial_{B_{\mu\nu}} J_{\mu\nu}^{B}(x,z)}{\mathcal{P}(G_{\mu\nu},B_{\mu\nu};z)} + \frac{2}{q} \frac{J^{B,\mu\nu}(x,z)}{\mathcal{P}(G_{\mu\nu},B_{\mu\nu};z)} [\partial_{z} B_{\mu\nu}(x,z)] \ . \end{align}
We must emphasize at this point that $s_{\text{tot}}$ is in fact not the thermodynamic entropy. 
			%
			%

Taking the ensemble average with the probability distribution $\mathcal{P}(G_{\mu\nu},B_{\mu\nu};z)$, we obtain,
\begin{align}
				\partial_{z} S_{\text{tot}}(z) & = \langle \partial_{z} s_{\text{tot}}(G_{\mu\nu},B_{\mu\nu};z) \rangle \nn & = \frac{2}{\lambda} \int_{G_{\mu\nu}(x,0)}^{G_{\mu\nu}(x,z_{f})} d G_{\mu\nu} \frac{J_{\mu\nu}^{G}(x,z) \mathcal{G}^{\mu\nu\rho\gamma}(x,z) J_{\rho\gamma}^{G}(x,z)}{\mathcal{P}(G_{\mu\nu},B_{\mu\nu};z)} + \frac{2}{q} \int_{B_{\mu\nu}(x,0)}^{B_{\mu\nu}(x,z_{f})} d B_{\mu\nu} \frac{J^{B}_{\mu\nu}(x,z)  J^{B,\mu\nu}(x,z)}{\mathcal{P}(G_{\mu\nu},B_{\mu\nu};z)} \geq 0 , \label{Entropy_Production_Monotonicity}   
			\end{align}
where we have used the following conservation equation, namely,
\begin{align} & \Big\langle \frac{\partial_{G_{\mu\nu}} J_{\mu\nu}^{G}(x,z)}{\mathcal{P}(G_{\mu\nu},B_{\mu\nu};z)} + \frac{\partial_{B_{\mu\nu}} J_{\mu\nu}^{B}(x,z)}{\mathcal{P}(G_{\mu\nu},B_{\mu\nu};z)} \Big\rangle = \int_{G_{\mu\nu}(x,0)}^{G_{\mu\nu}(x,z_{f})} d G_{\mu\nu} \partial_{G_{\mu\nu}} J_{\mu\nu}^{G}(x,z) + \int_{B_{\mu\nu}(x,0)}^{B_{\mu\nu}(x,z_{f})} d B_{\mu\nu} \partial_{B_{\mu\nu}} J_{\mu\nu}^{B}(x,z) = 0\ . \end{align}
%
%
In the line integral expression $\int_{G_{\mu\nu}(x,0)}^{G_{\mu\nu}(x,z_{f})} d G_{\mu\nu}$ ($\int_{B_{\mu\nu}(x,0)}^{B_{\mu\nu}(x,z_{f})} d B_{\mu\nu}$), 
$B_{\mu\nu}(x,z)$ and $\Phi(x,z)$ ($G_{\mu\nu}(x,z)$ and $\Phi(x,z)$) are given along the fixed path determined by the RG transformation.
Eq.\ \eqref{Entropy_Production_Monotonicity} indicates that the microscopic Gibbs-type entropy always increases during the RG flow. In particular, this expression is quite similar to Eq.\ (\ref{Entropy_Production}) based on the macroscopic definition of the entropy functional except for the integration in the metric and Kalb-Ramond field during the RG flow. We recall $J_{\mu\nu}^{G}(x,z) \sim \mathcal{P}(G_{\mu\nu},B_{\mu\nu};z) \partial_{z} G_{\mu\nu}(x,z)$ and $J_{\mu\nu}^{B}(x,z) \sim \mathcal{P}(G_{\mu\nu},B_{\mu\nu};z) \partial_{z} B_{\mu\nu}(x,z)$, where the RG flows of the metric and the Kalb-Ramond gauge field are modified by the introduction of noise. 
Based on the IR boundary condition to be discussed in the next section, we obtain 
		$J_{\mu\nu}^{G}(x,z) \sim \mathcal{P}(G_{\mu\nu},B_{\mu\nu};z) \beta_{\mu\nu}^{G}$ and $J_{\mu\nu}^{B}(x,z) \sim \mathcal{P}(G_{\mu\nu},B_{\mu\nu};z) \beta_{\mu\nu}^{B}$, where $\partial_{z} G_{\mu\nu}(x,z) \sim \beta_{\mu\nu}^{G}$ and $\partial_{z} B_{\mu\nu}(x,z) \sim \beta_{\mu\nu}^{B}$. As a result, we obtain
\begin{equation}
				\begin{aligned}
				 \frac{J_{\mu\nu}^{G}(x,z) \mathcal{G}^{\mu\nu\rho\gamma}(x,z) J_{\rho\gamma}^{G}(x,z)}{\mathcal{P}(G_{\mu\nu},B_{\mu\nu};z)} + \frac{J_{\mu\nu}^{B}(x,z) J^{B,\mu\nu}(x,z)}{\mathcal{P}(G_{\mu\nu},B_{\mu\nu};z)} \sim \mathcal{P}(G_{\mu\nu},B_{\mu\nu};z) \beta_{\mu\nu}^{G} \mathcal{G}^{\mu\nu\rho\gamma}(x,z)\beta_{\rho\gamma}^{G} + \mathcal{P}(G_{\mu\nu},B_{\mu\nu};z) \beta_{\mu\nu}^{B} \beta^{B,\mu\nu} \ ,
				\end{aligned}
			\end{equation}
which is essentially the same as the entropy production rate a la Perelman. 
			
			%
			%
			
One may ask the physical meaning of the total entropy, where the addition of the environmental entropy seems to be ambiguous. We suspect that the total entropy may correspond to the relative entropy since a background contribution (environmental entropy) has been `subtracted' from the system entropy. 
In this case the monotonicity of its gradient flow is guaranteed \cite{RG_Flow_Relative_Entropy_I,RG_Flow_Relative_Entropy_II,RG_Flow_Relative_Entropy_Gradient_Flow}. We discuss this issue in the last section.

\section{Holographic description of Perelman's entropy functional} \label{HDEFT_Entropy_Gradient_Flow}
			
In the previous section, we have introduced a microscopic entropy functional in terms of a probability distribution function to satisfy an effective Fokker-Planck equation analogous to the Wheeler–DeWitt equation \cite{DeWitt_Metric}. Subsequently, we investigated the monotonicity of this entropy functional. Our next goal is to compare this monotonicity behavior with that of a Perelman-type macroscopic entropy functional for the low-energy effective description of the worldsheet nonlinear $\sigma$ model \cite{Ricci_NLsM_Gradient_i,Ricci_NLsM_Gradient_ii, Yu_Nakamura_Ricci_Flow}. We will find that both share a similar functional form to support the holographic dual effective field theory.
			
Refs. \cite{Ricci_NLsM_Gradient_i,Ricci_NLsM_Gradient_ii} considered the RG flow equations for the target space metric and dilaton field 
(in addition, the two form Kalb–Ramond gauge field in ref.\ \cite{Ricci_NLsM_Gradient_i}) in the worldsheet nonlinear $\sigma$ model. They constructed an effective entropy functional for the target spacetime, 
which is the same as Perelman's $F$-functional \cite{Ricci_Flow_Monotonicity}, where a `volume' preserving constraint has been taken into account beyond the `conventional' low-energy effective theory description for the target manifold. Based on this entropy functional, they confirmed that the Ricci flow is nothing but a gradient flow with positive-definite metric. In particular, by realizing the beta function as a gradient of Perelman entropy functional satisfying the monotonicity condition, they showed that the Ricci flow is monotonous. We recall the correspondence between Perelman's entropy functional and Zamolodchikov's $c$-functional, where the Ricci flow and the RG flow are given by the gradient flows of the Perelman's $F$- and Zamolodchikov's $c$- functionals, respectively \cite{Witten_Perelman}. The interested reader can refer to appendix \ref{mono} for a brief review of the monotonicity of the Ricci flow.

			%
			%
			
Considering the essential information in the effective action for the emergent holographic dual Eq.\ (\ref{Full_Partition_Function}), we can generalize it as follows
			\begin{align}
				\mathcal{Z} &= \int D G_{\mu\nu}(x,z) D \Pi_{G}^{\mu\nu}(x,z) D B_{\mu\nu}(x,z) D \Pi_{B}^{\mu\nu}(x,z) D \Phi(x,z) D \Pi_{\Phi}(x,z) \notag\\& \exp\Big\{ - \frac{1}{\alpha'} \int d^{D} x \sqrt{G(x,z_{f})} e^{- 2 \Phi(x,z_{f})} ~ \mathcal{V}_{\text{eff}}[G_{\mu\nu}(x,z_{f}),B_{\mu\nu}(x,z_{f}),\Phi(x,z_{f})] \Big\} \notag\\& \exp\Big[ - \frac{1}{\alpha'} \int_{0}^{z_{f}} d z \int d^{D} x \sqrt{G(x,z)} e^{- 2 \Phi(x,z)} \Big\{ \notag\\& ~~~~ \Pi_{G}^{\mu\nu}(x,z) \Big(\partial_{z} G_{\mu\nu}(x,z) + \beta_{\mu\nu}^{G}[G_{\mu\nu}(x,z),B_{\mu\nu}(x,z),\Phi(x,z)] \Big) + \frac{\lambda}{2} \Pi_{G}^{\mu\nu}(x,z) \mathcal{G}_{\mu\nu\rho\lambda}(x,z) \Pi_{G}^{\rho\lambda}(x,z) \notag\\& + \Pi_{B}^{\mu\nu}(x,z) \Big(\partial_{z} B_{\mu\nu}(x,z) + \beta_{\mu\nu}^{B}[G_{\mu\nu}(x,z),B_{\mu\nu}(x,z),\Phi(x,z)] \Big) + \frac{q}{2} \Pi^{B}_{\mu\nu}(x,z) \Pi^{B,\mu\nu}(x,z) \notag\\& + \Pi_{\Phi}(x,z) \Big(\partial_{z} \Phi(x,z) + \beta^{\Phi}[G_{\mu\nu}(x,z),B_{\mu\nu}(x,z),\Phi(x,z)] \Big) + \mathcal{V}_{\text{eff}}[G_{\mu\nu}(x,z),B_{\mu\nu}(x,z),\Phi(x,z)] \Big\} \Big] .  
			\end{align}
Here, $\mathcal{V}_{\text{eff}}[G_{\mu\nu}(x,z),B_{\mu\nu}(x,z),\Phi(x,z)]$ is 
a Wilsonian effective potential, which results from quantum fluctuations of some matter fields in NLSMs. Moreover, the RG $\beta$-functions are given by gradients of this effective potential
\begin{align} \beta_{\mu\nu}^{G}[G_{\mu\nu}(x,z),B_{\mu\nu}(x,z),\Phi(x,z)] & = - \frac{\partial \mathcal{V}_{\text{eff}}[G_{\mu\nu}(x,z),B_{\mu\nu}(x,z),\Phi(x,z)]}{\partial G^{\mu\nu}(x,z)} , \label{Beta_ft_metric} \\ \beta_{\mu\nu}^{B}[G_{\mu\nu}(x,z),B_{\mu\nu}(x,z),\Phi(x,z)] & = - \frac{\partial \mathcal{V}_{\text{eff}}[G_{\mu\nu}(x,z),B_{\mu\nu}(x,z),\Phi(x,z)]}{\partial B^{\mu\nu}(x,z)} , \label{Beta_ft_KR_Gauge_Field}
				\\ \beta^{\Phi}[G_{\mu\nu}(x,z),B_{\mu\nu}(x,z),\Phi(x,z)] & = - \frac{\partial \mathcal{V}_{\text{eff}}[G_{\mu\nu}(x,z),B_{\mu\nu}(x,z),\Phi(x,z)]}{\partial \Phi(x,z)} . \label{Beta_ft_dilaton} \end{align}
			%
			%
Although one can easily verify the expressions for the Wilsonian effective potential and the corresponding RG $\beta-$functions compared to Eq.\ (\ref{Full_Partition_Function}), their concrete equations are not important for the following discussions.
			
We will construct the holographic Perelman's entropy functional and argue that its RG flow corresponding to the entropy production rate is nothing but the Weyl anomaly. In addition, we prove that the entropy production rate is always positive. 
We emphasize that our holographic entropy functional generalizes the 1-loop level Perelman's $F$-functional 
to all-loop order as discussed in the introduction.
			
Recall that the Hamilton-Jacobi equation is (from Eq.\ \eqref{HJ})
\begin{align} & \frac{d \ln \mathcal{Z}}{d z_{f}} = \Big( [\partial_{z_{f}} G_{\mu\nu}(x,z_{f})] \frac{\partial}{\partial G_{\mu\nu}(x,z_{f})} + [\partial_{z_{f}} B_{\mu\nu}(x,z_{f})] \frac{\partial}{\partial B_{\mu\nu}(x,z_{f})} + [\partial_{z_{f}} \Phi(x,z_{f})] \frac{\partial}{\partial \Phi(x,z_{f})} + \frac{\partial}{\partial z_{f}} \Big) \ln \mathcal{Z} = 0 . \end{align}
We reformulate the Hamilton-Jacobi equation to be compared with the local RG equation in the following way
\begin{align} 0 & = \Big( [\partial_{z_{f}} G_{\mu\nu}(x,z_{f})] \frac{\partial}{\partial G_{\mu\nu}(x,z_{f})} + [\partial_{z_{f}} B_{\mu\nu}(x,z_{f})] \frac{\partial}{\partial B_{\mu\nu}(x,z_{f})} + [\partial_{z_{f}} \Phi(x,z_{f})] \frac{\partial}{\partial \Phi(x,z_{f})} \nn & ~~~~~~~~~~~~~~~~~~~~~~~~~~~~~~~~~~~~~~~~~~~~~~~ ~~~~~~~~~~~~~~~~~~~~~~~~~~~~~~~~~~ + \frac{\partial}{\partial z_{f}} \Big) \mathcal{V}_{\text{eff}}[G_{\mu\nu}(x,z_{f}),B_{\mu\nu}(x,z_{f}),\Phi(x,z_{f})]
				\nn & + \Pi_{G}^{\mu\nu}(x,z_{f}) \Big(\partial_{z_{f}} G_{\mu\nu}(x,z_{f}) + \beta_{\mu\nu}^{G}[G_{\mu\nu}(x,z_{f}),B_{\mu\nu}(x,z_{f}),\Phi(x,z_{f})] \Big) + \frac{\lambda}{2} \Pi_{G}^{\mu\nu}(x,z_{f}) \mathcal{G}_{\mu\nu\rho\lambda}(x,z_{f}) \Pi_{G}^{\rho\lambda}(x,z_{f}) \notag\\& + \Pi_{B}^{\mu\nu}(x,z_{f}) \Big(\partial_{z} B_{\mu\nu}(x,z_{f}) + \beta_{\mu\nu}^{B}[G_{\mu\nu}(x,z_{f}),B_{\mu\nu}(x,z_{f}),\Phi(x,z_{f})] \Big) + \frac{q}{2} \Pi^{B}_{\mu\nu}(x,z_{f}) \Pi^{B,\mu\nu}(x,z_{f}) \notag\\& + \Pi_{\Phi}(x,z_{f}) \Big(\partial_{z_{f}} \Phi(x,z_{f}) + \beta^{\Phi}[G_{\mu\nu}(x,z_{f}),B_{\mu\nu}(x,z_{f}),\Phi(x,z_{f})] \Big) + \mathcal{V}_{\text{eff}}[G_{\mu\nu}(x,z_{f}),B_{\mu\nu}(x,z_{f}),\Phi(x,z_{f})] . \label{HJ_First_Form} \end{align}
			%
			
The three canonical momenta, $\Pi_{G}^{\mu\nu}(x,z_{f})$, $\Pi_{B}^{\mu\nu}(x,z_{f})$, and $\Pi_{\Phi}(x,z_{f})$ are determined by the IR boundary conditions as discussed in the previous section. From the on-shell IR boundary action \eqref{OS}, rewritten in terms of the effective potential,
\begin{align} \mathcal{S}_{\text{eff}} & = \frac{1}{\alpha'} \int d^{D} x \sqrt{G(x,z_{f})} e^{- 2 \Phi(x,z_{f})} \Big\{ \Pi_{G}^{\mu\nu}(x,z_{f}) G_{\mu\nu}(x,z_{f}) + \Pi_{B}^{\mu\nu}(x,z_{f}) B_{\mu\nu}(x,z_{f}) + \Pi_{\Phi}(x,z_{f}) \Phi(x,z_{f}) \nn &
				+ \mathcal{V}_{\text{eff}}[G_{\mu\nu}(x,z_{f}),B_{\mu\nu}(x,z_{f}),\Phi(x,z_{f})] \Big\} , \label{Renormalized_Free_Energy} \end{align}
			we obtain
			\begin{align} \Pi_{G}^{\mu\nu}(x,z_{f}) & = - \frac{\partial \mathcal{V}_{\text{eff}}[G_{\mu\nu}(x,z_{f}),B_{\mu\nu}(x,z_{f}),\Phi(x,z_{f})]}{\partial G_{\mu\nu}(x,z_{f})} = \beta^{G,\mu\nu}[G_{\mu\nu}(x,z_{f}),B_{\mu\nu}(x,z_{f}),\Phi(x,z_{f})] , \label{IR_BC_Metric} 
			\\ \Pi_{B}^{\mu\nu}(x,z_{f}) & = - \frac{\partial \mathcal{V}_{\text{eff}}[G_{\mu\nu}(x,z_{f}),B_{\mu\nu}(x,z_{f}),\Phi(x,z_{f})]}{\partial B_{\mu\nu}(x,z_{f})} = \beta^{B,\mu\nu}[G_{\mu\nu}(x,z_{f}),B_{\mu\nu}(x,z_{f}),\Phi(x,z_{f})] , \label{IR_BC_KR_Gauge_Field} 
				\\ \Pi_{\Phi}(x,z_{f}) & = - \frac{\partial \mathcal{V}_{\text{eff}}[G_{\mu\nu}(x,z_{f}),B_{\mu\nu}(x,z_{f}),\Phi(x,z_{f})]}{\partial \Phi(x,z_{f})} = \beta^{\Phi}[G_{\mu\nu}(x,z_{f}),B_{\mu\nu}(x,z_{f}),\Phi(x,z_{f})] . \label{IR_BC_Dilaton} \end{align}
As a result, we rewrite the above Hamilton-Jacobi equation \eqref{HJ_First_Form} as 
			\begin{align} 0 & = \frac{\lambda}{2} \frac{\partial \mathcal{V}_{\text{eff}}[G_{\mu\nu}(x,z_{f}),B_{\mu\nu}(x,z_{f}),\Phi(x,z_{f})]}{\partial G_{\mu\nu}(x,z_{f})} \mathcal{G}_{\mu\nu\rho\lambda}(x,z_{f}) \frac{\partial \mathcal{V}_{\text{eff}}[G_{\mu\nu}(x,z_{f}),B_{\mu\nu}(x,z_{f}),\Phi(x,z_{f})]}{\partial G_{\rho\lambda}(x,z_{f})} 
				\nn & + \beta_{\mu\nu}^{G}[G_{\mu\nu}(x,z_{f}),B_{\mu\nu}(x,z_{f}),\Phi(x,z_{f})] \frac{\partial \mathcal{V}_{\text{eff}}[G_{\mu\nu}(x,z_{f}),B_{\mu\nu}(x,z_{f}),\Phi(x,z_{f})]}{\partial G_{\mu\nu}(x,z_{f})} 
				\nn & + \frac{q}{2} \frac{\partial \mathcal{V}_{\text{eff}}[G_{\mu\nu}(x,z_{f}),B_{\mu\nu}(x,z_{f}),\Phi(x,z_{f})]}{\partial B_{\mu\nu}(x,z_{f})} \frac{\partial \mathcal{V}_{\text{eff}}[G_{\mu\nu}(x,z_{f}),B_{\mu\nu}(x,z_{f}),\Phi(x,z_{f})]}{\partial B^{\mu\nu}(x,z_{f})} 
				\nn & + \beta_{\mu\nu}^{B}[G_{\mu\nu}(x,z_{f}),B_{\mu\nu}(x,z_{f}),\Phi(x,z_{f})] \frac{\partial \mathcal{V}_{\text{eff}}[G_{\mu\nu}(x,z_{f}),B_{\mu\nu}(x,z_{f}),\Phi(x,z_{f})]}{\partial B_{\mu\nu}(x,z_{f})}
				\nn & + \beta^{\Phi}[G_{\mu\nu}(x,z_{f}),B_{\mu\nu}(x,z_{f}),\Phi(x,z_{f})] \frac{\partial \mathcal{V}_{\text{eff}}[G_{\mu\nu}(x,z_{f}),B_{\mu\nu}(x,z_{f}),\Phi(x,z_{f})]}{\partial \Phi(x,z_{f})} 
				\nn & + \frac{\partial}{\partial z_{f}} \mathcal{V}_{\text{eff}}[G_{\mu\nu}(x,z_{f}),B_{\mu\nu}(x,z_{f}),\Phi(x,z_{f})] + \mathcal{V}_{\text{eff}}[G_{\mu\nu}(x,z_{f}),B_{\mu\nu}(x,z_{f}),\Phi(x,z_{f})] . 
			\end{align}
			
To find the Weyl anomaly, we consider the local RG equation \cite{Local_RG_I,Local_RG_II},
			\begin{align} & \frac{d \mathcal{S}_{\text{eff}}}{d z_{f}} = \Big( [\partial_{z_{f}} G_{\mu\nu}(x,z_{f})] \frac{\partial}{\partial G_{\mu\nu}(x,z_{f})} + [\partial_{z_{f}} B_{\mu\nu}(x,z_{f})] \frac{\partial}{\partial B_{\mu\nu}(x,z_{f})} + [\partial_{z_{f}} \Phi(x,z_{f})] \frac{\partial}{\partial \Phi(x,z_{f})} + \frac{\partial}{\partial z_{f}} \Big) \mathcal{S}_{\text{eff}} = 0 . 
			\end{align}
Here, the renormalized free energy $\mathcal{S}_{\text{eff}}$ is given by the IR boundary on-shell effective action, 
Eq.\ (\ref{Renormalized_Free_Energy}). Inserting the IR boundary effective action into this local RG equation, we obtain the Callan–Symanzik equation \cite{QFT_textbook} for the renormalized effective potential $\mathcal{V}_{\text{eff}}[G_{\mu\nu}(x,z_{f}),B_{\mu\nu}(x,z_{f}),\Phi(x,z_{f})]$ as
			\begin{align}
				\Big( [\partial_{z_{f}} G_{\mu\nu}(x,z_{f})] \frac{\partial}{\partial G_{\mu\nu}(x,z_{f})} + [\partial_{z_{f}} B_{\mu\nu}(x,z_{f})] \frac{\partial}{\partial B_{\mu\nu}(x,z_{f})} & + [\partial_{z_{f}} \Phi(x,z_{f})] \frac{\partial}{\partial \Phi(x,z_{f})} \nn & + \frac{\partial}{\partial z_{f}} \Big) \mathcal{V}_{\text{eff}}[G_{\mu\nu}(x,z_{f}),B_{\mu\nu}(x,z_{f}),\Phi(x,z_{f})] = \mathcal{A} . \label{Local_RG_Equation}
			\end{align}
Here, we identify $\mathcal{A}$ with the Weyl anomaly \cite{Local_RG_I,Local_RG_II}, given by 
			\begin{align} & \mathcal{A} = - [\partial_{z_{f}} G_{\mu\nu}(x,z_{f})] \Pi_{G}^{\mu\nu}(x,z_{f}) - [\partial_{z_{f}} B_{\mu\nu}(x,z_{f})] \Pi_{B}^{\mu\nu}(x,z_{f}) - [\partial_{z_{f}} \Phi(x,z_{f})] \Pi_{\Phi}(x,z_{f}) , \label{Weyl_Anomaly_Local_RG} \end{align}
in the sense that this anomaly functional is nonvanishing during the RG flow. 
On the other hand, it vanishes 
as the theory approaches the IR fixed point in the $z_{f} \rightarrow \infty$ limit.
			
Comparing the Hamilton-Jacobi equation (\ref{HJ_First_Form}) with the local RG equation (\ref{Local_RG_Equation}), we obtain
\begin{align} \mathcal{A} & = - \Pi_{G}^{\mu\nu}(x,z_{f}) \Big(\partial_{z_{f}} G_{\mu\nu}(x,z_{f}) + \beta_{\mu\nu}^{G}[G_{\mu\nu}(x,z_{f}),B_{\mu\nu}(x,z_{f}),\Phi(x,z_{f})] \Big) - \frac{\lambda}{2} \Pi_{G}^{\mu\nu}(x,z_{f}) \mathcal{G}_{\mu\nu\rho\lambda}(x,z_{f}) \Pi_{G}^{\rho\lambda}(x,z_{f}) \notag\\& - \Pi_{B}^{\mu\nu}(x,z_{f}) \Big(\partial_{z} B_{\mu\nu}(x,z_{f}) + \beta_{\mu\nu}^{B}[G_{\mu\nu}(x,z_{f}),B_{\mu\nu}(x,z_{f}),\Phi(x,z_{f})] \Big) - \frac{q}{2} \Pi_{B,\mu\nu}(x,z_{f}) \Pi_{B}^{\mu\nu}(x,z_{f}) \notag\\& - \Pi_{\Phi}(x,z_{f}) \Big(\partial_{z_{f}} \Phi(x,z_{f}) + \beta^{\Phi}[G_{\mu\nu}(x,z_{f}),B_{\mu\nu}(x,z_{f}),\Phi(x,z_{f})] \Big) - \mathcal{V}_{\text{eff}}[G_{\mu\nu}(x,z_{f}),B_{\mu\nu}(x,z_{f}),\Phi(x,z_{f})] . \end{align}
This Weyl anomaly from the Hamilton-Jacobi equation is compatible with that of the local RG equation (\ref{Weyl_Anomaly_Local_RG}) as long as $\frac{\partial}{\partial z_{f}} \mathcal{V}_{\text{eff}}[G_{\mu\nu}(x,z_{f}),B_{\mu\nu}(x,z_{f}),\Phi(x,z_{f})] = 0$ holds. This implies that the IR boundary on-shell renormalized effective action $\mathcal{S}_{eff}$ satisfies both the Hamilton-Jacobi equation and the local RG equation.
			
We can find a better expression for the Weyl anomaly. First, we recall the Hamilton's equation of motion for the canonical momentum as 
\begin{align} & \Pi_{G}^{\mu\nu}(x,z) = - \frac{1}{\lambda} \mathcal{G}^{\mu\nu\rho\lambda}(x,z) \Big(\partial_{z} G_{\rho\lambda}(x,z) - \beta_{\rho\lambda}^{G}[G_{\mu\nu}(x,z_{f}),B_{\mu\nu}(x,z_{f}),\Phi(x,z_{f})]\Big) , \\ & \Pi_{B}^{\mu\nu}(x,z) = - \frac{1}{q} \Big(\partial_{z} B^{\mu\nu}(x,z) - \beta^{B,\mu\nu}[G_{\mu\nu}(x,z_{f}),B_{\mu\nu}(x,z_{f}),\Phi(x,z_{f})]\Big) . \end{align}
Introducing the IR boundary conditions, Eq.\ (\ref{IR_BC_Metric}) 
and Eq.\ (\ref{IR_BC_KR_Gauge_Field}) into these equations, we obtain
			\begin{align} & \partial_{z_{f}} G_{\mu\nu}(x,z_{f}) = (1 - \lambda) \beta_{\mu\nu}^{G}[G_{\mu\nu}(x,z_{f}),B_{\mu\nu}(x,z_{f}),\Phi(x,z_{f})] , \\ & \partial_{z_{f}} B_{\mu\nu}(x,z_{f}) = (1 - q) \beta_{\mu\nu}^{B}[G_{\mu\nu}(x,z_{f}),B_{\mu\nu}(x,z_{f}),\Phi(x,z_{f})] . \end{align} 
Recalling $\partial_{z_{f}} \Phi(x,z_{f}) = \beta^{\Phi}[G_{\mu\nu}(x,z_{f}),B_{\mu\nu}(x,z_{f}),\Phi(x,z_{f})]$ 
and Eq.\ (\ref{IR_BC_Dilaton}) for the dilaton field, we find the Weyl anomaly as follows 
			%
			%
			\begin{align} \mathcal{A} = &- (1 - \lambda) \beta_{\mu\nu}^{G}[G_{\mu\nu}(x,z_{f}),B_{\mu\nu}(x,z_{f}),\Phi(x,z_{f})] \mathcal{G}^{\mu\nu\rho\lambda}(x,z_{f}) \beta^{G}_{\rho\lambda}[G_{\mu\nu}(x,z_{f}),B_{\mu\nu}(x,z_{f}),\Phi(x,z_{f})]
			\nn & - (1 - q) \beta_{\mu\nu}^{B}[G_{\mu\nu}(x,z_{f}),B_{\mu\nu}(x,z_{f}),\Phi(x,z_{f})] \beta^{B,\mu\nu}[G_{\mu\nu}(x,z_{f}),B_{\mu\nu}(x,z_{f}),\Phi(x,z_{f})]
				\nn & - \Big( \beta^{\Phi}[G_{\mu\nu}(x,z_{f}),B_{\mu\nu}(x,z_{f}),\Phi(x,z_{f})] \Big)^{2} . \end{align}
			
This expression for the Weyl anomaly can be justified physically. 
We introduce 
			\begin{align} & S = \frac{1}{\alpha'} \int d^{D} x \sqrt{G(x,z_{f})} e^{- 2 \Phi(x,z_{f})} \Big\{ \Pi_{G}^{\mu\nu}(x,z_{f}) G_{\mu\nu}(x,z_{f}) + \Pi_{B}^{\mu\nu}(x,z_{f}) B_{\mu\nu}(x,z_{f}) + \Pi_{\Phi}(x,z_{f}) \Phi(x,z_{f}) \Big\} \end{align}
from the IR boundary effective action, where the Wilsonian effective potential has been cut out. 
We claim that this quantity is analogous to the Perelman's entropy functional, thus calling it the holographic Perelman's entropy functional for the monotonicity of the RG flow. 
Compared to the previous 1-loop level construction, this entropy functional is fully renormalized and 
non-perturbative to 
all-loop order as discussed in detail in appendix A. 
In this respect we claim that our entropy functional is a generalized version of the Perelman's $F$-functional.
			
First of all, all the RG $\beta$-functions are given by the gradients of this holographic Perelman's entropy functional, respectively, as follows
			\begin{align} \frac{\partial S}{\partial G_{\mu\nu}(x,z_{f})} & = \Pi_{G}^{\mu\nu}(x,z_{f}) = \beta^{G,\mu\nu}[G_{\mu\nu}(x,z_{f}),B_{\mu\nu}(x,z_{f}),\Phi(x,z_{f})] , 
			\\ \frac{\partial S}{\partial B_{\mu\nu}(x,z_{f})} & = \Pi_{B}^{\mu\nu}(x,z_{f}) = \beta^{B,\mu\nu}[G_{\mu\nu}(x,z_{f}),B_{\mu\nu}(x,z_{f}),\Phi(x,z_{f})] , 
				\\ \frac{\partial S}{\partial \Phi(x,z_{f})} & = \Pi_{\Phi}(x,z_{f}) = \beta^{\Phi}[G_{\mu\nu}(x,z_{f}),B_{\mu\nu}(x,z_{f}),\Phi(x,z_{f})] . \end{align}
We emphasize that the generalized volume constraint
\bqa && \int d^{D} x \sqrt{G(x,z_{f})} e^{- 2 \Phi(x,z_{f})} = const. \eqa
has been used to obtain the first equality as the case of the Perelman's entropy functional while we used the IR boundary conditions for the second one. In addition, it is straightforward to check out the monotonicity of the RG flow as
\begin{align} \frac{d S}{d z_{f}} & = \Big( [\partial_{z_{f}} G_{\mu\nu}(x,z_{f})] \frac{\partial S}{\partial G_{\mu\nu}(x,z_{f})} + [\partial_{z_{f}} B_{\mu\nu}(x,z_{f})] \frac{\partial S}{\partial B_{\mu\nu}(x,z_{f})} + [\partial_{z_{f}} \Phi(x,z_{f})] \frac{\partial S}{\partial \Phi(x,z_{f})} + \frac{\partial S}{\partial z_{f}} \Big) 
				\nn & = [\partial_{z_{f}} G_{\mu\nu}(x,z_{f})] \Pi_{G}^{\mu\nu}(x,z_{f}) + [\partial_{z_{f}} B_{\mu\nu}(x,z_{f})] \Pi_{B}^{\mu\nu}(x,z_{f}) + [\partial_{z_{f}} \Phi(x,z_{f})] \Pi_{\Phi}(x,z_{f}) 
				\nn & = (1 - \lambda) \beta_{\mu\nu}^{G}[G_{\mu\nu}(x,z_{f}),B_{\mu\nu}(x,z_{f}),\Phi(x,z_{f})] \mathcal{G}^{\mu\nu\rho\lambda}(x,z_{f}) \beta^{G}_{\rho\lambda}[G_{\mu\nu}(x,z_{f}),B_{\mu\nu}(x,z_{f}),\Phi(x,z_{f})] \nn & + (1 - q) \beta_{\mu\nu}^{B}[G_{\mu\nu}(x,z_{f}),B_{\mu\nu}(x,z_{f}),\Phi(x,z_{f})] \beta^{B,\mu\nu}[G_{\mu\nu}(x,z_{f}),B_{\mu\nu}(x,z_{f}),\Phi(x,z_{f})] \nn & + \Big( \beta^{\Phi}[G_{\mu\nu}(x,z_{f}),B_{\mu\nu}(x,z_{f}),\Phi(x,z_{f})] \Big)^{2} 
				\nn & = - \mathcal{A} , \end{align}
which turns out to coincide with the Weyl anomaly discussed above. It is quite interesting to observe that this expression is almost identical to the RG flow of the Perelman's entropy functional except for the dilaton contribution.
			
			%
			%
			
Two remarks are in order. If we consider random noise fluctuations for the dilaton sector, we would obtain $(1 - \gamma) \Big( \beta^{\Phi}[G_{\mu\nu}(x,z_{f}),B_{\mu\nu}(x,z_{f}),\Phi(x,z_{f})] \Big)^{2}$, where $\gamma$ is the strength of the noise in the dilaton sector. More importantly, there exists a sign change in the RG flow of this holographic entropy functional. At present, the physical origin of this sign change is not clear. We suspect that $\gamma > 1$ gives rise to an instability in the RG flow.

\section{Summary and Discussion} \label{Summary_Discussion}
			
In the present study, we proposed a general prescription on how to construct a low-energy effective field theory expected to work at strong couplings. Previously, we derived or more precisely, constructed an effective holographic dual field theory, applying the Wilsonian RG transformation into a quantum field theory (QFT) \cite{RG_Flow_Holography_Monotonicity,Emergent_AdS2_BH_RG,Nonperturbative_Wilson_RG_Disorder,Nonperturbative_Wilson_RG, Einstein_Klein_Gordon_RG_Kim,Einstein_Dirac_RG_Kim,RG_GR_Geometry_I_Kim,RG_GR_Geometry_II_Kim,Kondo_Holography_Kim,Kitaev_Entanglement_Entropy_Kim,RG_Holography_First_Kim}. Here, we decomposed effective two-particle interactions into one-particle propagation coupled to dual collective fields or order parameter fields. Carrying out the Wilsonian RG transformation explicitly, we found that these background dual fields evolve by the RG transformation. 
As a result, not only the single-particle dynamics but also two-particle one become renormalized self-consistently and non-perturbatively through iterative RG transformations. These collective dual fields are determined by the variational principle to minimize the resulting effective action, referred to as an emergent dual holographic description, where the extra dimension is identified with an RG scale parameter. In this study, we formalized and reorganized our previous brute-force calculations in a general prescription.
			
Given the 1-loop RG $\beta$-functions, we modify them to introduce random noise fluctuations into them. One may think that these noise-corrected RG $\beta$-functions are analogous to the Langevin equation. Such random noise fluctuations correspond to irrelevant $T \bar{T}$ like deformations in the UV QFT. It is straightforward to reformulate these coupled RG-flow equations in the path integral representation, referred to as a generating functional or an effective partition function. Following the Faddeev-Popov procedure and taking the ensemble average for random configurations of noise, we obtain an effective holographic dual description for strongly interacting QFTs. Here, one crucial point is that an effective potential of the 1-loop order has been introduced explicitly to confirm that the RG flow is given by the gradient of this effective potential. If the Wilsonian effective potential resulting from high-energy quantum fluctuations is not taken into account, the construction of the holographic dual effective field theory is nothing but the cohomological-type topological field theory formulation a la Witten \cite{TQFT_Cohomology,TQFT_Witten_Type}. Here, random noise fluctuations correspond to the path integral of the auxiliary field in the superspace formulation and Faddeev–Popov ghosts describe the Jacobian factor to constrain the paths of all the coupling fields into the RG flow equations. In this case we cannot go beyond the 1-loop order. The introduction of the Wilsonian effective potential gives rise to non-perturbative resummation for a class of singular quantum fluctuations beyond the purely topological construction. 
The resulting holographic dual effective field theory is non-perturbative in nature, discussed in detail in appendix A.
			
Recently, we showed that BRST symmetries in the emergent holographic dual effective field theory give rise to novel Ward identities, resulting in a generalized fluctuation-dissipation theorem in the RG flow and constraining the many-body dynamics \cite{RG_Flow_Holography_Monotonicity}. Here, the effective potential responsible for the gradient flow breaks the BRST symmetry, which causes additional contributions in the Ward identity to modify the `equilibrium' fluctuation-dissipation theorem of the coupling fields during the RG flow. In this study, the role of the emergent BRST symmetries has not been investigated. Moreover, the superspace formulation \cite{Horava_I,Horava_II} has not been achieved.

To verify this general structure, we have constructed an effective holographic dual field theory for a worldsheet nonlinear $\sigma$ model, where not only RG flow equations but also the low energy effective action is well known to describe the RG flow as a gradient flow. Introducing random noise fluctuations into the RG flow of the $D-$dimensional target spacetime metric, we obtain an effective $(D+1)-$dimensional holographic dual field theory after noise averaging, where the low-energy dilaton-gravity-gauge effective action has been taken into account with all the RG flow equations. Although the first order differential equation of the RG flow is promoted to be the second order one of the bulk evolution in the presence of the Wilsonian effective potential, it turns out that both the UV and IR boundary conditions are essentially given by the same form of the original RG flow equation. These boundary conditions allow a nontrivial dilaton configuration $\Phi(x) = v_{\mu} x^{\mu}$ in $D-$dimensional flat spacetimes. As a result, the conformal anomaly cancellation can be realized in any spacetime dimensions as far as $v_{\mu} v^{\mu} = \frac{26 - D}{6 \alpha'} $ is satisfied. This implies that our $(D+1)-$dimensional holographic dual effective field theory can describe an RG flow between two noncritical string theories of any $D-$spacetime dimensions in a non-perturbative fashion.

Recently, the Ricci flow equation has been reformulated as a cohomological type topological field theory \cite{Horava_I,Horava_II}. Here, the effective potential has not been taken into account and thus, the resulting effective field theory is purely topological. 
			
To understand this holographic effective field theory formulation, we considered the entropy production rate, following ref. \cite{Entropy_Production}. Introducing a probability distribution function for both target spacetime metric and two-form gauge-field configurations, we defined a microscopically constructed entropy functional in terms of the probability distribution function. Based on the intuitively derived effective Fokker-Planck equation for the probability distribution function, we could show the monotonicity of the entropy functional. In particular, we observed that this microscopically constructed entropy functional has essentially the same mathematical expression for the entropy production rate as that derived from the so-called Perelman's macroscopically constructed entropy functional \cite{Ricci_NLsM_Gradient_i,Ricci_NLsM_Gradient_ii} for the critical bosonic string theory except for the integration along the RG flow in the target space. 
			
In addition to this microscopic entropy functional, we considered a holographic entropy functional. We point out that this holographic entropy functional differs from the microscopically constructed `thermodynamic' entropy functional although one may be related to the other somehow. The holographic entropy functional is given by the renormalized IR boundary effective action, where the effective potential resulting from quantum fluctuations of matters is cut out. It turns out that the RG flow of this entropy or the entropy production rate along the RG flow can be identified with the Weyl anomaly, argued from the consistency between the Hamilton-Jacobi equation and the local RG equation in our holographic dual effective field theory. Interestingly, we found that this Weyl anomaly is essentially the same as the RG flow of the Perelman's entropy functional. In this respect, we called this entropy functional the holographic Perelman's entropy. We pointed out that this holographic Perelman's entropy functional generalizes the original Perelman's $F$-functional in a non-perturbative way.
			
One may ask more precise relation between the holographic Perelman's entropy and the Weyl anomaly, i.e., the a-theorem for example. Introducing the RG $\beta$-function of the Ricci flow equation into the Weyl anomaly, one finds that it does not give rise to the square of the Riemann tensor while the Ricci-tensor square and the Ricci-scalar square both are well reproduced. We recall that both the a- and c-coefficients involve the square of the Riemann tensor in the Weyl anomaly of 4 spacetime dimensions \cite{Local_RG_I,Local_RG_II,Local_RG_III}. However, we point out that the holographic dual description gives rise to $a = c$ in the large $N_{c}$ limit \cite{Local_RG_III,a_c_Holography_I,a_c_Holography_II}, where $N_{c}$ corresponds to the number of color degrees of freedom. As a result, the square of the Riemann tensor is canceled between the a- and c-term, and only the Ricci-tensor square and the Ricci-scalar square terms remain, where their relative ratio is given by the de Witt metric. On the other hand, we point out that all these quantities are defined at IR, i.e., $z = z_{f}$, which would have divergent contributions. To extract out correct anomaly coefficients, we have to perform the holographic renormalization. This is quite an important issue, to be addressed near future.
			
Considering recent discussions on the connection between the relative entropy and the RG flow \cite{RG_Flow_Relative_Entropy_I,RG_Flow_Relative_Entropy_II,RG_Flow_Relative_Entropy_Gradient_Flow}, it seems natural to suggest that either the Perelman's $F-$functional or the Zamolodchikov's $c-$functional would correspond to a quantum field theoretic relative entropy. We recall that the relative entropy shows its monotonicity. In addition, it has been shown that a functional RG equation is given by a gradient flow of a field-theoretic relative entropy \cite{RG_Flow_Relative_Entropy_Gradient_Flow}. We point out that recent quantum information theoretic discussions on dual holography \cite{Relative_Entropy_JLMS,Fisher_Info,EW_Reconstruction} may give geometric interpretation on the monotonicity of the RG flow in addition to the entanglement entropy perspectives \cite{a_f_theorem_EE_i,a_f_theorem_EE_ii,a_f_theorem_EE_iii,a_f_theorem_EE_iv,a_f_theorem_EE_v,c_theorem_holography_i,c_theorem_holography_ii}.
			
			%
			%
An important issue is how to understand the fate of tachyons of the bosonic string theory in our holographic dual effective field theory. To figure out this fundamental issue, we have to solve three coupled nonlinear partial differential equations of the metric tensor, the two-form gauge field, and the dilaton field with an appropriate UV boundary condition. This is really a fascinating problem, which may allow us to reinterpret the tachyon condensation of the string field theory \cite{SFT1,SFT2} in the context of our theoretical framework.

\begin{acknowledgments}
K.-S. Kim was supported by the Ministry of Education, Science, and Technology (RS-2024-00337134) of the National Research Foundation of Korea (NRF) and by TJ Park Science Fellowship of the POSCO TJ Park Foundation. The work of A.M. is supported by POSTECH BK21 postdoctoral fellowship. The work of D.M. is supported by the \emph{Young Scientist Training (YST) Fellowship} from Asia Pacific Center for Theoretical Physics. A.M. and D.M. acknowledge support by the National Research Foundation of Korea (NRF) grant funded by the Korean government (MSIT) (No. 2022R1A2C1003182). S.R.~is supported by a Simons Investigator Grant from the Simons Foundation (Award No.~566116). 
K.-S. K. appreciates insightful discussions with P. Yi. 
			\end{acknowledgments}

			\appendix

			\section{Non-perturbative nature of the holographic dual effective field theory}\label{NP}
			
			It is important to figure out how renormalization effects are introduced into the holographic dual effective field theory beyond the cohomological type topological field theory construction. In this appendix, we show how our construction of the holographic dual effective field theory is consistent with the recursive Wilsonian RG transformation scheme, which has been discussed in section \ref{Cohomological_Construction_HDEFT}. In particular, we discuss how such renormalization effects are taken into account in a non-perturbative way based on the two building blocks, 1-loop RG $\beta-$functions and 1-loop Wilsonian effective potential.
			
			First, we consider the $z_{f} \rightarrow 0$ limit. We recall the holographic dual effective field theory with an IR boundary action as follows
			\begin{align} Z & = \int D G_{\mu\nu}(x,z) D B_{\mu\nu}(x,z) D \Phi(x,z) ~ \mathcal{J}\Big(\frac{\partial}{\partial G_{\mu\nu}(x,z)},\frac{\partial}{\partial B_{\mu\nu}(x,z)},\frac{\partial}{\partial \Phi(x,z)}\Big) \nn & \delta\Big(\partial_{z} \Phi(x,z) + \frac{D - 26}{6} - \frac{\alpha'}{2} \nabla^{2} \Phi(x,z) + \alpha' \partial_{\mu} \Phi(x,z) \partial^{\mu} \Phi(x,z) - \frac{\alpha'}{24} H_{\mu\nu\lambda}(x,z) H^{\mu\nu\lambda}(x,z)\Big) \nn & \int D x^{\mu}(\sigma) D b_{ab}(\sigma) D c^{a}(\sigma) 
				\nn & \exp\Big[ - \frac{1}{4 \pi \alpha'} \int_{M} d^{2} \sigma \sqrt{g(\sigma)} \Big\{ \Big( g^{ab}(\sigma) G_{\mu\nu}(x,z_{f}) + i \epsilon^{ab}B_{\mu\nu}(x,z_{f}) \Big) \partial_{a} x^{\mu}(\sigma) \partial_{b} x^{\nu}(\sigma) + \alpha' R^{(2)}(\sigma) \Phi(x,z_{f}) \Big\} 
				\nn & \hspace*{6.5cm}\quad \quad \quad \quad \quad \quad \quad \quad \quad \quad \quad ~ - \frac{1}{2 \pi} \int_{M} d^{2} \sigma \sqrt{g(\sigma)} b_{ab}(\sigma) \nabla^{a} c^{b}(\sigma) \Big] 
				\nn & \exp\Big[ - \frac{1}{\alpha'} \int_{0}^{z_{f}} d z \int d^{D} x \sqrt{G(x,z)} e^{- 2 \Phi(x,z)} \Big\{ - \frac{1}{2 \lambda} \Big(\partial_{z} G_{\mu\nu}(x,z) + \alpha' R_{\mu\nu}(x,z) + 2 \alpha' \nabla_{\mu} \nabla_{\nu} \Phi(x,z) \nn & - \frac{\alpha'}{4} H_{\mu\chi\eta}(x,z) H_{\nu}^{\chi\eta}(x,z)\Big)
				\mathcal{G}^{\mu\nu\rho\lambda}(x,z) \Big(\partial_{z} G_{\rho\lambda}(x,z) + \alpha' R_{\rho\lambda}(x,z) + 2 \alpha' \nabla_{\rho} \nabla_{\lambda} \Phi(x,z) - \frac{\alpha'}{4} H_{\rho\gamma\omega}(x,z) H_{\lambda}^{\gamma\omega}(x,z)\Big) \nn & - \frac{1}{2 q} \Big(\partial_{z} B_{\mu\nu}(x,z) - \frac{\alpha'}{2} \nabla^{\omega} H_{\omega\mu\nu}(x,z) + \alpha' \nabla^{\omega} \Phi(x,z) H_{\omega\mu\nu}(x,z)\Big) \Big(\partial_{z} B^{\mu\nu}(x,z) - \frac{\alpha'}{2} \nabla^{\omega} H_{\omega}^{\mu\nu}(x,z) \nn & + \alpha' \nabla^{\omega} \Phi(x,z) H_{\omega}^{\mu\nu}(x,z)\Big)
				+ \frac{\alpha'}{4} R(x,z) - \frac{D - 26}{6} - \frac{\alpha'}{48} H_{\mu\nu\lambda}(x,z) H^{\mu\nu\lambda}(x,z) + \alpha' \partial_{\mu} \Phi(x,z) \partial^{\mu} \Phi(x,z) \Big\} \Big] . \end{align}
			Here, all the canonical momentum fields have been integrated out to give the Lagrangian formulation. 
			
			In the $z_{f} \rightarrow 0$ limit, we expand the metric, the two-form gauge field, and the dilaton field as 
			\begin{align} G_{\mu\nu}(x,z_{f}) & = G^{(0)}_{\mu\nu}(x) + z_{f} G^{(1)}_{\mu\nu}(x) , \\ B_{\mu\nu}(x,z_{f}) & = B^{(0)}_{\mu\nu}(x) + z_{f} B^{(1)}_{\mu\nu}(x) , \\ \Phi(x,z_{f}) & = \Phi^{(0)}(x) + z_{f} \Phi^{(1)}(x) . \end{align}
			Introducing these equations into the above partition function with the $z_{f} \rightarrow 0$ limit, we obtain
			\begin{align}
				& Z[G^{(0)}_{\mu\nu}(x),B^{(0)}_{\mu\nu}(x),\Phi^{(0)}(x)] \nn = & \int D G^{(1)}_{\mu\nu}(x) D B^{(1)}_{\mu\nu}(x) D \Phi^{(1)}(x) ~ \mathcal{J}\Big(\frac{\partial}{\partial G_{\mu\nu}^{(0)}(x)},\frac{\partial}{\partial B_{\mu\nu}^{(0)}(x)},\frac{\partial}{\partial \Phi^{(0)}(x)}\Big) \nn & \delta\Big(\Phi^{(1)}(x) + \frac{D - 26}{6} - \frac{\alpha'}{2} \nabla^{2} \Phi^{(0)}(x) + \alpha' \partial_{\mu} \Phi^{(0)}(x) \partial^{\mu} \Phi^{(0)}(x) - \frac{\alpha'}{24} H_{\mu\nu\lambda}^{(0)}(x) H^{\mu\nu\lambda (0)}(x)\Big) \nn & \int D x^{\mu}(\sigma) D b_{ab}(\sigma) D c^{a}(\sigma) \exp\Big[ - \frac{1}{4 \pi \alpha'} \int_{M} d^{2} \sigma \sqrt{g(\sigma)} \Big\{ g^{ab}(\sigma) \Big( G^{(0)}_{\mu\nu}(x) + z_{f} G^{(1)}_{\mu\nu}(x) \Big) \partial_{a} x^{\mu}(\sigma) \partial_{b} x^{\nu}(\sigma) \nn & + i \epsilon^{ab} \Big( B^{(0)}_{\mu\nu}(x) + z_{f} B^{(1)}_{\mu\nu}(x) \Big) \partial_{a} x^{\mu}(\sigma) \partial_{b} x^{\nu}(\sigma) + \alpha' R^{(2)}(\sigma) \Big( \Phi^{(0)}(x) + z_{f} \Phi^{(1)}(x) \Big) \Big\} - \frac{1}{2 \pi} \int_{M} d^{2} \sigma \sqrt{g(\sigma)} b_{ab}(\sigma) \nabla^{a} c^{b}(\sigma) \Big] \nn & \exp\Big[ - \frac{1}{\alpha'} z_{f} \int d^{D} x \sqrt{G^{(0)}(x)} e^{- 2 \Phi^{(0)}(x)} \Big\{ - \frac{1}{2 \lambda} \Big(G_{\mu\nu}^{(1)}(x) + \alpha' R_{\mu\nu}^{(0)}(x) + 2 \alpha' \nabla_{\mu} \nabla_{\nu} \Phi^{(0)}(x) \nn & - \frac{\alpha'}{4} H_{\mu\chi\eta}^{(0)}(x) H_{\nu}^{\chi\eta (0)}(x)\Big)
				\mathcal{G}^{\mu\nu\rho\lambda (0)}(x) \Big(G_{\rho\lambda}^{(1)}(x) + \alpha' R_{\rho\lambda}^{(0)}(x) + 2 \alpha' \nabla_{\rho} \nabla_{\lambda} \Phi^{(0)}(x) - \frac{\alpha'}{4} H_{\rho\gamma\omega}^{(0)}(x) H_{\lambda}^{\gamma\omega (0)}(x)\Big) \nn & - \frac{1}{2 q} \Big(B_{\mu\nu}^{(1)}(x) - \frac{\alpha'}{2} \nabla^{\omega} H_{\omega\mu\nu}^{(0)}(x) + \alpha' \nabla^{\omega} \Phi^{(0)}(x) H_{\omega\mu\nu}^{(0)}(x)\Big) \Big(B^{\mu\nu (1)}(x) - \frac{\alpha'}{2} \nabla^{\omega} H_{\omega}^{\mu\nu (0)}(x) + \alpha' \nabla^{\omega} \Phi^{(0)}(x) H_{\omega}^{\mu\nu (0)}(x)\Big) \nn &
				+ \frac{\alpha'}{4} R^{(0)}(x) - \frac{D - 26}{6} - \frac{\alpha'}{48} H_{\mu\nu\lambda}^{(0)}(x) H^{\mu\nu\lambda (0)}(x) + \alpha' \partial_{\mu} \Phi^{(0)}(x) \partial^{\mu} \Phi^{(0)}(x) \Big\} \Big] .  
			\end{align}			
			%
			%
			Here, we note that the Jacobian factor is given by
			\begin{align} \mathcal{J}\Big(\frac{\partial}{\partial G_{\mu\nu}^{(0)}(x)},\frac{\partial}{\partial B_{\mu\nu}^{(0)}(x)},\frac{\partial}{\partial \Phi^{(0)}(x)}\Big) & 
				= \mbox{Det} \begin{pmatrix} \frac{\partial [G_{\omega\lambda}^{(0)} + z_{f} \beta_{\omega\lambda}^{G (0)}]}{\partial G_{\mu\nu}^{(0)}} & \frac{\partial [G_{\omega\lambda}^{(0)} + z_{f} \beta_{\omega\lambda}^{G (0)}]}{\partial B_{\mu\nu}^{(0)}} & \frac{\partial [G_{\omega\lambda}^{(0)} + z_{f} \beta_{\omega\lambda}^{G (0)}]}{\partial \Phi^{(0)}} \\ \frac{\partial [B_{\omega\lambda}^{(0)} + z_{f} \beta_{\omega\lambda}^{B (0)}]}{\partial G_{\mu\nu}^{(0)}} & \frac{\partial [B_{\omega\lambda}^{(0)} + z_{f} \beta_{\omega\lambda}^{B (0)}]}{\partial B_{\mu\nu}^{(0)}} & \frac{\partial [B_{\omega\lambda}^{(0)} + z_{f} \beta_{\omega\lambda}^{B (0)}]}{\partial \Phi^{(0)}} \\ \frac{\partial [\Phi^{(0)} + z_{f} \beta^{\Phi (0)}]}{\partial G_{\mu\nu}^{(0)}} & \frac{\partial [\Phi^{(0)} + z_{f} \beta^{\Phi (0)}]}{\partial B_{\mu\nu}^{(0)}} & \frac{\partial [\Phi^{(0)} + z_{f} \beta^{\Phi (0)}]}{\partial \Phi^{(0)}}\end{pmatrix} . \end{align}
			
			Performing the path integral $\int D G^{(1)}_{\mu\nu}(x) D B^{(1)}_{\mu\nu}(x) D \Phi^{(1)}(x)$, we obtain			
			\begin{align}
				& Z[G^{(0)}_{\mu\nu}(x),B^{(0)}_{\mu\nu}(x),\Phi^{(0)}(x)] = \mathcal{J}\Big(\frac{\partial}{\partial G_{\mu\nu}^{(0)}(x)},\frac{\partial}{\partial B_{\mu\nu}^{(0)}(x)},\frac{\partial}{\partial \Phi^{(0)}(x)}\Big) \nn & \exp\Bigg[ - \frac{1}{\alpha'} z_{f} \int d^{D} x \sqrt{G^{(0)}(x)} e^{- 2 \Phi^{(0)}(x)} \Bigg\{ \frac{\alpha'}{4} R^{(0)}(x) - \frac{D - 26}{6} - \frac{\alpha'}{48} H_{\mu\nu\lambda}^{(0)}(x) H^{\mu\nu\lambda (0)}(x) + \alpha' \partial_{\mu} \Phi^{(0)}(x) \partial^{\mu} \Phi^{(0)}(x) \Bigg\} \Bigg] \nn & \int D x^{\mu}(\sigma) D b_{ab}(\sigma) D c^{a}(\sigma) \exp\Bigg[ - \frac{1}{4 \pi \alpha'} \int_{M} d^{2} \sigma \sqrt{g(\sigma)} \Bigg\{ g^{ab}(\sigma) \Bigg( G^{(0)}_{\mu\nu}(x) - z_{f} \Big(\alpha' R_{\mu\nu}^{(0)}(x) + 2 \alpha' \nabla_{\mu} \nabla_{\nu} \Phi^{(0)}(x) \nn & - \frac{\alpha'}{4} H_{\mu\chi\eta}^{(0)}(x) H_{\nu}^{\chi\eta (0)}(x)\Big) \Bigg) \partial_{a} x^{\mu}(\sigma) \partial_{b} x^{\nu}(\sigma) + i \epsilon^{ab} \Bigg( B^{(0)}_{\mu\nu}(x) + z_{f} \Big(\frac{\alpha'}{2} \nabla^{\omega} H_{\omega\mu\nu}^{(0)}(x) - \alpha' \nabla^{\omega} \Phi^{(0)}(x) H_{\omega\mu\nu}^{(0)}(x) \Big) \Bigg) \partial_{a} x^{\mu}(\sigma) \partial_{b} x^{\nu}(\sigma) \nn & + \alpha' R^{(2)}(\sigma) \Bigg( \Phi^{(0)}(x) - z_{f} \Big(\frac{D - 26}{6} - \frac{\alpha'}{2} \nabla^{2} \Phi^{(0)}(x) + \alpha' \partial_{\mu} \Phi^{(0)}(x) \partial^{\mu} \Phi^{(0)}(x) - \frac{\alpha'}{24} H_{\mu\nu\lambda}^{(0)}(x) H^{\mu\nu\lambda (0)}(x)\Big) \Bigg) \Bigg\} \nn & - \frac{1}{2 \pi} \int_{M} d^{2} \sigma \sqrt{g(\sigma)} b_{ab}(\sigma) \nabla^{a} c^{b}(\sigma) \Bigg] ~ \exp\Bigg[ - \frac{1}{\alpha'} z_{f} \int d^{D} x \sqrt{G^{(0)}(x)} e^{- 2 \Phi^{(0)}(x)} \Bigg\{ \nn & \frac{\lambda}{32 \pi^{2}} \Bigg( \frac{\int_{M} d^{2} \sigma \sqrt{g(\sigma)} g^{ab}(\sigma) \partial_{a} x^{\mu}(\sigma) \partial_{b} x^{\nu}(\sigma)}{\sqrt{G^{(0)}(x)} e^{- 2 \Phi^{(0)}(x)}} \Bigg) \mathcal{G}_{\mu\nu\rho\lambda}^{(0)}(x) \Bigg( \frac{\int_{M} d^{2} \sigma' \sqrt{g(\sigma')} g^{a'b'}(\sigma') \partial_{a'} x^{\rho}(\sigma') \partial_{b'} x^{\lambda}(\sigma')}{\sqrt{G^{(0)}(x)} e^{- 2 \Phi^{(0)}(x)}} \Bigg) \nn & + \frac{q}{32 \pi^{2}} \Bigg( \frac{\int_{M} d^{2} \sigma \sqrt{g(\sigma)} i \epsilon^{ab} \partial_{a} x^{\mu}(\sigma) \partial_{b} x^{\nu}(\sigma)}{\sqrt{G^{(0)}(x)} e^{- 2 \Phi^{(0)}(x)}} \Bigg) \Bigg( \frac{\int_{M} d^{2} \sigma' \sqrt{g(\sigma')} i \epsilon^{a'b'} \partial_{a'} x_{\mu}(\sigma') \partial_{b'} x_{\nu}(\sigma')}{\sqrt{G^{(0)}(x)} e^{- 2 \Phi^{(0)}(x)}} \Bigg) \Bigg\} \Bigg] . \label{TT_bar_Deform_f1}
			\end{align}			
			%
			%
			Here, we point out that there appears an Wilsonian effective potential $\frac{1}{\alpha'} z_{f} \int d^{D} x \sqrt{G^{(0)}(x)} e^{- 2 \Phi^{(0)}(x)} \Big\{ \frac{\alpha'}{4} R^{(0)}(x) - \frac{D - 26}{6} - \frac{\alpha'}{48} H_{\mu\nu\lambda}^{(0)}(x) H^{\mu\nu\lambda (0)}(x) + \alpha' \partial_{\mu} \Phi^{(0)}(x) \partial^{\mu} \Phi^{(0)}(x) \Big\}$ for the dynamics of the target spacetime in addition to the Jacobian factor. The string dynamics is realized in this curved spacetime self-consistently.
			It is interesting to observe that the introduction of random noise fluctuations in the RG flow corresponds to the $T \bar{T}$ and $J J$ deformation, respectively, \cite{TTbar_Deformation} in the worldsheet nonlinear $\sigma$ model, given by $\sim T^{\mu\nu}_{(0)}(x) \mathcal{G}_{\mu\nu\rho\lambda}^{(0)}(x) T^{\rho\lambda}_{(0)}(x)$ and $\sim J_{\mu\nu}^{(0)}(x) J^{\mu\nu (0)}(x)$, respectively, where $T^{\mu\nu}_{(0)}(x) \sim \frac{1}{\sqrt{G^{(0)}(x)} e^{- 2 \Phi^{(0)}(x)}} \int_{M} d^{2} \sigma \sqrt{g(\sigma)} g^{ab}(\sigma) \partial_{a} x^{\mu}(\sigma) \partial_{b} x^{\nu}(\sigma)$ and $J^{\mu\nu (0)}(x) \sim \frac{1}{\sqrt{G^{(0)}(x)} e^{- 2 \Phi^{(0)}(x)}} \int_{M} d^{2} \sigma \sqrt{g(\sigma)} i \epsilon^{ab} \partial_{a} x^{\mu}(\sigma) \partial_{b} x^{\nu}(\sigma)$.
			
			Taking the $\lambda \rightarrow 0$ limit, we reach the following expression
			\begin{align}
				& Z[G^{(0)}_{\mu\nu}(x),B^{(0)}_{\mu\nu}(x),\Phi^{(0)}(x)] = \mathcal{J}\Big(\frac{\partial}{\partial G_{\mu\nu}^{(0)}(x)},\frac{\partial}{\partial B_{\mu\nu}^{(0)}(x)},\frac{\partial}{\partial \Phi^{(0)}(x)}\Big) \nn & \exp\Bigg[ - \frac{1}{\alpha'} z_{f} \int d^{D} x \sqrt{G^{(0)}(x)} e^{- 2 \Phi^{(0)}(x)} \Bigg\{ \frac{\alpha'}{4} R^{(0)}(x) - \frac{D - 26}{6} - \frac{\alpha'}{48} H_{\mu\nu\lambda}^{(0)}(x) H^{\mu\nu\lambda (0)}(x) + \alpha' \partial_{\mu} \Phi^{(0)}(x) \partial^{\mu} \Phi^{(0)}(x) \Bigg\} \Bigg] \nn & \int D x^{\mu}(\sigma) D b_{ab}(\sigma) D c^{a}(\sigma) \exp\Bigg[ - \frac{1}{4 \pi \alpha'} \int_{M} d^{2} \sigma \sqrt{g(\sigma)} \Bigg\{ g^{ab}(\sigma) \Bigg( G^{(0)}_{\mu\nu}(x) - z_{f} \Big(\alpha' R_{\mu\nu}^{(0)}(x) + 2 \alpha' \nabla_{\mu} \nabla_{\nu} \Phi^{(0)}(x) \nn & - \frac{\alpha'}{4} H_{\mu\chi\eta}^{(0)}(x) H_{\nu}^{\chi\eta (0)}(x)\Big) \Bigg) \partial_{a} x^{\mu}(\sigma) \partial_{b} x^{\nu}(\sigma) + i \epsilon^{ab} \Bigg( B^{(0)}_{\mu\nu}(x) + z_{f} \Big(\frac{\alpha'}{2} \nabla^{\omega} H_{\omega\mu\nu}^{(0)}(x) - \alpha' \nabla^{\omega} \Phi^{(0)}(x) H_{\omega\mu\nu}^{(0)}(x) \Big) \Bigg) \partial_{a} x^{\mu}(\sigma) \partial_{b} x^{\nu}(\sigma) \nn & + \alpha' R^{(2)}(\sigma) \Bigg( \Phi^{(0)}(x) - z_{f} \Big(\frac{D - 26}{6} - \frac{\alpha'}{2} \nabla^{2} \Phi^{(0)}(x) + \alpha' \partial_{\mu} \Phi^{(0)}(x) \partial^{\mu} \Phi^{(0)}(x) - \frac{\alpha'}{24} H_{\mu\nu\lambda}^{(0)}(x) H^{\mu\nu\lambda (0)}(x)\Big) \Bigg) \Bigg\} \nn & - \frac{1}{2 \pi} \int_{M} d^{2} \sigma \sqrt{g(\sigma)} b_{ab}(\sigma) \nabla^{a} c^{b}(\sigma) \Bigg] . \label{HDEFT_Near_UV} 
			\end{align}
			This partition function is consistent with the 1-loop RG result for the worldsheet nonlinear $\sigma$ model \cite{Polchinski}, where the RG $\beta-$functions of the metric, Kalb-Ramond gauge field, and dilaton are given by the gradients of the Wilsonian effective potential, respectively. More precisely, if we require 
			\begin{align}
			0 & = \alpha' R_{\mu\nu}^{(0)}(x) + 2 \alpha' \nabla_{\mu} \nabla_{\nu} \Phi^{(0)}(x) - \frac{\alpha'}{4} H_{\mu\chi\eta}^{(0)}(x) H_{\nu}^{\chi\eta (0)}(x) , \\ 0 & = \frac{\alpha'}{2} \nabla^{\omega} H_{\omega\mu\nu}^{(0)}(x) - \alpha' \nabla^{\omega} \Phi^{(0)}(x) H_{\omega\mu\nu}^{(0)}(x) , \\ 0 & = \frac{D - 26}{6} - \frac{\alpha'}{2} \nabla^{2} \Phi^{(0)}(x) + \alpha' \partial_{\mu} \Phi^{(0)}(x) \partial^{\mu} \Phi^{(0)}(x) - \frac{\alpha'}{24} H_{\mu\nu\lambda}^{(0)}(x) H^{\mu\nu\lambda (0)}(x) , 
			\end{align}
			the solution describes an IR fixed point of the target spacetime in the 1-loop level. In addition, these equations of motion are given by the variational principle of the target spacetime effective 1-loop action 
			\begin{align}
			S_{eff}^{(1)} & = \frac{1}{\alpha'} \int d^{D} x \sqrt{G^{(0)}(x)} e^{- 2 \Phi^{(0)}(x)} \Big\{ \frac{\alpha'}{4} R^{(0)}(x) - \frac{D - 26}{6} - \frac{\alpha'}{48} H_{\mu\nu\lambda}^{(0)}(x) H^{\mu\nu\lambda (0)}(x) + \alpha' \partial_{\mu} \Phi^{(0)}(x) \partial^{\mu} \Phi^{(0)}(x) \Big\} ,
			\end{align}
			as well known \cite{Polchinski}.
			
			%
			%
			
			To take higher order quantum corrections beyond this one loop level, we make the integral for the extra-dimensional space in a discrete form as follows
			\begin{align}
				& Z[G^{(0)}_{\mu\nu}(x),B^{(0)}_{\mu\nu}(x),\Phi^{(0)}(x)] \nn = & \int \Pi_{k = 1}^{f} ~ D G_{\mu\nu}^{(k)}(x) D B_{\mu\nu}^{(k)}(x) D \Phi^{(k)}(x) ~ \mathcal{J}\Big(\frac{\partial}{\partial G_{\mu\nu}^{(k-1)}(x)},\frac{\partial}{\partial B_{\mu\nu}^{(k-1)}(x)},\frac{\partial}{\partial \Phi^{(k-1)}(x)}\Big) \nn & \Pi_{k = 1}^{f} ~ \delta\Big(\frac{\Phi^{(k)}(x) - \Phi^{(k-1)}(x)}{\Delta z} + \frac{D - 26}{6} - \frac{\alpha'}{2} \nabla^{2} \Phi^{(k-1)}(x) + \alpha' \partial_{\mu} \Phi^{(k-1)}(x) \partial^{\mu} \Phi^{(k-1)}(x) - \frac{\alpha'}{24} H_{\mu\nu\lambda}^{(k-1)}(x) H^{\mu\nu\lambda (k-1)}(x)\Big) \nn & \int D x^{\mu}(\sigma) D b_{ab}(\sigma) D c^{a}(\sigma) 
				\nn & \exp\Big[ - \frac{1}{4 \pi \alpha'} \int_{M} d^{2} \sigma \sqrt{g(\sigma)} \Big\{ \Big( g^{ab}(\sigma) G_{\mu\nu}^{(f)}(x) + i \epsilon^{ab}B_{\mu\nu}^{(f)}(x) \Big) \partial_{a} x^{\mu}(\sigma) \partial_{b} x^{\nu}(\sigma) + \alpha' R^{(2)}(\sigma) \Phi^{(f)}(x) \Big\} 
				\nn & \hspace*{6.5cm}\quad \quad \quad \quad \quad \quad \quad \quad ~~ - \frac{1}{2 \pi} \int_{M} d^{2} \sigma \sqrt{g(\sigma)} b_{ab}(\sigma) \nabla^{a} c^{b}(\sigma) \Big] \nn & \exp\Big[ - \frac{1}{\alpha'} \Delta z \sum_{k = 1}^{f} \int d^{D} x \sqrt{G^{(k-1)}(x)} e^{- 2 \Phi^{(k-1)}(x)} \Big\{ - \frac{1}{2 \lambda} \Big(\frac{G_{\mu\nu}^{(k)}(x) - G_{\mu\nu}^{(k-1)}(x)}{\Delta z} + \alpha' R_{\mu\nu}^{(k-1)}(x) \nn & + 2 \alpha' \nabla_{\mu} \nabla_{\nu} \Phi^{(k-1)}(x) - \frac{\alpha'}{4} H_{\mu\chi\eta}^{(k-1)}(x) H_{\nu}^{\chi\eta (k-1)}(x)\Big)
				\mathcal{G}^{\mu\nu\rho\lambda}_{(k-1)}(x) \Big(\frac{G_{\rho\lambda}^{(k)}(x) - G_{\rho\lambda}^{(k-1)}(x)}{\Delta z} + \alpha' R_{\rho\lambda}^{(k-1)}(x) \nn & + 2 \alpha' \nabla_{\rho} \nabla_{\lambda} \Phi^{(k-1)}(x) - \frac{\alpha'}{4} H_{\rho\gamma\omega}^{(k-1)}(x) H_{\lambda}^{\gamma\omega (k-1)}(x)\Big)  - \frac{1}{2 q} \Big(\frac{B_{\mu\nu}^{(k)}(x) - B_{\mu\nu}^{(k-1)}(x)}{\Delta z} - \frac{\alpha'}{2} \nabla^{\omega} H_{\omega\mu\nu}^{(k-1)}(x) \nn & + \alpha' \nabla^{\omega} \Phi^{(k-1)}(x) H_{\omega\mu\nu}^{(k-1)}(x)\Big) \Big(\frac{B^{\mu\nu (k)}(x) - B^{\mu\nu (k-1)}(x)}{\Delta z} - \frac{\alpha'}{2} \nabla^{\omega} H_{\omega}^{\mu\nu (k-1)}(x)  + \alpha' \nabla^{\omega} \Phi^{(k-1)}(x) H_{\omega}^{\mu\nu (k-1)}(x)\Big)
				\nn & + \frac{\alpha'}{4} R^{(k-1)}(x) - \frac{D - 26}{6} - \frac{\alpha'}{48} H_{\mu\nu\lambda}^{(k-1)}(x) H^{\mu\nu\lambda (k-1)}(x) + \alpha' \partial_{\mu} \Phi^{(k-1)}(x) \partial^{\mu} \Phi^{(k-1)}(x) \Big\} \Big] .
			\end{align} 
			Here, the Jacobian factor is given by
			\begin{align} \mathcal{J}\Big(\frac{\partial}{\partial G_{\mu\nu}^{(k-1)}(x)},\frac{\partial}{\partial B_{\mu\nu}^{(k-1)}(x)},\frac{\partial}{\partial \Phi^{(k-1)}(x)}\Big) & \equiv \frac{\partial \Big(G_{\mu\nu}^{(k)}(x), B_{\mu\nu}^{(k)}(x), \Phi^{(k)}(x)\Big)}{\partial \Big(G_{\mu\nu}^{(k-1)}(x), B_{\mu\nu}^{(k-1)}(x), \Phi^{(k-1)}(x)\Big)} \nn & 
				= \mbox{Det} \begin{pmatrix} \frac{\partial [G_{\omega\lambda}^{(k-1)} + \Delta z \beta_{\omega\lambda}^{G (k-1)}]}{\partial G_{\mu\nu}^{(k-1)}} & \frac{\partial [G_{\omega\lambda}^{(k-1)} + \Delta z \beta_{\omega\lambda}^{G (k-1)}]}{\partial B_{\mu\nu}^{(k-1)}} & \frac{\partial [G_{\omega\lambda}^{(k-1)} + \Delta z \beta_{\omega\lambda}^{G (k-1)}]}{\partial \Phi^{(k-1)}} \\ \frac{\partial [B_{\omega\lambda}^{(k-1)} + \Delta z \beta_{\omega\lambda}^{B (k-1)}]}{\partial G_{\mu\nu}^{(k-1)}} & \frac{\partial [B_{\omega\lambda}^{(k-1)} + \Delta z \beta_{\omega\lambda}^{B (k-1)}]}{\partial B_{\mu\nu}^{(k-1)}} & \frac{\partial [B_{\omega\lambda}^{(k-1)} + \Delta z \beta_{\omega\lambda}^{B (k-1)}]}{\partial \Phi^{(k-1)}} \\ \frac{\partial [\Phi^{(k-1)} + \Delta z \beta^{\Phi (k-1)}]}{\partial G_{\mu\nu}^{(k-1)}} & \frac{\partial [\Phi^{(k-1)} + \Delta z \beta^{\Phi (k-1)}]}{\partial B_{\mu\nu}^{(k-1)}} & \frac{\partial [\Phi^{(k-1)} + \Delta z \beta^{\Phi (k-1)}]}{\partial \Phi^{(k-1)}}\end{pmatrix} . \end{align}

			If we limit the discrete sum up to $f = 1$, we reproduce Eq. (\ref{HDEFT_Near_UV}) in the following way:
			\begin{align}
				& Z[G^{(0)}_{\mu\nu}(x),B^{(0)}_{\mu\nu}(x),\Phi^{(0)}(x)] \nn = & \int D G_{\mu\nu}^{(1)}(x) D B_{\mu\nu}^{(1)}(x) D \Phi^{(1)}(x) ~ \mathcal{J}\Big(\frac{\partial}{\partial G_{\mu\nu}^{(0)}(x)},\frac{\partial}{\partial B_{\mu\nu}^{(0)}(x)},\frac{\partial}{\partial \Phi^{(0)}(x)}\Big) \nn & \delta\Big(\frac{\Phi^{(1)}(x) - \Phi^{(0)}(x)}{\Delta z} + \frac{D - 26}{6} - \frac{\alpha'}{2} \nabla^{2} \Phi^{(0)}(x) + \alpha' \partial_{\mu} \Phi^{(0)}(x) \partial^{\mu} \Phi^{(0)}(x) - \frac{\alpha'}{24} H_{\mu\nu\lambda}^{(0)}(x) H^{\mu\nu\lambda (0)}(x)\Big) \nn & \int D x^{\mu}(\sigma) D b_{ab}(\sigma) D c^{a}(\sigma) 
				\nn & \exp\Big[ - \frac{1}{4 \pi \alpha'} \int_{M} d^{2} \sigma \sqrt{g(\sigma)} \Big\{ \Big( g^{ab}(\sigma) G_{\mu\nu}^{(1)}(x) + i \epsilon^{ab}B_{\mu\nu}^{(1)}(x) \Big) \partial_{a} x^{\mu}(\sigma) \partial_{b} x^{\nu}(\sigma) + \alpha' R^{(2)}(\sigma) \Phi^{(1)}(x) \Big\} 
				\nn & \hspace*{6.5cm}\quad \quad \quad \quad \quad \quad \quad \quad ~~ - \frac{1}{2 \pi} \int_{M} d^{2} \sigma \sqrt{g(\sigma)} b_{ab}(\sigma) \nabla^{a} c^{b}(\sigma) \Big] \nn & \exp\Big[ - \frac{1}{\alpha'} \Delta z \int d^{D} x \sqrt{G^{(0)}(x)} e^{- 2 \Phi^{(0)}(x)} \Big\{ - \frac{1}{2 \lambda} \Big(\frac{G_{\mu\nu}^{(1)}(x) - G_{\mu\nu}^{(0)}(x)}{\Delta z} + \alpha' R_{\mu\nu}^{(0)}(x) \nn & + 2 \alpha' \nabla_{\mu} \nabla_{\nu} \Phi^{(0)}(x) - \frac{\alpha'}{4} H_{\mu\chi\eta}^{(0)}(x) H_{\nu}^{\chi\eta (0)}(x)\Big)
				\mathcal{G}^{\mu\nu\rho\lambda}_{(0)}(x) \Big(\frac{G_{\rho\lambda}^{(1)}(x) - G_{\rho\lambda}^{(0)}(x)}{\Delta z} + \alpha' R_{\rho\lambda}^{(0)}(x) \nn & + 2 \alpha' \nabla_{\rho} \nabla_{\lambda} \Phi^{(0)}(x) - \frac{\alpha'}{4} H_{\rho\gamma\omega}^{(0)}(x) H_{\lambda}^{\gamma\omega (0)}(x)\Big)  - \frac{1}{2 q} \Big(\frac{B_{\mu\nu}^{(1)}(x) - B_{\mu\nu}^{(0)}(x)}{\Delta z} - \frac{\alpha'}{2} \nabla^{\omega} H_{\omega\mu\nu}^{(0)}(x) \nn & + \alpha' \nabla^{\omega} \Phi^{(0)}(x) H_{\omega\mu\nu}^{(0)}(x)\Big) \Big(\frac{B^{\mu\nu (1)}(x) - B^{\mu\nu (0)}(x)}{\Delta z} - \frac{\alpha'}{2} \nabla^{\omega} H_{\omega}^{\mu\nu (0)}(x)  + \alpha' \nabla^{\omega} \Phi^{(0)}(x) H_{\omega}^{\mu\nu (0)}(x)\Big)
				\nn & + \frac{\alpha'}{4} R^{(0)}(x) - \frac{D - 26}{6} - \frac{\alpha'}{48} H_{\mu\nu\lambda}^{(0)}(x) H^{\mu\nu\lambda (0)}(x) + \alpha' \partial_{\mu} \Phi^{(0)}(x) \partial^{\mu} \Phi^{(0)}(x) \Big\} \Big] ,
			\end{align}
			resulting in 
			\begin{align}
				& Z[G^{(0)}_{\mu\nu}(x),B^{(0)}_{\mu\nu}(x),\Phi^{(0)}(x)] = \mathcal{J}\Big(\frac{\partial}{\partial G_{\mu\nu}^{(0)}(x)},\frac{\partial}{\partial B_{\mu\nu}^{(0)}(x)},\frac{\partial}{\partial \Phi^{(0)}(x)}\Big) \nn & \exp\Bigg[ - \frac{1}{\alpha'} \Delta z \int d^{D} x \sqrt{G^{(0)}(x)} e^{- 2 \Phi^{(0)}(x)} \Bigg\{ \frac{\alpha'}{4} R^{(0)}(x) - \frac{D - 26}{6} - \frac{\alpha'}{48} H_{\mu\nu\lambda}^{(0)}(x) H^{\mu\nu\lambda (0)}(x) + \alpha' \partial_{\mu} \Phi^{(0)}(x) \partial^{\mu} \Phi^{(0)}(x) \Bigg\} \Bigg] \nn & \int D x^{\mu}(\sigma) D b_{ab}(\sigma) D c^{a}(\sigma) \exp\Bigg[ - \frac{1}{4 \pi \alpha'} \int_{M} d^{2} \sigma \sqrt{g(\sigma)} \Bigg\{ g^{ab}(\sigma) \Bigg( G^{(0)}_{\mu\nu}(x) - \Delta z \Big(\alpha' R_{\mu\nu}^{(0)}(x) + 2 \alpha' \nabla_{\mu} \nabla_{\nu} \Phi^{(0)}(x) \nn & - \frac{\alpha'}{4} H_{\mu\chi\eta}^{(0)}(x) H_{\nu}^{\chi\eta (0)}(x)\Big) \Bigg) \partial_{a} x^{\mu}(\sigma) \partial_{b} x^{\nu}(\sigma) + i \epsilon^{ab} \Bigg( B^{(0)}_{\mu\nu}(x) + \Delta z \Big(\frac{\alpha'}{2} \nabla^{\omega} H_{\omega\mu\nu}^{(0)}(x) - \alpha' \nabla^{\omega} \Phi^{(0)}(x) H_{\omega\mu\nu}^{(0)}(x) \Big) \Bigg) \partial_{a} x^{\mu}(\sigma) \partial_{b} x^{\nu}(\sigma) \nn & + \alpha' R^{(2)}(\sigma) \Bigg( \Phi^{(0)}(x) - \Delta z \Big(\frac{D - 26}{6} - \frac{\alpha'}{2} \nabla^{2} \Phi^{(0)}(x) + \alpha' \partial_{\mu} \Phi^{(0)}(x) \partial^{\mu} \Phi^{(0)}(x) - \frac{\alpha'}{24} H_{\mu\nu\lambda}^{(0)}(x) H^{\mu\nu\lambda (0)}(x)\Big) \Bigg) \Bigg\} \nn & - \frac{1}{2 \pi} \int_{M} d^{2} \sigma \sqrt{g(\sigma)} b_{ab}(\sigma) \nabla^{a} c^{b}(\sigma) \Bigg] ,
			\end{align} 
			where the $\lambda \rightarrow 0$ limit has been taken to recover the low energy effective action \cite{Polchinski}.
			
			Considering $f = 2$, we expect that the next leading correction would be introduced into the worldsheet nonlinear $\sigma$ model. The $f = 2$ partition function is given by
			\begin{align}
				& Z[G^{(0)}_{\mu\nu}(x),B^{(0)}_{\mu\nu}(x),\Phi^{(0)}(x)] \nn = & \int D G_{\mu\nu}^{(1)}(x) D G_{\mu\nu}^{(2)}(x) D B_{\mu\nu}^{(1)}(x) D B_{\mu\nu}^{(2)}(x) D \Phi^{(1)}(x) D \Phi^{(2)}(x) \nn & \mathcal{J}\Big(\frac{\partial}{\partial G_{\mu\nu}^{(0)}(x)},\frac{\partial}{\partial B_{\mu\nu}^{(0)}(x)},\frac{\partial}{\partial \Phi^{(0)}(x)}\Big) ~ \mathcal{J}\Big(\frac{\partial}{\partial G_{\mu\nu}^{(1)}(x)},\frac{\partial}{\partial B_{\mu\nu}^{(1)}(x)},\frac{\partial}{\partial \Phi^{(1)}(x)}\Big) \nn & \delta\Big(\frac{\Phi^{(1)}(x) - \Phi^{(0)}(x)}{\Delta z} + \frac{D - 26}{6} - \frac{\alpha'}{2} \nabla^{2} \Phi^{(0)}(x) + \alpha' \partial_{\mu} \Phi^{(0)}(x) \partial^{\mu} \Phi^{(0)}(x) - \frac{\alpha'}{24} H_{\mu\nu\lambda}^{(0)}(x) H^{\mu\nu\lambda (0)}(x)\Big) \nn & \delta\Big(\frac{\Phi^{(2)}(x) - \Phi^{(1)}(x)}{\Delta z} + \frac{D - 26}{6} - \frac{\alpha'}{2} \nabla^{2} \Phi^{(1)}(x) + \alpha' \partial_{\mu} \Phi^{(1)}(x) \partial^{\mu} \Phi^{(1)}(x) - \frac{\alpha'}{24} H_{\mu\nu\lambda}^{(1)}(x) H^{\mu\nu\lambda (1)}(x)\Big) \nn & \int D x^{\mu}(\sigma) D b_{ab}(\sigma) D c^{a}(\sigma) 
				\nn & \exp\Big[ - \frac{1}{4 \pi \alpha'} \int_{M} d^{2} \sigma \sqrt{g(\sigma)} \Big\{ \Big( g^{ab}(\sigma) G_{\mu\nu}^{(2)}(x) + i \epsilon^{ab}B_{\mu\nu}^{(2)}(x) \Big) \partial_{a} x^{\mu}(\sigma) \partial_{b} x^{\nu}(\sigma) + \alpha' R^{(2)}(\sigma) \Phi^{(2)}(x) \Big\} 
				\nn & \hspace*{6.5cm}\quad \quad \quad \quad \quad \quad \quad \quad ~~ - \frac{1}{2 \pi} \int_{M} d^{2} \sigma \sqrt{g(\sigma)} b_{ab}(\sigma) \nabla^{a} c^{b}(\sigma) \Big] \nn & \exp\Big[ - \frac{1}{\alpha'} \Delta z \int d^{D} x \sqrt{G^{(0)}(x)} e^{- 2 \Phi^{(0)}(x)} \Big\{ - \frac{1}{2 \lambda} \Big(\frac{G_{\mu\nu}^{(1)}(x) - G_{\mu\nu}^{(0)}(x)}{\Delta z} + \alpha' R_{\mu\nu}^{(0)}(x) \nn & + 2 \alpha' \nabla_{\mu} \nabla_{\nu} \Phi^{(0)}(x) - \frac{\alpha'}{4} H_{\mu\chi\eta}^{(0)}(x) H_{\nu}^{\chi\eta (0)}(x)\Big)
				\mathcal{G}^{\mu\nu\rho\lambda}_{(0)}(x) \Big(\frac{G_{\rho\lambda}^{(1)}(x) - G_{\rho\lambda}^{(0)}(x)}{\Delta z} + \alpha' R_{\rho\lambda}^{(0)}(x) \nn & + 2 \alpha' \nabla_{\rho} \nabla_{\lambda} \Phi^{(0)}(x) - \frac{\alpha'}{4} H_{\rho\gamma\omega}^{(0)}(x) H_{\lambda}^{\gamma\omega (0)}(x)\Big)  - \frac{1}{2 q} \Big(\frac{B_{\mu\nu}^{(1)}(x) - B_{\mu\nu}^{(0)}(x)}{\Delta z} - \frac{\alpha'}{2} \nabla^{\omega} H_{\omega\mu\nu}^{(0)}(x) \nn & + \alpha' \nabla^{\omega} \Phi^{(0)}(x) H_{\omega\mu\nu}^{(0)}(x)\Big) \Big(\frac{B^{\mu\nu (1)}(x) - B^{\mu\nu (0)}(x)}{\Delta z} - \frac{\alpha'}{2} \nabla^{\omega} H_{\omega}^{\mu\nu (0)}(x)  + \alpha' \nabla^{\omega} \Phi^{(0)}(x) H_{\omega}^{\mu\nu (0)}(x)\Big)
				\nn & + \frac{\alpha'}{4} R^{(0)}(x) - \frac{D - 26}{6} - \frac{\alpha'}{48} H_{\mu\nu\lambda}^{(0)}(x) H^{\mu\nu\lambda (0)}(x) + \alpha' \partial_{\mu} \Phi^{(0)}(x) \partial^{\mu} \Phi^{(0)}(x) \Big\} \nn & - \frac{1}{\alpha'} \Delta z \int d^{D} x \sqrt{G^{(1)}(x)} e^{- 2 \Phi^{(1)}(x)} \Big\{ - \frac{1}{2 \lambda} \Big(\frac{G_{\mu\nu}^{(2)}(x) - G_{\mu\nu}^{(1)}(x)}{\Delta z} + \alpha' R_{\mu\nu}^{(1)}(x) \nn & + 2 \alpha' \nabla_{\mu} \nabla_{\nu} \Phi^{(1)}(x) - \frac{\alpha'}{4} H_{\mu\chi\eta}^{(1)}(x) H_{\nu}^{\chi\eta (1)}(x)\Big)
				\mathcal{G}^{\mu\nu\rho\lambda}_{(1)}(x) \Big(\frac{G_{\rho\lambda}^{(2)}(x) - G_{\rho\lambda}^{(1)}(x)}{\Delta z} + \alpha' R_{\rho\lambda}^{(1)}(x) \nn & + 2 \alpha' \nabla_{\rho} \nabla_{\lambda} \Phi^{(1)}(x) - \frac{\alpha'}{4} H_{\rho\gamma\omega}^{(1)}(x) H_{\lambda}^{\gamma\omega (1)}(x)\Big)  - \frac{1}{2 q} \Big(\frac{B_{\mu\nu}^{(2)}(x) - B_{\mu\nu}^{(1)}(x)}{\Delta z} - \frac{\alpha'}{2} \nabla^{\omega} H_{\omega\mu\nu}^{(1)}(x) \nn & + \alpha' \nabla^{\omega} \Phi^{(1)}(x) H_{\omega\mu\nu}^{(1)}(x)\Big) \Big(\frac{B^{\mu\nu (2)}(x) - B^{\mu\nu (1)}(x)}{\Delta z} - \frac{\alpha'}{2} \nabla^{\omega} H_{\omega}^{\mu\nu (1)}(x)  + \alpha' \nabla^{\omega} \Phi^{(1)}(x) H_{\omega}^{\mu\nu (1)}(x)\Big)
				\nn & + \frac{\alpha'}{4} R^{(1)}(x) - \frac{D - 26}{6} - \frac{\alpha'}{48} H_{\mu\nu\lambda}^{(1)}(x) H^{\mu\nu\lambda (1)}(x) + \alpha' \partial_{\mu} \Phi^{(1)}(x) \partial^{\mu} \Phi^{(1)}(x) \Big\} \Big] .
			\end{align} 
			
			Performing the path integral $\int D G_{\mu\nu}^{(2)}(x) D B_{\mu\nu}^{(2)}(x) D \Phi^{(2)}(x)$ and taking the $\lambda \rightarrow 0$ limit, we obtain
			\begin{align}
				& Z[G^{(0)}_{\mu\nu}(x),B^{(0)}_{\mu\nu}(x),\Phi^{(0)}(x)] = \mathcal{J}\Big(\frac{\partial}{\partial G_{\mu\nu}^{(0)}(x)},\frac{\partial}{\partial B_{\mu\nu}^{(0)}(x)},\frac{\partial}{\partial \Phi^{(0)}(x)}\Big) \nn & \exp\Bigg[ - \frac{1}{\alpha'} \Delta z \int d^{D} x \sqrt{G^{(0)}(x)} e^{- 2 \Phi^{(0)}(x)} \Bigg\{ \frac{\alpha'}{4} R^{(0)}(x) - \frac{D - 26}{6} - \frac{\alpha'}{48} H_{\mu\nu\lambda}^{(0)}(x) H^{\mu\nu\lambda (0)}(x) + \alpha' \partial_{\mu} \Phi^{(0)}(x) \partial^{\mu} \Phi^{(0)}(x) \Bigg\} \Bigg] \nn & \int D G_{\mu\nu}^{(1)}(x) D B_{\mu\nu}^{(1)}(x) D \Phi^{(1)}(x) ~ \mathcal{J}\Big(\frac{\partial}{\partial G_{\mu\nu}^{(1)}(x)},\frac{\partial}{\partial B_{\mu\nu}^{(1)}(x)},\frac{\partial}{\partial \Phi^{(1)}(x)}\Big) \nn & \delta\Bigg(\frac{G_{\mu\nu}^{(1)}(x) - G_{\mu\nu}^{(0)}(x)}{\Delta z} + \Delta z \Big(\alpha' R_{\mu\nu}^{(0)}(x) + 2 \alpha' \nabla_{\mu} \nabla_{\nu} \Phi^{(0)}(x) - \frac{\alpha'}{4} H_{\mu\chi\eta}^{(0)}(x) H_{\nu}^{\chi\eta (0)}(x)\Big)\Bigg) \nn & \delta\Bigg(\frac{B_{\mu\nu}^{(1)}(x) - B_{\mu\nu}^{(0)}(x)}{\Delta z} - \Delta z \Big(\frac{\alpha'}{2} \nabla^{\omega} H_{\omega\mu\nu}^{(0)}(x) - \alpha' \nabla^{\omega} \Phi^{(0)}(x) H_{\omega\mu\nu}^{(0)}(x) \Big)\Bigg) \nn & \delta\Bigg(\frac{\Phi^{(1)}(x) - \Phi^{(0)}(x)}{\Delta z} + \Delta z \Big(\frac{D - 26}{6} - \frac{\alpha'}{2} \nabla^{2} \Phi^{(0)}(x) + \alpha' \partial_{\mu} \Phi^{(0)}(x) \partial^{\mu} \Phi^{(0)}(x) - \frac{\alpha'}{24} H_{\mu\nu\lambda}^{(0)}(x) H^{\mu\nu\lambda (0)}(x)\Big)\Bigg) \nn & \exp\Bigg[ - \frac{1}{\alpha'} \Delta z \int d^{D} x \sqrt{G^{(1)}(x)} e^{- 2 \Phi^{(1)}(x)} \Bigg\{ \frac{\alpha'}{4} R^{(1)}(x) - \frac{D - 26}{6} - \frac{\alpha'}{48} H_{\mu\nu\lambda}^{(1)}(x) H^{\mu\nu\lambda (1)}(x) + \alpha' \partial_{\mu} \Phi^{(1)}(x) \partial^{\mu} \Phi^{(1)}(x) \Bigg\} \Bigg] \nn & \int D x^{\mu}(\sigma) D b_{ab}(\sigma) D c^{a}(\sigma) \exp\Bigg[ - \frac{1}{4 \pi \alpha'} \int_{M} d^{2} \sigma \sqrt{g(\sigma)} \Bigg\{ g^{ab}(\sigma) \Bigg( G^{(1)}_{\mu\nu}(x) - \Delta z \Big(\alpha' R_{\mu\nu}^{(1)}(x) + 2 \alpha' \nabla_{\mu} \nabla_{\nu} \Phi^{(1)}(x) \nn & - \frac{\alpha'}{4} H_{\mu\chi\eta}^{(1)}(x) H_{\nu}^{\chi\eta (1)}(x)\Big) \Bigg) \partial_{a} x^{\mu}(\sigma) \partial_{b} x^{\nu}(\sigma) + i \epsilon^{ab} \Bigg( B^{(1)}_{\mu\nu}(x) + \Delta z \Big(\frac{\alpha'}{2} \nabla^{\omega} H_{\omega\mu\nu}^{(1)}(x) - \alpha' \nabla^{\omega} \Phi^{(1)}(x) H_{\omega\mu\nu}^{(1)}(x) \Big) \Bigg) \partial_{a} x^{\mu}(\sigma) \partial_{b} x^{\nu}(\sigma) \nn & + \alpha' R^{(2)}(\sigma) \Bigg( \Phi^{(1)}(x) - \Delta z \Big(\frac{D - 26}{6} - \frac{\alpha'}{2} \nabla^{2} \Phi^{(1)}(x) + \alpha' \partial_{\mu} \Phi^{(1)}(x) \partial^{\mu} \Phi^{(1)}(x) - \frac{\alpha'}{24} H_{\mu\nu\lambda}^{(1)}(x) H^{\mu\nu\lambda (1)}(x)\Big) \Bigg) \Bigg\} \nn & - \frac{1}{2 \pi} \int_{M} d^{2} \sigma \sqrt{g(\sigma)} b_{ab}(\sigma) \nabla^{a} c^{b}(\sigma) \Bigg] . \label{Nonperturbative_RG_f2}
			\end{align} 
			This path integral representation shows how quantum corrections are taken into account in a non-perturbative way for $\alpha'$. There is an $\alpha'$ correction in $- \frac{1}{\alpha'} \Delta z \int d^{D} x \sqrt{G^{(0)}(x)} e^{- 2 \Phi^{(0)}(x)} \Bigg\{ \frac{\alpha'}{4} R^{(0)}(x) - \frac{D - 26}{6} - \frac{\alpha'}{48} H_{\mu\nu\lambda}^{(0)}(x) H^{\mu\nu\lambda (0)}(x) + \alpha' \partial_{\mu} \Phi^{(0)}(x) \partial^{\mu} \Phi^{(0)}(x) \Bigg\}$. In addition to this $\alpha'$ correction, we observe that an $\alpha'^{2}$ correction appears naturally as 
			$- \frac{1}{\alpha'} \Delta z \int d^{D} x \sqrt{G^{(1)}(x)} e^{- 2 \Phi^{(1)}(x)} \Bigg\{ \frac{\alpha'}{4} R^{(1)}(x) - \frac{D - 26}{6} - \frac{\alpha'}{48} H_{\mu\nu\lambda}^{(1)}(x) H^{\mu\nu\lambda (1)}(x) + \alpha' \partial_{\mu} \Phi^{(1)}(x) \partial^{\mu} \Phi^{(1)}(x) \Bigg\}$, where $R_{\mu\nu}^{(1)}(x)$ is given by $G^{(1)}_{\mu\nu}(x)$ which has an $\alpha'$ correction. Interestingly, the $\alpha'^{2}$ correction arises in the $(\Delta z)^{2}$ order. 
			
			Performing the path integral $\int D G_{\mu\nu}^{(1)}(x) D B_{\mu\nu}^{(1)}(x) D \Phi^{(1)}(x)$, we introduce  
			\begin{align}
				G^{(1)}_{\mu\nu}(x) & = G^{(0)}_{\mu\nu}(x) - \Delta z \Big(\alpha' R_{\mu\nu}^{(0)}(x) + 2 \alpha' \nabla_{\mu} \nabla_{\nu} \Phi^{(0)}(x) - \frac{\alpha'}{4} H_{\mu\chi\eta}^{(0)}(x) H_{\nu}^{\chi\eta (0)}(x)\Big) , 
				\\ B^{(1)}_{\mu\nu}(x) & = B^{(0)}_{\mu\nu}(x) + \Delta z \Big(\frac{\alpha'}{2} \nabla^{\omega} H_{\omega\mu\nu}^{(0)}(x) - \alpha' \nabla^{\omega} \Phi^{(0)}(x) H_{\omega\mu\nu}^{(0)}(x) \Big) ,  
				\\ \Phi^{(1)}(x) & = \Phi^{(0)}(x) - \Delta z \Big(\frac{D - 26}{6} - \frac{\alpha'}{2} \nabla^{2} \Phi^{(0)}(x) + \alpha' \partial_{\mu} \Phi^{(0)}(x) \partial^{\mu} \Phi^{(0)}(x) - \frac{\alpha'}{24} H_{\mu\nu\lambda}^{(0)}(x) H^{\mu\nu\lambda (0)}(x)\Big) ,
			\end{align}
			into the above effective action. Then, the resulting effective action is given by quite a complicated functional of $G^{(0)}_{\mu\nu}(x)$, $B_{\mu\nu}^{(0)}(x)$, and $\Phi^{(0)}(x)$. These fields are determined by the variational principle of the resulting Wilsonian effective action with respect to $G^{(0)}_{\mu\nu}(x)$, $B_{\mu\nu}^{(0)}(x)$, and $\Phi^{(0)}(x)$. This indicates that all-loop order $\alpha'$ corrections are re-summed in the level of an effective action for target spacetime, which shows the non-perturbative nature of our holographic dual effective field theory \cite{RG_Flow_Holography_Monotonicity,Emergent_AdS2_BH_RG,Nonperturbative_Wilson_RG_Disorder,Nonperturbative_Wilson_RG, Einstein_Klein_Gordon_RG_Kim,Einstein_Dirac_RG_Kim,RG_GR_Geometry_I_Kim,RG_GR_Geometry_II_Kim,Kondo_Holography_Kim,Kitaev_Entanglement_Entropy_Kim,RG_Holography_First_Kim}. This aspect serves as our fundamental motivation of the present theoretical framework.
			
			To figure out the role of the Wilsonian effective potential in the non-perturbative description, we neglect the effective potential part. Then, we obtain
			\begin{align}
				& Z[G^{(0)}_{\mu\nu}(x),B^{(0)}_{\mu\nu}(x),\Phi^{(0)}(x)] = \mathcal{J}\Big(\frac{\partial}{\partial G_{\mu\nu}^{(0)}(x)},\frac{\partial}{\partial B_{\mu\nu}^{(0)}(x)},\frac{\partial}{\partial \Phi^{(0)}(x)}\Big) \nn & \int D G_{\mu\nu}^{(1)}(x) D B_{\mu\nu}^{(1)}(x) D \Phi^{(1)}(x) ~ \mathcal{J}\Big(\frac{\partial}{\partial G_{\mu\nu}^{(1)}(x)},\frac{\partial}{\partial B_{\mu\nu}^{(1)}(x)},\frac{\partial}{\partial \Phi^{(1)}(x)}\Big) \nn & \delta\Bigg(\frac{G_{\mu\nu}^{(1)}(x) - G_{\mu\nu}^{(0)}(x)}{\Delta z} + \Delta z \Big(\alpha' R_{\mu\nu}^{(0)}(x) + 2 \alpha' \nabla_{\mu} \nabla_{\nu} \Phi^{(0)}(x) - \frac{\alpha'}{4} H_{\mu\chi\eta}^{(0)}(x) H_{\nu}^{\chi\eta (0)}(x)\Big)\Bigg) \nn & \delta\Bigg(\frac{B_{\mu\nu}^{(1)}(x) - B_{\mu\nu}^{(0)}(x)}{\Delta z} - \Delta z \Big(\frac{\alpha'}{2} \nabla^{\omega} H_{\omega\mu\nu}^{(0)}(x) - \alpha' \nabla^{\omega} \Phi^{(0)}(x) H_{\omega\mu\nu}^{(0)}(x) \Big)\Bigg) \nn & \delta\Bigg(\frac{\Phi^{(1)}(x) - \Phi^{(0)}(x)}{\Delta z} + \Delta z \Big(\frac{D - 26}{6} - \frac{\alpha'}{2} \nabla^{2} \Phi^{(0)}(x) + \alpha' \partial_{\mu} \Phi^{(0)}(x) \partial^{\mu} \Phi^{(0)}(x) - \frac{\alpha'}{24} H_{\mu\nu\lambda}^{(0)}(x) H^{\mu\nu\lambda (0)}(x)\Big)\Bigg) \nn & \int D x^{\mu}(\sigma) D b_{ab}(\sigma) D c^{a}(\sigma) \exp\Bigg[ - \frac{1}{4 \pi \alpha'} \int_{M} d^{2} \sigma \sqrt{g(\sigma)} \Bigg\{ g^{ab}(\sigma) \Bigg( G^{(1)}_{\mu\nu}(x) - \Delta z \Big(\alpha' R_{\mu\nu}^{(1)}(x) + 2 \alpha' \nabla_{\mu} \nabla_{\nu} \Phi^{(1)}(x) \nn & - \frac{\alpha'}{4} H_{\mu\chi\eta}^{(1)}(x) H_{\nu}^{\chi\eta (1)}(x)\Big) \Bigg) \partial_{a} x^{\mu}(\sigma) \partial_{b} x^{\nu}(\sigma) + i \epsilon^{ab} \Bigg( B^{(1)}_{\mu\nu}(x) + \Delta z \Big(\frac{\alpha'}{2} \nabla^{\omega} H_{\omega\mu\nu}^{(1)}(x) - \alpha' \nabla^{\omega} \Phi^{(1)}(x) H_{\omega\mu\nu}^{(1)}(x) \Big) \Bigg) \partial_{a} x^{\mu}(\sigma) \partial_{b} x^{\nu}(\sigma) \nn & + \alpha' R^{(2)}(\sigma) \Bigg( \Phi^{(1)}(x) - \Delta z \Big(\frac{D - 26}{6} - \frac{\alpha'}{2} \nabla^{2} \Phi^{(1)}(x) + \alpha' \partial_{\mu} \Phi^{(1)}(x) \partial^{\mu} \Phi^{(1)}(x) - \frac{\alpha'}{24} H_{\mu\nu\lambda}^{(1)}(x) H^{\mu\nu\lambda (1)}(x)\Big) \Bigg) \Bigg\} \nn & - \frac{1}{2 \pi} \int_{M} d^{2} \sigma \sqrt{g(\sigma)} b_{ab}(\sigma) \nabla^{a} c^{b}(\sigma) \Bigg] .
			\end{align} 
			Now, we can perform the path integral $\int D G_{\mu\nu}^{(1)}(x) D B_{\mu\nu}^{(1)}(x) D \Phi^{(1)}(x)$ to obtain
			\begin{align}
				& Z[G^{(0)}_{\mu\nu}(x),B^{(0)}_{\mu\nu}(x),\Phi^{(0)}(x)] = \mathcal{J}\Big(\frac{\partial}{\partial G_{\mu\nu}^{(0)}(x)},\frac{\partial}{\partial B_{\mu\nu}^{(0)}(x)},\frac{\partial}{\partial \Phi^{(0)}(x)}\Big) \nn & \int D x^{\mu}(\sigma) D b_{ab}(\sigma) D c^{a}(\sigma) \exp\Bigg[ - \frac{1}{4 \pi \alpha'} \int_{M} d^{2} \sigma \sqrt{g(\sigma)} \Bigg\{ g^{ab}(\sigma) \Bigg( G^{(0)}_{\mu\nu}(x) - \Delta z \Big(\alpha' R_{\mu\nu}^{(0)}(x) + 2 \alpha' \nabla_{\mu} \nabla_{\nu} \Phi^{(0)}(x) \nn & - \frac{\alpha'}{4} H_{\mu\chi\eta}^{(0)}(x) H_{\nu}^{\chi\eta (0)}(x)\Big) - \mathcal{O}\Big( (\Delta z)^{2} \Big) \Bigg) \partial_{a} x^{\mu}(\sigma) \partial_{b} x^{\nu}(\sigma) + i \epsilon^{ab} \Bigg( B^{(0)}_{\mu\nu}(x) + \Delta z \Big(\frac{\alpha'}{2} \nabla^{\omega} H_{\omega\mu\nu}^{(0)}(x) - \alpha' \nabla^{\omega} \Phi^{(0)}(x) H_{\omega\mu\nu}^{(0)}(x) \Big) \nn & + \mathcal{O}\Big( (\Delta z)^{2} \Big) \Bigg) \partial_{a} x^{\mu}(\sigma) \partial_{b} x^{\nu}(\sigma) + \alpha' R^{(2)}(\sigma) \Bigg( \Phi^{(0)}(x) - \Delta z \Big(\frac{D - 26}{6} - \frac{\alpha'}{2} \nabla^{2} \Phi^{(0)}(x) + \alpha' \partial_{\mu} \Phi^{(0)}(x) \partial^{\mu} \Phi^{(0)}(x) \nn & - \frac{\alpha'}{24} H_{\mu\nu\lambda}^{(0)}(x) H^{\mu\nu\lambda (0)}(x)\Big) - \mathcal{O}\Big( (\Delta z)^{2} \Big) \Bigg) \Bigg\} - \frac{1}{2 \pi} \int_{M} d^{2} \sigma \sqrt{g(\sigma)} b_{ab}(\sigma) \nabla^{a} c^{b}(\sigma) \Bigg] .
			\end{align} 
			This is completely the same as the 1-loop RG result. The $\mathcal{O}\Big( (\Delta z)^{2} \Big)$ terms arise from the 1-loop RG $\beta-$functions when we try to solve the first order differential equations in an iterative way. In this respect introduction of the Wilsonian effective potential is crucial for the description of non-perturbative physics.

			\section{Stochastic thermodynamics in the Langevin system: a brief review}\label{Review:Langevin}
			
			%
			%

			Here we review stochastic thermodynamics 
			of Brownian motion which is also incorporated within supersymmetric field theory method \cite{MSR_Formulation_SUSY_i,MSR_Formulation_SUSY_ii,MSR_Formulation_SUSY_iii,MSR_Formulation_SUSY_iv,MSR_Formulation_SUSY_v,MSR_Formulation_SUSY_vi}.
			We introduce the Langevin equation,
			\begin{align}
				\partial_{t} x(t) = \mu F(x(t),\lambda(t)) + \xi(t) , \label{Langevin_Equation} 
			\end{align}
			describing the overdamped dynamics of Brownian motion. Here, 
			\begin{equation}
				F(x(t),\lambda(t)) = - \partial_{x} V(x(t),\lambda(t)) + f(x(t), \lambda(t))    
			\end{equation}
			is the force, where $V(x(t),\lambda(t))$ is a conservative potential and $f(x(t),\lambda(t))$ is an external force. 
			These force sources may be time-dependent through an external control parameter $\lambda(t)$ varied according to some prescribed experimental protocol from $\lambda(0) = \lambda_{0}$ to $\lambda(t_{f}) = \lambda_{f}$. $\mu$ is the mobility of the particle. $\xi(t)$ serves as stochastic increments modeled as Gaussian white noise,
			\begin{align} 
				\langle \xi(t) \xi(t') \rangle = 2 D \delta(t-t') , \label{White_Noise} 
			\end{align}
			where $D$ is the diffusion constant, given by the Einstein relation $D = \beta^{-1} \mu$ at temperature $T = \beta^{-1}$ in equilibrium. 
			
			%
			%
			
			One can reformulate this Langevin equation in the path integral representation, introducing a generating functional analogous to the partition function in equilibrium. To translate the equation of motion into a generating functional, we consider the following identity 
			\begin{align} 1 = \int_{x_{i}}^{x_{f}} D x(t) \delta\Big(\partial_{t} x(t) - \mu F(x(t),\lambda(t)) - \xi(t)\Big) \mbox{det}\Big(\partial_{t} - \mu \partial_{x} F(x(t),\lambda(t))\Big) , 
			\end{align}
			and implement the Faddeev-Popov procedure \cite{QFT_textbook}. $\mbox{det}\Big(\partial_{t} - \mu \partial_{x} F(x(t),\lambda(t))\Big)$ is a Jacobian factor to describe the change of an integration measure. This is nothing but the identity
			$1 = \int_{x_{i}}^{x_{f}} d x \delta(f(x)) f'(x)$, where $f'(x) = \frac{d f(x)}{d x}$. Then, the generating functional is given by 
			\begin{align}
				\mathcal{W}(x_{f},t_{f};x_{i},t_{i}) &= \mathcal{N} \int D \xi(t) \exp\Big( - \frac{1}{4 D} \int_{t_{i}}^{t_{f}} d t \xi^{2}(t) \Big) \nn & \int_{x_{i}}^{x_{f}} D x(t) \delta\Big(\partial_{t} x(t) - \mu F(x(t),\lambda(t)) - \xi(t)\Big) \mbox{det}\Big(\partial_{t} - \mu \partial_{x} F(x(t),\lambda(t))\Big) .  \label{Generating_Functional_Faddeev_Popov}  
			\end{align}
			Here, $\mathcal{N}$ is a normalization constant to reproduce Eq.\ \eqref{White_Noise}.
			
			Introducing a Lagrange multiplier field $p(t)$ and a ghost field $c(t)$, one can exponentiate this expression as follows
			\begin{align}
				\mathcal{W}(x_{f},t_{f};x_{i},t_{i}) & = \mathcal{N} \int_{x_{i}}^{x_{f}} D x(t) D p(t) D c(t) D \bar{c}(t) \int D \xi(t) \exp\Big( - \frac{1}{4 D} \int_{t_{i}}^{t_{f}} d t \xi^{2}(t) \Big) \nn & \exp\Big[- \int_{t_{i}}^{t_{f}} d t \Big\{ i p(t) \Big( \partial_{t} x(t) - \mu F(x(t),\lambda(t)) - \xi(t) \Big) + \bar{c}(t) \Big(\partial_{t} - \mu \partial_{x} F(x(t),\lambda(t))\Big) c(t) \Big\} \Big] . \label{Generating_Functional_Noise}  
			\end{align}
			Here, the Lagrange multiplier $p(t)$ may be regarded as the canonical momentum of the position $x(t)$. The ghost $c(t)$ is a fermionic variable while $\bar{c}(t)$ is its canonical conjugate partner which on integration reproduces the Jacobian factor in Eq. (\ref{Generating_Functional_Faddeev_Popov}). Carrying out the Gaussian integral for random noise fluctuations, we obtain an effective `partition function' as follows \cite{MSR_Formulation_SUSY_i,MSR_Formulation_SUSY_ii,MSR_Formulation_SUSY_iii,MSR_Formulation_SUSY_iv,MSR_Formulation_SUSY_v,MSR_Formulation_SUSY_vi} 
			\begin{align}
				\mathcal{W}(x_{f},t_{f};x_{i},t_{i}) 
				& = \mathcal{N} \int_{x_{i}}^{x_{f}} D x(t) D p(t) D c(t) D \bar{c}(t) \exp\Big[- \int_{t_{i}}^{t_{f}} d t \Big\{ i p(t) \Big( \partial_{t} x(t) - \mu F(x(t),\lambda(t)) \Big) + D p^{2}(t) 
				\nn & + \bar{c}(t) \Big(\partial_{t} - \mu \partial_{x} F(x(t),\lambda(t))\Big) c(t) \Big\} \Big] . \label{Generating_Functional_Langevin}    
			\end{align}

			%
			%
			
			Seifert investigated the entropy production in this overdamped Langevin system. To show the monotonicity of entropy, ref.\ \cite{Entropy_Production} considered a probability distribution function as follows
			\begin{align} & p(x,t) \equiv \langle \delta(x-x(t)) \rangle \equiv \mathcal{N} \int D \xi(t') \exp\Big( - \frac{1}{4 D} \int_{t_{i}}^{t} d t' \xi^{2}(t') \Big) \delta(x-x(t)) . \end{align} 
			It is not difficult to prove that this probability distribution function satisfies the Fokker-Planck equation. Here, we present our intuitive derivation of the Fokker-Planck equation.
			
			The generating functional 
			Eq.\ \eqref{Generating_Functional_Langevin} 
			indicates that the corresponding effective Lagrangian is given by
			\begin{align} & \mathcal{L}_{{\it eff}} = i p(t) \Big( \partial_{t} x(t) - \mu F(x(t),\lambda(t)) \Big) + D p^{2}(t) . \end{align}
			Recalling the Legendre transformation, 
			\begin{align} & \mathcal{L}_{{\it eff}} = i p(t) \partial_{t} x(t) - \mathcal{H}_{eff} , \end{align}
			we obtain the following effective Hamiltonian,
			\begin{align} & \mathcal{H}_{{\it eff}} = i \mu p(t) F(x(t),\lambda(t)) - D p^{2}(t) . \end{align}
			Introducing $p = i \partial_{x}$ into this effective Hamiltonian, we obtain
			\begin{align} & \mathcal{H}_{{\it eff} } = - \mu \partial_{x} F(x,\lambda) + D \partial_{x}^{2} . \end{align}
			Considering $\partial_{t} p(x,t) = \mathcal{H}_{{\it eff}} p(x,t)$ in the overdamped system, we construct a current conservation equation as follows
			\begin{align} \partial_{t} p(x,t) = - \partial_{x} j(x,t) = - \partial_{x} [ (\mu F(x,\lambda) - D \partial_{x}) p(x,t) ] , \end{align}
			where the initial condition is $p(x,0) = p_{0}(x) = \delta(x-x_{i})$. The conserved current is given by
			\begin{align} j(x,t) = (\mu F(x,\lambda) - D \partial_{x}) p(x,t) . \end{align} 
			In Eq. (\ref{Generating_Functional_Noise}), we obtain $- i \xi(t) = 2D p(t)$. Then, this conserved current may be interpreted as the velocity of the Brownian motion. In other words, we have $j(x,t) \sim p(x,t) \partial_{t} x(t)$ as expected.
			
			This conservation equation is identical to the Fokker-Planck equation. We benchmark this strategy in constructing an effective Fokker-Planck equation for the holographic dual effective field theory.

			%
			%
			
			It is not so difficult to show the formal path integral expression for the probability distribution function as follows
			\begin{align} 
				p(x,t;x_{i},t_{i}) &= \frac{\mathcal{N}}{\mathcal{W}} \int_{x_{i}}^{x} D x(t') D p(t') D \bar{c}(t') D c(t') \exp\Big[- \int_{t_{i}}^{t} d t' \Big\{ i p(t') \Big( \partial_{t'} x(t') - \mu F(x(t'),\lambda(t')) \Big) + D p^{2}(t') 
				\nn & + \bar{c}(t') \Big(\partial_{t'} - \mu \partial_{x} F(x(t'),\lambda(t'))\Big) c(t') \Big\} \Big] . \end{align}
			
			This probability distribution function essentially reproduces the path integral representation of the generating functional. One can verify
			\begin{align}  \int_{x_{i}}^{x_{f}} d x p(x,t) = 1 . \end{align} 
			
			Based on this probability distribution function, 
			ref.\ \cite{Entropy_Production} introduced a path-dependent microscopic entropy functional as
			\begin{align}  s_{sys}(x,t) = - \ln p(x,t) , \end{align}
			referred to as the system entropy. The observed system entropy is given by the ensemble average with respect to the probability as follows
			\begin{align} & S_{sys}(t) = \langle s_{sys}(x,t) \rangle= - \int_{x_{i}}^{x_{f}} d x p(x,t) \ln p(x,t) , \end{align}
			analogous to the Gibbs or Shannon entropy. This microscopic definition of the entropy reproduces the macroscopic thermodynamic entropy at equilibrium (fixed $\lambda$) in the following way
			\begin{align} s_{sys}(x,t) = \beta [V(x,\lambda) - \mathcal{F}(\lambda)], \end{align} 
			where the equilibrium free energy $\mathcal{F}(\lambda)$ is $\mathcal{F}(\lambda) = - \beta^{-1} \ln \int_{x_{i}}^{x_{f}} d x e^{- \beta V(x,\lambda)}$ with the conserved potential $V(x,\lambda)$ introduced before. 
			
			It is natural to consider the rate of heat dissipation in the environment as 
			\begin{align} & \partial_{t} q(x,t) = F(x,\lambda) \partial_{t} x(t) = \beta^{-1} \partial_{t} s_{env}(x,t) . \end{align}
			We recall that $F(x,\lambda)$ is the force. Accordingly, one may identify the exchanged heat with an increase in the environment entropy $s_{env}(x,t)$ at temperature $\beta^{-1} = D / \mu$.
			
			Combining these two contributions, ref. \cite{Entropy_Production} found the trajectory-dependent total entropy production rate as follows
			\begin{align} & \partial_{t} s_{tot}(x,t) = \partial_{t} s_{env}(x,t) + \partial_{t} s_{sys}(x,t) = \frac{\partial_{x} j(x,t)}{p(x,t)} + \frac{j(x,t)}{D p(x,t)} \partial_{t} x(t) . \end{align}
			We emphasize that this total entropy differs from the thermodynamic entropy while the system entropy corresponds to it. 
			Taking the ensemble average, 
			ref.\ \cite{Entropy_Production} showed that the averaged total entropy production rate is always positive, given by
			\begin{align} \partial_{t} S_{tot}(t) = \langle \partial_{t} s_{tot}(x,t) \rangle = \int_{x_{i}}^{x_{f}} d x \frac{j^{2}(x,t)}{D p(x,t)} \geq 0 , \end{align} 
			where the equality holds in equilibrium only. The ensemble-averaged entropy production rate of the environment is given by
			\begin{align} \partial_{t} S_{env}(x,t) = \langle \partial_{t} s_{env}(x,t) \rangle = \beta \int_{x_{i}}^{x_{f}} d x F(x,t) j(x,t) , \end{align}
			where the force $F(x,t)$ and the conserved current $j(x,t)$ have been introduced above. Following this procedure, we define the entropy production involved with the RG flow in the holographic dual effective field theory and show its monotonicity along this line of thought \cite{RG_Flow_Holography_Monotonicity}.

			\section{Monotonicity of the RG flow in the worldsheet NLSM}\label{mono}

			%
			%
			Here we review the monotonicity of the RG flow for the target space metric $G_{\mu\nu}(x)$ in a NLSM following \cite{Ricci_NLsM_Gradient_i,Ricci_NLsM_Gradient_ii}.
			We introduce the worldsheet nonlinear $\sigma$ model to describe the dynamics of a bosonic string \cite{Polchinski},
			\begin{align} I = \frac{1}{4 \pi \alpha'} \int_{M} d^{2} \sigma \sqrt{g(\sigma)} \Big( g^{ab}(\sigma) \partial_{a} x^{\mu}(\sigma) \partial_{b} x^{\nu}(\sigma) G_{\mu\nu}(x) + \alpha' R^{(2)}(\sigma) \phi(x) \Big) . \label{Classical_String_Action} \end{align}
			Here, $\sigma = (\sigma_{1}, \sigma_{2})$ is a two-dimensional worldsheet coordinate, and the subscript $M$ of the integral is a worldsheet manifold. $g^{ab}(\sigma)$ is a worldsheet metric with $a, b = 1, 2$, and $\sqrt{g(\sigma)}$ is its determinant. $x^{\mu}(\sigma)$ describes the dynamics of a bosonic string, which maps the two-dimensional worldsheet manifold to an emergent $D$-dimensional target spacetime. Here, $\mu, \nu$ span from $0$ to $D-1$, indexing the target spacetime coordinate. $G_{\mu\nu}(x)$ is the corresponding metric for the target spacetime. $R^{(2)}(\sigma)$ is the worldsheet Ricci scalar, and $\phi(x)$ is a dilaton field. $\alpha'$ is the string coupling constant in the worldsheet nonlinear $\sigma$ model.
			
			%
			%
			
			This worldsheet nonlinear $\sigma$ model has two types of gauge symmetries, diffeomorphism and Weyl transformations at the classical level. 
			To perform the path integral quantization a la Polyakov, one has to introduce gauge fixing, where a Jacobian factor is taken into account a la Faddeev–Popov \cite{Polchinski}. Considering the gauge symmetries, the path integral for the worldsheet metric is carried out and the infinite gauge volume is factored out. Furthermore, the Jacobian factor involved with the gauge fixing introduces an action of ghost fields into Eq.\ (\ref{Classical_String_Action}), where the worldsheet metric is gauge-fixed. The resulting effective action has the BRST symmetry only when the dimension of the target spacetime is $D = 26$ \cite{Polchinski}.
			
			One may perform the path integral with respect to 
			the string field in a perturbative way. 
			Carrying out the RG transformation at one-loop level, one finds the RG flow equations for the target spacetime metric and the dilaton field, respectively, as follows
			\begin{align} 
				\frac{d G_{\mu\nu}(x)}{d t} = - \beta_{\mu\nu}^{G}[G_{\alpha\beta}(x),\phi(x)],
				\quad \frac{d \phi(x)}{d t} = - \beta_{\phi}[\phi(x),G_{\alpha\beta}(x)]. \label{RG_Flows} \end{align}
			Here, $t$ is an RG scale. The one-loop RG $\beta$ functions are given by \cite{Polchinski}
			\begin{align} 
				\beta_{\mu\nu}^{G}[G_{\alpha\beta}(x),\phi(x)] & = \alpha' \Big( R_{\mu\nu}(x) + 2 \nabla_{\mu} \nabla_{\nu} \phi(x) \Big) , 
				\\ {\beta}_{\phi}[\phi(x),G_{\alpha\beta}(x)] & = c_{0} - \alpha' \Big( \frac{1}{2} \nabla^{2} \phi(x) - \partial^{\mu} \phi(x) \partial_{\mu} \phi(x) \Big) , 
			\end{align}
			respectively. $c_{0} = \frac{D - 26}{6}$ is the central charge, which vanishes at $D = 26$.
			
			%
			%
			
			The existence of RG flows indicates that the bosonic string theory is not conformal at the one-loop level. In other words, there must be a conformal anomaly, given by \cite{Polchinski}
			\begin{align} & 2 \pi \alpha' T_{a}^{a}(\sigma) = - \partial_{a} x^{\mu}(\sigma) \partial^{a} x^{\nu}(\sigma) \beta_{\mu\nu}^{G}[G_{\alpha\beta}(x),\phi(x)] - \alpha' R^{(2)}(\sigma) \beta_{\phi}[\phi(x),G_{\alpha\beta}(x)] . \end{align} 
			This trace anomaly is cancelled only when both RG $\beta$ functions vanish, i.e., $\beta_{\mu\nu}^{G}[G_{\alpha\beta}(x),\phi(x)] = 0$ and ${\beta}_{\phi}[\phi(x),G_{\alpha\beta}(x)] = 0$. As a result, we obtain
			%
			%
			\begin{align} R_{\mu\nu}(x) + 2 \nabla_{\mu} \nabla_{\nu} \phi(x) & = 0 , \label{RG_Flow_Metric}
				\\ \frac{1}{2} \nabla^{2} \phi(x) - \partial^{\mu} \phi(x) \partial_{\mu} \phi(x) & = 0 , \end{align}
			which we interpret to be semi-classical equations of motion for the target spacetime metric and the dilaton field. It is not difficult to construct the following low-energy effective action \cite{Polchinski}
			\begin{align} & S_{eff} = \frac{\alpha'}{4} \int d^{26} x \sqrt{G(x)} e^{- 2 \phi(x)} \Big( R(x) + 4 \partial^{\mu} \phi(x) \partial_{\mu} \phi(x) \Big) , \label{26D_LEFT} \end{align} 
			extremizing which with respect to $G_{\mu\nu}(x)$ and $\phi(x)$ gives rise to these semi-classical equations of motion.

			Unfortunately, the low-energy effective action 
			(\ref{26D_LEFT}) cannot describe the monotonicity of the RG flow equation (\ref{RG_Flows})
			for the target spacetime metric $G_{\mu\nu}(x)$ because the resulting semi-classical equation of motion for $G_{\mu\nu}(x)$,
			Eq.\ (\ref{RG_Flow_Metric}),
			coincides with the vanishing RG $\beta$ function condition. In this respect refs. \cite{Ricci_NLsM_Gradient_i,Ricci_NLsM_Gradient_ii} introduced the following generating functional
			\begin{align} 
				& S_{eff} = \frac{\alpha'}{4} \int d^{26} x \sqrt{G(x)} e^{- 2 \phi(x)} \Big( R(x) + 4 \nabla^{2} \phi(x) - 4 \partial^{\mu} \phi(x) \partial_{\mu} \phi(x) \Big) \nn & \qquad + \lambda \Big( \int d^{26} x \sqrt{G(x)} e^{- 2 \phi(x)} - 1 \Big) . \label{Perelman_W_Entropy} \end{align} 
			The main difference between Eq.\ (\ref{26D_LEFT}) and Eq.\ (\ref{Perelman_W_Entropy}) is that Eq.\ (\ref{Perelman_W_Entropy}) takes into account a `generalized' volume-preserving constraint,
			\begin{align} \int d^{26} x \sqrt{G(x)} e^{- 2 \phi(x)} = 1 , \end{align}
			introduced by the Lagrange multiplier $\lambda$. Depending on the gauge choice, the first action part can be changed.
			Indeed, the first part in ref. \cite{Ricci_NLsM_Gradient_i} is different from that in ref. \cite{Ricci_NLsM_Gradient_ii}.
			
			Eq.\ (\ref{Perelman_W_Entropy}) is sometimes referred to as the Perelman's $F$-functional \cite{Ricci_Flow_Monotonicity}. However, it is not clear whether or not Eq.\ (\ref{Perelman_W_Entropy}) can be really related with the thermodynamic entropy given by the Legendre transformation of the free energy, for example, the low-energy effective action Eq. (\ref{26D_LEFT}). 
			See refs.\ \cite{Yu_Nakamura_Ricci_Flow,Gibbs_Entropy_for_Perelman_I,Gibbs_Entropy_for_Perelman_II,Gibbs_Entropy_for_Perelman_III} for related discussions.
			
			Extremizing Eq. (\ref{Perelman_W_Entropy}) with respect to the dilaton field, i.e., $\frac{\delta S_{eff}}{\delta \phi(x)} = 0$, one obtains
			%
			%
			\begin{align} & \frac{1}{4} \alpha' \Big( R(x) + 4 \nabla^{2} \phi(x) - 4 \partial^{\mu} \phi(x) \partial_{\mu} \phi(x) \Big) + \lambda = 0 . \label{Perelman_Dilaton_Equation} \end{align}
			As a result, the saddle-point value of $S_{eff}$ in Eq. (\ref{Perelman_W_Entropy}) is given by
			%
			%
			\begin{align} S_{eff} = - \lambda \end{align}
			under this extremization condition. 
			
			The next procedure is to extremize $S_{eff} = - \lambda$ under the volume-preserving constraint. Rewriting $\phi(x)$ with $\Phi(x) = e^{- \phi(x)}$ in Eq. (\ref{Perelman_Dilaton_Equation}), one obtains
			\begin{align} \Big( - \nabla^{2} + \frac{1}{4} R(x) \Big) \Phi(x) = - \frac{\lambda}{\alpha'} \Phi(x) , \label{Dilaton_Kernel} \end{align}
			where the volume-preserving constraint is given by
			\begin{align} \int d^{26} x \sqrt{G(x)} \Phi^{2}(x) = 1 . \end{align}
			As a result, extremizing $S_{eff}$ in $\phi(x)$ (Eq. (\ref{Perelman_W_Entropy})) translates into finding $- \lambda$ as the lowest eigenvalue of the Laplacian $- \nabla^{2} + \frac{1}{4} R(x)$. Indeed, the existence of a solution of this equation with $\Phi(x) = e^{- \phi(x)} > 0$ requires that $- \lambda$ is the lowest eigenvalue, which always exists on a compact space \cite{Ricci_Flow_Monotonicity}. The corresponding eigenfunction will not have any zeros and can be chosen positive, which is required to identify the eigenfunction with $e^{- \phi(x)}$ \cite{Ricci_NLsM_Gradient_ii}. 
			
			%
			%
			
			Now, we define the lowest eigenvalue of $- \lambda$ as $\mathcal{S}[G_{\alpha\beta}(x)]$ referred to as the Perelman's entropy functional. Then, it is straightforward to show that the RG $\beta$ function of the metric tensor is given by a gradient of the entropy functional,
			\begin{align} \beta_{\mu\nu}^{G}[G_{\alpha\beta}(x),\phi(x)] = \kappa_{\mu\nu,\rho\sigma}[G_{\alpha\beta}(x),\phi(x)] \frac{\delta \mathcal{S}[G_{\alpha\beta}(x)]}{\delta G_{\rho\sigma}(x)} . \label{Gradient_Flow} \end{align} 
			Here, we have
			\begin{align} \kappa_{\mu\nu,\rho\sigma}[G_{\alpha\beta}(x),\phi(x)] = \frac{4 G_{\mu\rho}(x) G_{\nu\sigma}(x)}{\sqrt{G(x)} e^{-2 \phi(x)}} . \end{align}
			This confirms that the RG flow of the target spacetime metric is a gradient flow with positive definite metric.
			
			To show the monotonicity of the RG flow of the metric tensor, one may consider the entropy production rate, given by
			\begin{align} & \frac{d}{d t} \mathcal{S}= - \beta_{\mu\nu}^{G}[G_{\alpha\beta}(x),\phi(x)] \frac{\delta \mathcal{S}}{\delta G_{\mu\nu}} \nn &\qquad= - \beta_{\mu\nu}^{G}[G_{\alpha\beta}(x),\phi(x)] \kappa^{\mu\nu, \rho\sigma}[G_{\alpha\beta}(x),\phi(x)] \beta_{\rho\sigma}^{G}[G_{\alpha\beta}(x),\phi(x)] \leq 0 . \label{Entropy_Production} \end{align}
			Here, the first equality results from the application of the chain rule, and the second one comes from Eq. (\ref{Gradient_Flow}). The inverse of $\kappa_{\mu\nu,\rho\sigma}[G_{\alpha\beta}(x),\phi(x)]$ is given by
			%
			%
			\begin{align} & \kappa^{\mu\nu,\rho\sigma}[G_{\alpha\beta}(x),\phi(x)] = \frac{1}{4} \sqrt{G(x)} e^{- 2 \phi(x)} G^{\mu\rho}(x) G^{\nu\sigma}(x) , \label{Positive_Definite_Metric} \end{align}
			which is always positive definite. 
			Eq.\ \eqref{Entropy_Production} 
			with Eq.\ \eqref{Positive_Definite_Metric}
			confirms that the RG flow of the target spacetime metric is monotonic.

\end{document}